%% file: main.tex
\documentclass{article}

\usepackage{arxiv}
\usepackage{microtype}      % microtypography
\usepackage{booktabs}       % professional-quality tables

\usepackage{amssymb}
\usepackage{latexsym}
\usepackage[T1]{fontenc}    % use 8-bit T1 fonts

\usepackage{subfig}
\usepackage{placeins}
\usepackage{url}
\usepackage{xcolor}
\definecolor{newcolor}{rgb}{.8,.349,.1}
\usepackage[utf8]{inputenc}
\usepackage{pgf}
\usepackage{diagbox}
\usepackage{amsmath}
\usepackage{bm}
\usepackage{hyperref}
\usepackage{color}
\usepackage{listings}
\usepackage{tikz}
\usepackage{graphicx}
\usepackage{amssymb,amsmath,amsthm}
\usepackage{tikz}
\usepackage{appendix}
\usepackage{framed,multirow}
\usepackage[boxed,linesnumbered]{algorithm2e}
\usetikzlibrary{matrix,chains,positioning,decorations.pathreplacing,arrows}
\usetikzlibrary{positioning,calc}

\newtheorem{theorem}{Remark}

\newcommand{\realspace}{{\rm I\!R}}
\makeatletter
\renewcommand{\@algocf@capt@boxed}{above}% formerly {bottom}
\makeatother

\author{ \href{https://pierrejacquier.com}{Pierre Jacquier} \\
	Department of Mechanical Engineering\\
	École de Technologie Supérieure\\
	Montréal, QC H3C 1K3, Canada  \\
	\texttt{pierre.jacquier.1@ens.etsmtl.ca} \\
	\And
	Azzedine Abdedou \\
	Department of Mechanical Engineering\\
	École de Technologie Supérieure\\
	Montréal, QC H3C 1K3, Canada  \\
	\texttt{azzedine.abdedou.1@ens.etsmtl.ca} \\
	\And
	Vincent Delmas \\
	Department of Mechanical Engineering\\
	École de Technologie Supérieure\\
	Montréal, QC H3C 1K3, Canada  \\
	\texttt{vincent.delmas.1@ens.etsmtl.ca} \\
	\And
	\href{https://www.etsmtl.ca/en/research/professors/asoulaimani/}{Azzeddine Soulaïmani}\thanks{Corresponding author}  \\
	Department of Mechanical Engineering\\
	École de Technologie Supérieure\\
	Montréal, QC H3C 1K3, Canada  \\
	\texttt{azzeddine.soulaimani@etsmtl.ca} \\
}

\renewcommand{\headeright}{Preprint}
\renewcommand{\undertitle}{Preprint}

\title{Non-Intrusive Reduced-Order Modeling Using Uncertainty-Aware Deep Neural Networks and Proper Orthogonal Decomposition: Application to Flood Modeling}

\begin{document}

\maketitle

\begin{abstract}
  \input{content/abstract}
\end{abstract}


\input{content/intro}
\input{content/reducingorder}
\input{content/learning}
\input{content/podbnn}

\input{content/benchmarks-uq}

\input{content/applications}
\clearpage
\input{content/wrapup}

\clearpage
\section*{Acknowledgments}
This research was enabled in part by funding from the Natural Sciences and Engineering Research Council of Canada and Hydro-Québec, by bathymetry data from the \href{https://cmm.qc.ca/}{Montreal Metropolitan Community} (Communauté métropolitaine de Montréal), and by computational support from \href{www.calculquebec.ca}{Calcul Québec} and \href{www.computecanada.ca}{Compute Canada}.
\bibliographystyle{unsrt}
\bibliography{refs/library}

\appendix
\section{Low-level details and results for each use case}\label{apx:table}
% \usepackage{multirow}
\begin{table}[h!]
\begin{tabular}{ll|c|c|c|c|c|c|c|}
\cline{3-9}
                                               &       & Architecture      & $N_e$     & $\tau$  & $\lambda$ & $\zeta$   & $\kappa$ & $\bm{\pi}$        \\ \hline
\multicolumn{1}{|l|}{\multirow{2}{*}{2d\_ack}} & EnsNN & $[128, 128, 128]$ & $120,000$  & $0.001$ & $0.005$ & $0$       & $0.01$   & N/A               \\ \cline{2-9} 
\multicolumn{1}{|l|}{}                         & BNN   & $[40, 40, 40]$    & $120,000$ & $0.01$  & N/A       & $0$       & $0.01$   & $[0.5, 4.0, 0.1]$ \\ \hline
\multicolumn{1}{|l|}{\multirow{2}{*}{1dt\_sw}} & EnsNN & $[256, 256, 256]$ & $100,000$ & $0.005$ & $10^{-4}$ & $0.001$   & $1.00$   & N/A               \\ \cline{2-9} 
\multicolumn{1}{|l|}{}                         & BNN   & $[256, 256, 256]$ & $70,000$  & $0.01$  & N/A       & $0.001$   & $0.01$   & $[0.5, 0.2, 0.1]$ \\ \hline
\multicolumn{1}{|l|}{\multirow{2}{*}{2d\_sw}}  & EnsNN & $[128, 128, 128]$ & $120,000$ & $0.03$  & $10^{-8}$ & $0$       & $1.00$   & N/A               \\ \cline{2-9} 
\multicolumn{1}{|l|}{}                         & BNN   & $[40, 40, 40]$    & $300,000$ & $0.01$  & N/A       & $0$       & $0.01$   & $[0.5, 4.0, 0.1]$ \\ \hline
\multicolumn{1}{|l|}{\multirow{2}{*}{2dt\_sw}} & EnsNN & $[128, 128, 128]$ & $70,000$  & $0.01$  & $0.001$ & $0.001$   & $0.01$   & N/A               \\ \cline{2-9} 
\multicolumn{1}{|l|}{}                         & BNN   & $[128, 128, 128]$ & $150,000$ & $0.003$ & N/A       & $10^{-5}$ & $0.01$   & $[0.5, 0.2, 0.1]$ \\ \hline
\end{tabular}
\end{table}
\noindent Documented code source for these experiments and others at \href{https://github.com/pierremtb/POD-UQNN}{https://github.com/pierremtb/POD-UQNN}.

\end{document}

%% file: content/abstract.tex
Deep Learning research is advancing at a fantastic rate, and there is much to gain from transferring this knowledge to older fields like Computational Fluid Dynamics in practical engineering contexts. This work compares state-of-the-art methods that address uncertainty quantification in Deep Neural Networks, pushing forward the reduced-order modeling approach of Proper Orthogonal Decomposition-Neural Networks (POD-NN) with Deep Ensembles and Variational Inference-based Bayesian Neural Networks on two-dimensional problems in space. These are first tested on benchmark problems, and then applied to a real-life application: flooding predictions in the Mille Îles river in the Montreal, Quebec, Canada metropolitan area. Our setup involves a set of input parameters, with a potentially noisy distribution, and accumulates the simulation data resulting from these parameters. The goal is to build a non-intrusive surrogate model that is able to know when it {does not} know, which is still an open research area in Neural Networks (and in AI in general). With the help of this model, probabilistic flooding maps are generated, aware of the model uncertainty. These insights on the unknown are also utilized for an uncertainty propagation task, allowing for flooded area predictions that are broader and safer than those made with a regular uncertainty-uninformed surrogate model. Our study of the time-dependent and highly nonlinear case of a dam break is also presented. Both the ensembles and the Bayesian approach lead to reliable results for multiple smooth physical solutions, providing the correct warning when going out-of-distribution. However, the former, referred to as POD-EnsNN, proved much easier to implement and showed greater flexibility than the latter in the case of discontinuities, where standard algorithms may oscillate or fail to converge.

%% file: content/intro.tex
\section{Introduction}
Machine Learning and other forms of Artificial Intelligence (AI) have been at the epicenter of massive breakthroughs in the notoriously difficult fields of computer vision, language modeling and content generation, as presented in \cite{szegedy2017inception}, \cite{Mikolov2013word2vec}, and \cite{karras2019analyzing}. Still, there are many other domains where robust and well-tested methods could be significantly improved by the modern computational tools associated with AI: antibiotic discovery is just one very recent example \cite{Stokes2020}. In the realm of high-fidelity computational mechanics, simulation time is tightly linked to the size of the mesh and the number of time-steps; in other words, to its accuracy, which could make it impractical to be used in real-time contexts for new parameters.  \par
Much research has been performed to address this large-size problem and to create Reduced-Ordered Models (ROM) that can effectively replace their heavier counterpart for tasks like design and optimization, or for real-time predictions. The most common way to build a ROM is to go through a compression phase into a \textit{reduced space}, defined by a set of \emph{reduced basis} (RB) {vectors}, which is at the root of many methods, according to \cite{Benner2015}. For the most part, RB methods involve an \textit{offline-online} paradigm, where the former is more computationally-heavy, and the latter should be fast enough to allow for real-time predictions. The idea is to collect data points called \emph{snapshots} from simulations, or any high-fidelity source, and extract the information that has the most significance on the dynamics of the system, the \textit{modes}, via a reduction method in the \textit{offline} stage. \par
Proper Orthogonal Decomposition (POD), as introduced in \cite{Holmes1997,sirovich1987turbulence}, coupled with the Singular Value Decomposition (SVD) algorithm \cite{Burkardt2006}, is by far the most popular method to reach a \textit{low-rank} approximation. Subsequently, the \textit{online} stage involves recovering the \textit{expansion coefficients}, projecting back into the uncompressed, real-life space. This recovery is where the separation between intrusive and non-intrusive methods appears, where the former use techniques based on the problem's formulation, such as the Galerkin procedure \cite{Couplet2005, zok2012cmame, zokagoa}. At the same time, the latter (non-intrusive methods) try to statistically infer the mapping by considering the snapshots as a dataset. In this non-intrusive context, the POD-NN framework proposed by \cite{Hestaven2018} and extended for time-dependent problems in \cite{Wang2019} aims at training an artificial Neural Network to perform the mapping. These time-dependent problems can also benefit from approaching the POD on a temporal subdomain level, which has proved useful to prevent long-term error propagation, as first detailed in \cite{Ijzerman2000} and {later assessed} in \cite{zokagoa}. \par
Conventionally, laws of physics are expressed as well-defined Partial Differential Equations (PDEs), with boundary/initial conditions as constraints. Still, lately, pure data-driven methods have led to new approaches in PDE discovery \cite{Brunton2016DiscoveringGov}. The explosive growth of this new field of Deep Learning in Computational Fluid Dynamics was predicted in \cite{Kutz2017}. Its flexibility allows for multiple applications, such as the recovery of missing CFD data \cite{Carlberg2019}, or aerodynamic design optimization \cite{Tao2019}. The cost associated with a fine mesh is high, but this has been overcome with a Machine Learning (ML) approach aimed at assessing errors and correcting quantities in a coarser setting \cite{Hanna2020}. New research in the field of numerical schemes was performed in \cite{Despres2020}, presenting the Volume of Fluid-Machine Learning (VOF-ML) approach applied in bi-material settings. A review of the vast landscape of possibilities is {offered} in \cite{Brunton2019}. The constraints of \emph{small data} also led researchers to try to balance the need for data in AI contexts with expert knowledge, as with governing equations. Presented in \cite{owhadi2015bayesian} and \cite{Raissi2017MLODE}, this was then extended to neural networks in \cite{Raissi2019PINNs} with applications in Computational Fluid Dynamics, as well as in vibration analysis \cite{Raissi2019DeepVIV}.
When modeling data organized in sequence, Recurrent Neural Networks \cite{rumelhart1985learning} are often predominant, especially the Long Short Term Memory (LSTM) variant \cite{lstm1997}. LSTM neural networks have recently been applied in the context of time-dependent flooding prediction in \cite{Hu2019}, with the promise of providing real-time results. A recent contribution by \cite{McDermott2019} even allows for an embedded Bayesian treatment. Finally, an older but thorough study of available Machine Learning methods applied to environmental sciences and hydrology is presented in \cite{williamhsieh2009}.
\par 
While their regression power is impressive, Deep Neural Networks are still, in their standard state, only able to predict a mean value, and do not provide any guidance on how much trust one can put into that value. To address this, recent additions to the Machine Learning landscape include Deep Ensembles \cite{lakshminarayanan2017simple} which suggest the training of an ensemble of specific, variance-informed deep neural networks, to obtain a complete uncertainty treatment. That work was subsequently extended to sub-ensembles for faster implementation {in }\cite{Valdenegro-Toro2019} and later reviewed in \cite{snoek2019trust}. Earlier, other works had successfully encompassed the Bayesian view of probabilities within a Deep Neural Network, with the work of \cite{Mackay1995}, \cite{Barber1998}, \cite{NIPS2011_4329}, \cite{pmlr-v37-hernandez-lobatoc15} ultimately leading to the backpropagation-compatible Bayesian Neural Networks defined in \cite{Blundell2015}, making use of Variational Inference \cite{Hinton1993varinf}, and paving the way for trainable Bayesian Neural Networks, also reviewed in \cite{snoek2019trust}. {Notable applications are surrogate modeling for inverse problems \cite{reviewer1-a}, and Bayesian physics-informed neural networks \cite{reviewer1-b}.} \par
In this work, we aim at transferring the recent breakthroughs in Deep Learning to Computational Fluid Dynamics, by extending the concept of POD-NN with state-of-the-art methods for uncertainty quantification in Deep Neural Networks. After setting up the POD approach in Section \ref{sec:podensnn-reducing}, the methodologies of Deep Ensembles and Variational Inference for Bayesian Neural Networks are presented in Sections \ref{sec:podensnn-learning} and \ref{sec:podensnn-learning-bnn}, respectively. Their performances are assessed according to {a} benchmark in Section \ref{sec:podensnn-bench}.
Our context of interest, flood modeling, is addressed in Section \ref{sec:podensnn-applications}.
A dam break scenario is presented in Section \ref{sec:app-riemann}, first in a 1D Riemann analytically tractable example in order to obtain a reproducible problem in this context and to validate the numerical solver used in higher-dimension problems.
The primary engineering aim is the training of a model capable of producing probabilistic flooding maps of the river presented in Section \ref{sec:app-river-setup}, with its results reported in Section \ref{sec:app-river-results}. A contribution to standard uncertainty propagation is offered in \ref{sec:app-river-up}, while Section \ref{sec:app-river-damb} uses the same river environment for a fictitious dam break simulation.
% For these real-life application examples, the case of the Mille Îles river located in the Greater Montreal area is considered. \par
The Mille Îles river located in the Greater Montreal area is considered for these real-life application examples. We summarize our conclusions on this successful application of Deep Ensembles and Variational Inference for Bayesian Neural Networks in Section \ref{sec:wrapup}, along with our recommendations for the most promising future work in this area.

%% file: content/reducingorder.tex
\section{Reduced Basis Generation with Proper Orthogonal Decomposition}
\label{sec:podensnn-reducing}

\subsection{Objective and setup}
\par We start by defining $u$, our $\realspace^{D}$-valued function of interest
\begin{align}\label{eq:podensnn-u-base}
    u:\ \realspace^{n + P} &\rightarrow \realspace^{D} \\
                         (\bm{x}, \bm{s}) &\mapsto u(\bm{x}, \bm{s}), \nonumber
\end{align}
with $\bm{x} \in \realspace^{n}$ as the spatial parameters and $\bm{s} \in \realspace^{P}$ as the additional non-spatial parameters, for anything from a fluid viscosity to the time variable. \par
Computing this function is costly, so only a finite number $S$ of solutions called \textit{snapshots} can be realized. These are obtained over a discretized space, which can either be a uniform grid or an unstructured mesh, with $n$ representing the number of dimensions and $D$ {the number of components of the vector-valued function $u$}. $N_s$ is the number of non-spatial parameters sampled, and $N_t$ counts the considered time-steps, which would be higher than one in a time-dependent setting, leading the total number of snapshots to be $S=N_sN_t$. \par
In our applications, the spatial mesh of $N_D$ nodes is considered fixed in time, and since it is known and defined upfront, it can be incorporated in (\ref{eq:podensnn-u-base}), removing $\bm{x}$ as a parameter in $u$, and making $H=N_D\times D$ {be} the total number of degrees of freedom (DOFs) of the mesh 
\begin{align}\label{eq:podensnn-u-D}
    u_D:\ \realspace^{P} &\rightarrow \realspace^{H} \\
                         \bm{s} &\mapsto u_D(\bm{s}). \nonumber
\end{align}
\par The simulation data, obtained from computing the function $u$ with $S$ parameter sets $\bm{s}^{(i)}$, is stored in a matrix of snapshots $\bm{U} = [u_D(\bm{s}^{(1)})|\ldots|u_D(\bm{s}^{(S)})] \in \realspace^{H \times S}$.
% While this new regression objective $u_D$ allows us to have space taken care of, having an $\realspace^H$-valued output is not practical since it's size can quickly become a problem.  \par
Proper Orthogonal Decomposition (POD) is used to build a Reduced-Order Model (ROM) and produce a \textit{low-rank approximation}, which will be much more efficient to compute and use when rapid multi-query simulations are required. With the snapshots method \cite{sirovich1987turbulence}, a reduced POD basis can be efficiently extracted in a finite-dimension context.
In our case, we begin with the $\bm{U}$ matrix, and use the Singular Value Decomposition algorithm \cite{Burkardt2006} to extract $\bm{W} \in \realspace^{H \times H}$, $\bm{Z} \in \realspace^{S \times S}$ and the $r$ descending-ordered positive singular values matrix $\bm{D} = \text{diag}(\xi_1, \xi_2, \ldots, \xi_r)$ such that
\begin{equation}\label{eq:podensnn-svd}
    \bm{U} = \bm{W} \begin{bmatrix} \bm{D} & 0 \\ 0 & 0 \end{bmatrix} \bm{Z}^\intercal.
\end{equation}
\par For the finite truncation of the first $L$ modes, the following criterion on the singular values is imposed, with a hyperparameter $\epsilon$ given as
\begin{equation}\label{eq:podensnn-svd-criterion}
    \dfrac{\sum_{l=L+1}^{r} \xi_l^2}{\sum_{l=1}^{r} \xi_l^2} \leq \epsilon,
\end{equation}
and then each mode vector $\bm{V}_j \in \realspace^{S}$ can be found from $\bm{U}$ and the $j$-th column of $\bm{Z}$, $\bm{Z}_j$, with
\begin{equation}\label{eq:podensnn-pod-bases-i}
    \bm{V}_j = \dfrac{1}{\xi_j} \bm{U} \bm{Z}_j,
\end{equation}
so that we can finally construct our POD mode matrix 
\begin{equation}\label{eq:podensnn-svd-bases-group}
    \bm{V} = \left[\bm{V}_1 | \ldots | \bm{V}_j | \ldots | \bm{V}_L\right] \in \realspace^{H \times L}.
\end{equation}

\subsection{Projections}
{Projecting} to and from the low-rank approximation requires {some} projection coefficients; those \emph{corresponding} to the matrix of snapshots are obtained by the following
\begin{equation}\label{eq:podensnn-pod-red-basis}
    \bm{v} = \bm{V}^\intercal \bm{U},
\end{equation}
and then $\bm{U}_\textrm{POD}${,} the approximation of $\bm{U}$, can be projected back to the expanded space:
\begin{equation}\label{eq:podensnn-pod-exp-basis}
\bm{U}_\textrm{POD} = \bm{V}\bm{V}^\intercal\bm{U} = \bm{V} \bm{v}.
\end{equation}
The following relative projection error can be computed to assess the quality of the compression/expansion procedure,
% To assess the quality of the compression/expansion procedure, the relative projection error can be written as
\begin{equation}
RE_\textrm{POD} = \sum_{j=1}^{S} \dfrac{||(\bm{U})_j-(\bm{U}_{POD})_j||_2}{||(\bm{U})_j||_2},
\end{equation}
with the {$j$} subscript denoting the $j$-th column of the targeted matrix, and $||\cdot||_2$ the $L^2$-norm.
% , and the mean projection error over the samples writes as
% \begin{equation}\label{eq:podensnn-sig-pod}
% \bm{\sigma}_{POD} = \dfrac{1}{2S}\sum_{j=1}^{S} |(\bm{U})_j-(\bm{U}_{POD})_j| \in \realspace^H,
% \end{equation}

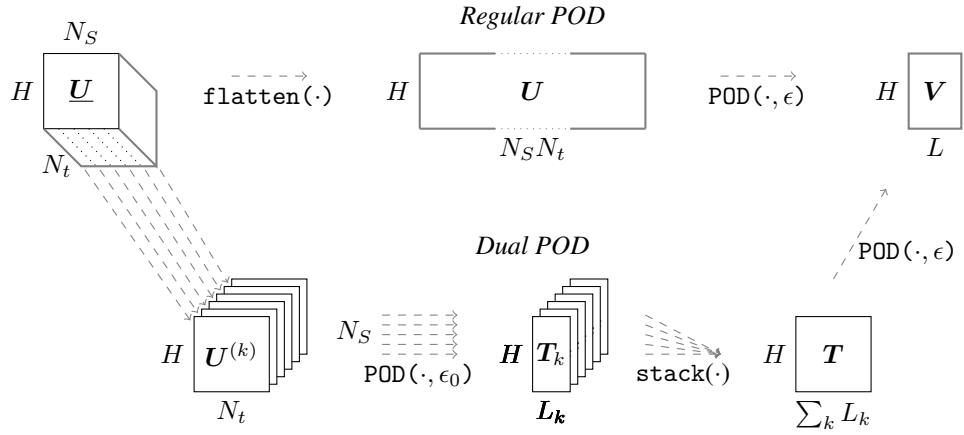
\begin{figure}
    \centering
    \begin{tikzpicture}
    \foreach \i in {1, ..., 4} {
        \draw[black, dotted] (0 - 2*\i/10, 0) -- (0.5 - 2*\i/10, -0.5);
    }
    \draw[black] (-1,0) rectangle (0,1) node[black, midway] {$\underline{\bm{U}}$};
    \node[align=center, above] at (-0.5, 1)  {$N_S$};
    \node[align=center, left] at (-1, 0.5) {$H$};
    \node[align=center, left] at (-0.5, -0.5) {$N_t$};
    \draw[gray, thick] (-1, 0) -- (-0.5, -0.5) -- (0.5, -0.5);
    \draw[gray, thick] (0, 1) -- (0.5, 0.5) -- (0.5, -0.5) -- (0, 0);

    \foreach \i in {0, ..., 5} {
        \draw[gray, dashed, ->] (0.55 - 2*\i/10, -0.55) -- (1.45 - \i/10, -2.05 - \i/10);
        \filldraw[color=black, fill=white] (1.5 - \i/10, -2 - \i/10) rectangle
                                           (1.5 - \i/10 + 1, -2 - \i/10 - 1)
                                           node[midway] {$\bm{U}^{(k)}$};
    }
    \node[align=center, left] at (1, -3) {$H$};
    \node[align=center, below] at (1.5, -3.5)  {$N_t$}; 
    \draw[gray, dashed, ->] (3.5, -3) -- (4.5, -3) node[black, midway, below] {\texttt{POD($\cdot, \epsilon_0$)}};
    \node[align=center, left] at (3.5, -2.7)  {$N_S$}; 
    \foreach \i in {1, ..., 4} {
        \draw[gray, dashed, ->] (3.5, -3 + 1.3*\i/10) -- (4.5, -3 + 1.3*\i/10);
    }

    \foreach \i in {0, ..., 5} {
        \filldraw[color=black, fill=white] (6 - \i/10, -2 - \i/10) rectangle
                                           (6 - \i/10 + 0.5, -2 - \i/10 - 1)
                                           node[midway] {$\bm{T}_{k}$};
    \node[align=center, left] at (5.5, -3) {$H$};
    \node[align=center, below] at (5.75, -3.5)  {$L_k$};
    }

    \foreach \i in {0, ..., 4} {
        \draw[gray, dashed, ->] (7, -3 + 1.3*\i/10) -- (8, -3);
    }
    \node[align=center, below] at (7.5, -3.0) {\texttt{stack}($\cdot$)};

    \draw (9, -2.5) rectangle (10, -3.5) node[midway] {$\bm{T}$};
    \node[align=center, below] at (9.5, -3.5) {$\sum_k L_k$};
    \node[align=center, left] at (9., -3.) {$H$};

    \draw[gray, dashed, ->] (9.5, -2) -- (10.2, -0.8) node[black, pos=0.3, right] {\texttt{POD($\cdot, \epsilon$)}};
    
    \draw[gray, dashed, ->] (1.5, 0.7) -- (2.5, 0.7) node[black, midway, below] {\texttt{flatten($\cdot$)}};
    
    \draw[white] (4,0) rectangle (7,1) node[black, midway] {$\bm{U}$};
    \draw[gray, thick] (4, 1) -- (4, 0) node[black, midway, left] {$H$};
    \draw[gray, thick] (4, 0) -- (5, 0);
    \node[align=center, below] at (5.5, 0) {$N_S N_t$};
    \draw[gray, thick] (4, 1) -- (5, 1);
    \draw[gray, dotted] (5, 0) -- (6, 0);
    \draw[gray, dotted] (5, 1) -- (6, 1);
    \draw[gray, thick] (6, 0) -- (7, 0);
    \draw[gray, thick] (6, 1) -- (7, 1);
    \draw[gray, thick] (7, 0) -- (7, 1);
    
    \draw[gray, dashed, ->] (8, 0.7) -- (9, 0.7) node[black, midway, below] {\texttt{POD($\cdot, \epsilon$)}};
    
    \node[align=center, left] at (10.5, 0.5) {$H$};
    \node[align=center, below] at (10.85, 0) {$L$};
    \draw[gray, thick] (10.5, 0) rectangle (11.2, 1) node[black, midway] {$\bm{V}$};

    \node[align=center, above] at (5.5, 1.2) {\emph{Regular POD}};
    \node[align=center, below] at (5.5, -1.3) {\emph{Dual POD}};

    \end{tikzpicture}
\caption{Representation of the two methods for POD order reduction in time-dependent problems}
\label{fig:dimensions}
\end{figure}

\subsection{Improving POD speed for time-dependent problems}\label{sec:podnn-reducing-dualpod}
While the SVD algorithm is well-known and widely used, it can quickly get overwhelmed by the dimensionality of the problem, especially in a time-dependent context, such as Burgers' equation and its variations (Euler, Shallow Water, etc.), which will be discussed later in Section {\ref{sec:app-riemann}}. Indeed, since time is being added as an input parameter, the matrix of snapshots $\bm{U} \in \realspace^{H \times S}$ can have a considerable width, making it very difficult and time-consuming to manipulate. One way to deal with this is the two-step POD algorithm introduced in \cite{Wang2019}. \par 
Instead of invoking the algorithm directly on the wide matrix $\bm{U}$, the idea is to perform the SVD first along the time axis for each parameter, as POD is usually used for standard space-time problems for a single parameter. We therefore consider the structured tensor $\underline{\bm{U}} \in \realspace^{H\times N_S \times N_t}$ as a starting point.\par
The workflow is as follows:
\begin{enumerate}
    \item The "time-trajectory of each parameter value," quoting directly from \cite{Wang2019}, is being fed to the SVD algorithm, and the subsequent process of reconstructing a POD basis $\bm{T}_k$ is performed for each time-trajectory $\bm{U}^{(k)}$, with $k \in [1, N_S]$. A specific stopping hyperparameter, $\epsilon_0$, is used here.
    \item Each basis $\bm{T}_k$ is collected in a new time-compressed matrix $\bm{T}$, in which the SVD algorithm is performed, along with the regular {hyperparameter} $\epsilon$, so the final POD basis construction to produce $\bm{V}$ can be achieved.
\end{enumerate}
{Fig.}\ \ref{fig:dimensions} offers a visual representation of this process, and a pseudo-code implementation is available in Algorithm \ref{alg:dualpod}.
\IncMargin{16pt}
\begin{algorithm}[h]
  \DontPrintSemicolon
  \SetKwFunction{FPod}{POD}
  \SetKwProg{Fn}{Function}{:}{\KwRet $\bm{V}$}
  \Fn{\FPod{$\bm{U}$, $\epsilon$}}{
        $\bm{D}, \bm{Z} \leftarrow SVD(\bm{U})$\;
        $\bm{\Lambda} \leftarrow \bm{D}^2$\;
        $\bm{\Lambda}_{trunc} \leftarrow \bm{\Lambda}\left[\dfrac{\sum_{i=0}^L \Lambda_i}{\sum_i \Lambda_i} \geq (1-\epsilon)\right]$\;
        $\bm{V}\leftarrow \bm{U}.\bm{Z}.\bm{\Lambda}_{trunc}^{-1/2}$\;
  }
  \;
  \SetKwFunction{FDualPod}{DualPOD}
  \SetKwProg{Un}{Function}{:}{\KwRet $\bm{V}$}
  \Fn{\FDualPod{$\underline{\bm{U}}$, $\epsilon$, $\epsilon_0$}}{
      $\bm{T} \leftarrow \bm{0}$\;
        \For{$k$ in $N_S$}{
            $\bm{T}_k \leftarrow \textrm{POD}(\bm{U}^{(k)}, \epsilon_0)$\;
        }
        $\bm{V} \leftarrow \textrm{POD}(\bm{T}, \epsilon)$\;
  }
\caption{{T}wo-step POD that allows for the management of large, time-dependent datasets{.}}
\label{alg:dualpod}
\end{algorithm}

%% file: content/learning.tex
\section{Learning Expansion Coefficients Distributions using Deep Ensembles}
\label{sec:podensnn-learning}
\input{content/fig-vardnn-pod.tex}
\subsection{Regression objective}
\par Building a non-intrusive ROM involves a statistical step to construct the function responsible for inferring the expansion parameters $\bm{v}$ from new non-spatial parameters $\bm{s}$. This regression step is performed offline, and as we have considered the spatial parameters $\bm{x}$ to be externally handled, it can be represented as a mapping $u_{DB}$  outputting the projection coefficients $\bm{v}(\bm{s})$, as in
\begin{align}\label{eq:podensnn-u-db}
    u_{DB}:\ \realspace^{P} &\to \realspace^{L}\\
    \bm{s} &\mapsto \bm{v}(\bm{s})\nonumber.
\end{align}

\subsection{Deep Neural Networks with built-in variance}
\par This statistical step is handled in the POD-NN framework by inferring the mapping with a Deep Neural Network, $\hat{u}_{DB}(\bm{s};\bm{w},\bm{b})$. The \emph{weights} and \emph{biases} of the network, $\bm{w}$ and $\bm{b}$, respectively, represent the model parameters and are learned during training (\textit{offline} phase), to be later reused to make predictions (\textit{online} phase). The network's number of hidden layers is called the \textit{depth}, $d$, which is chosen without accounting for the input and output layers. Each layer has a specific number of neurons that constitutes its \textit{width}, $l^{(j)}$. \par

The main difference here with an ordinary DNN architecture for regression resides in the dual output, first presented in \cite{nix1994estimating} and reused in \cite{lakshminarayanan2017simple}, where the final layer size is twice the number of expansion coefficients to project, $l^{(d+1)}=2L$, since it outputs both a \emph{mean} value $\bm{\mu}^v$ and a \emph{raw variance} $\bm{\rho}^v$, which will then be constrained for positiveness through a softplus function, finally outputting ${\bm{\sigma^v}}^2$ as
\begin{equation}\label{eq:softplus}
    {\bm{\sigma}^v}^2 = \textrm{softplus}(\bm{\rho}^v) := \log(1 + \exp(\bm{\bm{\rho}^v})).
\end{equation}
A representation of this DNN is pictured in {Fig.}\ \ref{fig:vardnn-reg}, with $d$ hidden layers, and therefore, $d+2$ layers in total. Each hidden layer state $\bm{h}^{(j)}$ gets computed from its input $\bm{h}^{(j-1)}$ alongside the layer's weights $\bm{w}^{(j)}$ and biases $\bm{b}^{(j)}$, and finally goes through an activation function $\phi$
\begin{equation}\label{eq:podensnn-hidden-states}
    \bm{h}^{(j)} = \phi\left(\bm{w}^{(j)}\bm{h}^{(j-1)} + \bm{b}^{(j)}\right),
\end{equation}
with $\bm{h}^{(0)} = \bm{s}$, {the} input of $\hat{u}_{DB}$, and $\bm{h}^{(d+1)} = [\bm{\mu}^v,\ \bm{\rho}^v]^\intercal$, {the} output of $\hat{u}_{DB}$.  \par
Since this predicted variance reports the spread, or noise, in data (the inputs' data are drawn from a distribution), and so it would not be reduced even if we were to grow our dataset larger, it accounts for the \emph{aleatoric uncertainty}, which is usually separated from \emph{epistemic uncertainty}. This latter form is inherent to the model \cite{UncertaintiesBayesianDV}. \par 
One can think about this concept of aleatoric uncertainty as a measurement problem with the goal of measuring a quantity $u$. The tool used for measurement has some inherent noise $n$, random and dependent upon the parameter $x$ in the measurable domain, making the measured quantity $u(x)+n(x)$. The model presented here, as introduced in \cite{nix1994estimating}, is designed to perform the regression on both components, with an estimated variance alongside the regular point-estimate of the mean. \par

\subsection{Ensemble training}
\par {An} $N$-sized training dataset $\mathcal{D}=\{\bm{X}_i, \bm{v}_i\}$ {is considered}, with $\bm{X}_i$ denoting the normalized non-spatial parameters $\bm{s}$, and $\bm{v}_i$ the corresponding expansion coefficients from a training/validation-split of the matrix of snapshots $\bm{U}${. An} \emph{optimizer} performs several \textit{training epochs} $N_e$ to minimize the following Negative Log-Likelihood loss function with respect to the network weights and biases parametrized by $\bm{\theta}=(\bm{w}, \bm{b})$
\begin{align}
    % \mathcal{L}(\bm{w}, \bm{b}; \bm{X}, \bm{v}) = \dfrac{1}{N} \sum_{i=1}^{N} \left[ \hat{u}_{DB}(\bm{X}; \bm{w}, \bm{b})_i - (\bm{v})_i\right]^2 + \lambda r(\bm{w}, \bm{b}),
   \mathcal{L}_{\textrm{NLL}}(\mathcal{D},\bm{\theta}):=\dfrac{1}{N} \sum_{i=1}^{N}\left[\dfrac{\log\ \bm{\sigma}_{\bm{\theta}}^v(\bm{X}_i)^2}{2}+ \dfrac{(\bm{v}_i-\bm{\mu}^v_{\bm{\theta}}(\bm{X}_i))^2}{2 \bm{\sigma}_{\bm{\theta}}^v(\bm{X}_i)^2}\right],
\end{align}
% with the subscript $i$ describing the $i$-th training point, $\lambda$ the regularization parameter, and $r(\cdot)$ a regularization function, such as L2 regularization, coined first as \textit{weight decay} (\cite{Krogh1992WeightDecay}).
with $\bm{\mu}^v_{\bm{\theta}}(\bm{X})$ and $\bm{\sigma}_{\bm{\theta}}^v(\bm{X})^2$ as the mean and variance, respectively, retrieved from the $\bm{\theta}$-parametrized network.  \par
% with $p$ denoting the likelihood of predicting the targets $\bm{v}$ provided the inputs $\bm{X}$, which represents a dataset of size $N$, as well as $\bm{\mu}_{\bm{\theta}}^v$, and ${\bm{\sigma}_{\bm{\theta}}^v}^2$, respectively, the mean and variance retrieved from the $\bm{\theta}$-parametrized network.  \par
In practice, this loss gets an L2 regularization as an additional term, commonly known as \emph{weight decay} in Neural Network contexts \cite{Krogh1992WeightDecay}, producing
\begin{align}
   \mathcal{L}^\lambda_{\textrm{NLL}}(\mathcal{D}, \bm{\theta}):=\mathcal{L}_\textrm{NLL}(\mathcal{D}, \bm{\theta})+\lambda ||\bm{w}||^2.
\end{align}

\par Non-convex optimizers, such as Adam \cite{kingma2014adam} or other Stochastic Gradient Descent variants, are needed to handle this loss function, often irregular and non-convex in a Deep Learning context. The derivative of the loss $\mathcal{L}_{\textrm{NLL}}$ with respect to the weights $\bm{w}$ and biases $\bm{b}$ is obtained through \emph{automatic differentiation} \cite{Rumelhart1986}, a technique that relies on monitoring the gradients during the forward pass of the network (\ref{eq:podensnn-hidden-states}). Using \textit{backpropagation} \cite{Linnainmaa1976}, the updated weights $\bm{w}^{n+1}$ and biases $\bm{b}^{n+1}$ corresponding to the epoch $n+1$ can be written as
\begin{equation}
    \left(\bm{w}^{n+1}, \bm{b}^{n+1}\right)=\left(\bm{w}^n, \bm{b}^n\right)
    -  \tau f\left(\dfrac{\partial{\mathcal{L}^\lambda_\textrm{NLL}(\mathcal{D}, (\bm{w}^n, \bm{b}^n))}}{\partial{\left(\bm{w}^n, \bm{b}^n\right)}}\right),
\end{equation}
where $f(\cdot)$ is a function of the loss derivative with respect to weights and biases that is dependent upon the {chosen} optimizer, and $\tau$ is the \textit{learning rate}, a hyperparameter defining the step size taken by the optimizer. \par

The idea behind Deep Ensembles, presented in \cite{lakshminarayanan2017simple} and recommended in \cite{snoek2019trust}, is to randomly initialize $M$ sets of $\bm{\theta}_m=(\bm{w}, \bm{b})$, thereby creating $M$ independent neural networks (NNs). Each NN is then subsequently trained.
Overall, the predictions moments in the reduced space $(\bm{\mu}^v_{\bm{\theta}_m},\bm{\sigma}^v_{\bm{\theta}_m})$ of each NN create a probability mixture, which, as suggested by the original authors, we can approximate in a single Gaussian distribution, leading to a mean expressed as
\begin{equation}
    \bm{\mu}^v_*(\bm{X}) = \dfrac{1}{M} \sum_{m=1}^{M}\bm{\mu}^v_{\bm{\theta}_m}(\bm{X}),\label{eq:article-deepens-mu}
\end{equation}
and a variance subsequently obtained as
\begin{equation}
    \bm{\sigma}^v_*(\bm{X})^2 = \dfrac{1}{M} \sum_{m=1}^{M} \left[\bm{\sigma}_{\bm{\theta}_m}^v(\bm{X})^2 + \bm{\mu}^v_{\bm{\theta}_m}(\bm{X})^2\right] - \bm{\mu}_*^v(\bm{X})^2.\label{eq:article-deepens-sig}
\end{equation}
The model is now accounting for the \emph{epistemic uncertainty} through random initialization and variability in the training step. This uncertainty is directly linked to the model and could be reduced if we had more data. The uncertainty is directly related to the data-fitting capabilities of the model and thus will snowball in the absence of such data since there are no more constraints. In our case, it has the highest value, compared to aleatoric uncertainty, since one of our objectives is to be warned when the model is making predictions that are out-of-distribution. \par
This model will be referred to as POD-EnsNN, and its training steps are listed in Algorithm \ref{alg:ensnn-train}. Since these networks are independent, parallelizing their training is relatively easy (see Algorithm \ref{alg:horovod}), with only the results needing to be averaged-over. 

\begin{algorithm}[h]
  \DontPrintSemicolon
  Prepare the dataset $\mathcal{D}=\left\{\bm{X}_i, \bm{v}_i\right\}$\;
  \For{each model in the ensemble $1\leq m \leq M$}{
    Train the model $m$:\;
    \For{each epoch $1\leq e \leq N_e$}{
          Retrieve the outputs $(\bm{\mu}^v_{\bm{\theta}_m}(\bm{X}), \bm{\rho}^v_{\bm{\theta}_m}(\bm{X}))$ from the forward pass $\hat{u}_D(\bm{X})$\;
          Perform the variance treatment, $\bm{\sigma}^{v}_{\bm{\theta}_m}(\bm{X})^2=\textrm{softplus}(\bm{\rho}^v_{\bm{\theta}_m}(\bm{X}))$\;
          Compute the loss $\mathcal{L}_{\textrm{NLL}}$\;
          Backpropagate the gradients to the parameters $\bm{\theta}_m$\;
    }
  }
  {
\For{each model in the ensemble $1\leq m \leq M$}{
  Retrieve statistical outputs $(\bm{\mu}^v_{\bm{\theta}_m}(\bm{X}_{\textrm{tst}, i}), \bm{\sigma}_{\bm{\theta}_m}^v(\bm{X}_{\textrm{tst}, i})^2)$ for the model $m$ for a test dataset $\mathcal{D}_\textrm{tst}=\left\{\bm{X}_{\textrm{tst}, i}, \bm{v}_{\textrm{tst}, i}\right\}$\;
  }
  }
  Approximate the predictions for the reduced space in a Gaussian $\mathcal{N}\left(\bm{\mu}_*^v(\bm{X}_{\textrm{tst}, i}), \bm{\sigma}_*^v(\bm{X}_{\textrm{tst}, i})^2\right)$\;
\caption{Deep Ensembles training and predictions{.}}
\label{alg:ensnn-train}
\end{algorithm}

\begin{algorithm}[h]
  \DontPrintSemicolon
  \SetKwFunction{FPod}{TrainOnOneDevice}
  \SetKwProg{Fn}{Function}{:}{\KwRet $\hat{u}_{DB}$}
  \Fn{\FPod{$\bm{X}$, $\bm{v}$, $\lambda$, $ \tau$, $N_e$}}{
    Import the TensorFlow library as $tf$\;
    Import the Horovod library as $hvd$ and initialize it with $hvd.\textrm{init}()$\;
    Get the assigned device id $i = hvd.\textrm{localRank}()$\;
    Get local devices $\bm{D} = tf.\textrm{getVisibleDevices}()$\;
    Force the device for TensorFlow: $tf.\textrm{setVisibleDevices}(\bm{D}_i)$\;
    Init the model: $\hat{u}_{DB} = \textrm{NeuralNetwork}( \tau, \lambda)$\;
    Train it: $\hat{u}_{DB}.\textrm{fit}(\bm{X}, \bm{v}, N_e)$\;
  }
  Run the meta-command:
  \texttt{horovodrun -np $M$ -H localhost:$M$ TrainOnOneDevice}$(\bm{X}, \bm{v}, \lambda,  \tau, N_e)$\;
\caption{Pseudo-code showing parallelization with Horovod \cite{Sergeev2018horovod}{.}}
\label{alg:horovod}
\end{algorithm}

\subsection{Predictions in the expanded space}
While embedding uncertainty quantification within Deep Neural Networks helps to obtain a confidence interval on the predicted expansion coefficients $\bm{v}$, it is still necessary to then perform the {expansion} step to retrieve the full solution, as presented in (\ref{eq:podensnn-pod-exp-basis}). It is defined as a dot product with the modes matrix $\bm{V}$. \par 
While this applies perfectly to the predicted mean {$\bm{\mu}^v$}, care must be taken when handling the predicted standard deviation {$\bm{\sigma}^v$}, as there is no theoretical guarantee for the statistical moments on the reduced basis to translate linearly in the expanded space. However, after the mixture approximation, the distribution over the coefficients $\bm{v}$ is known as follows:
\begin{equation}
    \hat{\bm{v}}(\bm{X})=\hat{u}_{DB}(\bm{X}; \bm{w}, \bm{b})\sim\mathcal{N}\left(\bm{\mu}_*^v(\bm{X}), {\bm{\sigma}_*^v}^2(\bm{X})\right).
\end{equation}
Therefore, unlimited samples $\hat{\bm{v}}^{(i)}$ can be drawn from this distribution, and individually decompressed into a corresponding full solution $\hat{u}_D^{(i)}=\bm{V}.\hat{\bm{v}}^{(i)}$, from (\ref{eq:podensnn-pod-exp-basis}). The following Monte-Carlo approximation of the full distribution on $\hat{u}_D$ is hence proposed, drawing $N_\textrm{ex}$ samples and using the rapid surrogate model to compute
\begin{align}
    \bm{\mu}_*(\bm{X}) &= \dfrac{1}{N_\textrm{ex}} \sum_{i=1}^{N_\textrm{ex}} \hat{u}_D^{(i)}
    =\dfrac{1}{N_\textrm{ex}} \sum_{i=1}^{N_\textrm{ex}} \bm{V}.\bm{v}^{(i)}, \label{eq:decompress-mu}\\
    \bm{\sigma}^2_*(\bm{X}) &= \dfrac{1}{N_\textrm{ex}} \sum_{i=1}^{N_\textrm{ex}} \left[\hat{u}_D^{(i)} - \bm{\mu}_*\right]^2 
    = \dfrac{1}{N_\textrm{ex}} \sum_{i=1}^{N_\textrm{ex}} \left[\bm{V}.\bm{v}^{(i)} - \bm{\mu}_*\right]^2
    \label{eq:decompress-var},
\end{align}
which represents the approximated statistical moments of the distribution on the predicted full solution $\hat{u}_D(\bm{X})$, also referred to as $\hat{u}_D^\mu$ and $\hat{u}_D^\sigma$.

\subsection{Metrics}
% \par Just as for the loss, we consider the widely-used \textit{mean squared error} to assess the quality of the predictions, either on a $N_{\textrm{val}}$-sized validation dataset $(\bm{X}_{\textrm{val}}, \bm{v}_{\textrm{val}})$, or on a $N_{\textrm{val}}$-sized validation dataset $(\bm{X}_{\textrm{val}}, \bm{v}_{\textrm{val}})$. 
% \begin{equation}
%     MSE(\mu^{v}_{\textrm{val}}, \bm{v}_{\textrm{val}}, N_{\textrm{val}}) = \dfrac{1}{N_{\textrm{val}}} \sum_{i=1}^{N_{\textrm{val}}} \left[(\mu^{v}_{\textrm{val}})_i- (\bm{v}_{\textrm{val}})_i\right]^2,
% \end{equation}
% with the inputs $\bm{X}_{\textrm{val}}$ being the non-spatial parameters $\bm{s}$ after normalization and the projection coefficients $\bm{v}_{\textrm{val}}$ coming from a validation-split matrix of snapshots $\bm{U}_{\textrm{val}}$, the \textit{validation error} $E_{\textrm{val}}=MSE(\mu^{v}_{\textrm{val}}, \bm{v}_{\textrm{val}}; N_{\textrm{val}})$ is computed with $(\mu^{v}_{\textrm{val}})_i =  \hat{u}_{DB}(\bm{X_{\textrm{val}}}; \bm{w}^{n+1}, \bm{b}^{n+1})_i$, $\bm{w}^{k}$ and $\bm{b}^{k}$ being the weights and biases of the network at a training epoch $k$. The same goes for the \textit{test error} $E_T$, wihch will be evaluated after the training ($k=n+1$).

\par In addition to the regularized loss $\mathcal{L}^\lambda_{\textrm{NLL}}$, we define a relative error $RE$ on the mean prediction as\begin{equation}
  \label{eq:re}
    % RE(\hat{\bm{U}}, \bm{U}; H, S) = \dfrac{1}{H S} \sum_{i=1}^{H}\sum_{j=1}^{S} \dfrac{|\hat{\bm{U}}_{ik} - \bm{U}_{ik}|}{\max(|\hat{\bm{U}}_{ij}|, |\bm{U}_{ij}|)},
% =\dfrac{||\bm{V}\left[\bm{v}_{\textrm{val}} - \bm{ \hat{u}_{DB}}\left(\bm{X}_{\textrm{val}}; \bm{w}^{n+1}, \bm{b}^{n+1}\right)\right]||}{||\bm{V}\bm{v}_{\textrm{val}}||},
    RE(\bm{\mu_*}, \bm{U}) = \dfrac{||\sum_{i=1}^{{S}}(\bm{\mu}_*(\bm{X_i}) - \bm{U}_i)||_2}{||\sum_{i=1}^{S}\bm{U}_i||_2},
\end{equation}
with $\bm{U}_i$ the $i$-th column of the snapshots matrix, corresponding to the input $X_i$. It can be applied for training, validation, or testing, as defined in Section \ref{sec:podensnn-reducing}. During the training, we report two metrics: the training loss $\mathcal{L}^\lambda_{\textrm{NLL}}$ and the validation relative error $RE_{\textrm{val}}$. \par
To quantify the uncertainty associated with the model predictions, we define the \emph{mean prediction interval width} (MPIW) \cite{Yao2019}, aimed at tracking the size of the 95\% confidence interval, i.e., $\pm 2 \sigma_*^2$, as follows 
\begin{equation}\label{eq:mpiw}
  MPIW(\bm{\sigma}_*)=
  \dfrac{1}{HN} \sum_{j=1}^H \sum_{i=1}^N
  \left[
    \hat{u}_{D}^\textrm{upper}(\bm{X}_i)_j - 
    \hat{u}_{D}^\textrm{lower}(\bm{X}_i)_j
  \right]
  =
  \dfrac{1}{HN} \sum_{j=1}^H \sum_{i=1}^N 4\bm{\sigma}^2_*(\bm{X}_i)_j,
\end{equation}
with the $j$ subscript denoting the $j$-th degree of freedom of a solution.

\subsection{Adversarial training}
First proposed in \cite{Szegedy2014advtra} and studied in \cite{Goodfellow2014advtra}, the concept of \emph{adversarial training}, not to be confused with Generative Adversarial Networks \cite{Goodfellow2014GANs}, aims at improving the robustness of Neural Networks when confronted with noisy data, which could potentially be intentionally created. \par 
In the Deep Ensembles framework, adversarial training is an optional component that, according to \cite{lakshminarayanan2017simple}, can help to smooth out the output. This technique can be particularly useful as shown in the subsequent test case, where the model is struggling with the highly-nonlinear wave being produced by {Shallow Water equations} (see Section {\ref{sec:app-riemann}}). \par
A simple implementation is the \emph{gradient sign} technique, which adds noise in the opposite direction of the gradient descent, scaled by a new hyperparameter $\zeta$, at each training epoch, {as} shown in Algorithm \ref{alg:adv-train}. The idea is to perform \emph{data augmentation} at each training epoch. The additional data comes from the generated adversarial samples that will help to train the network more robustly, given that these problematic samples are being inserted in the dataset.
\begin{algorithm}[h]
  \DontPrintSemicolon
  \SetKwFunction{FPod}{getAdversarialLoss}
  \SetKwProg{Fn}{Function}{:}{\KwRet $\mathcal{L}_T$}
  \Fn{\FPod{$\bm{X}$, $\bm{v}$, $\epsilon$}}{
    $\mathcal{L}_T \leftarrow \mathcal{L}_\textrm{NLL}^\lambda(\{\hat{u}_D(\bm{X}), \bm{v}\}, \bm{\theta})$\;
    $\bm{X}' \leftarrow \bm{X} + \zeta\ \textrm{sign}(\dfrac{{\partial}\mathcal{L}_T}{\partial \bm{X}})$\;
    $\mathcal{L}_T \leftarrow \mathcal{L}_T + \mathcal{L}_\textrm{NLL}^\lambda(\{\hat{u}_D(\bm{X}'), \bm{v}\}, \bm{\theta})$\;
  }
\caption{{A}dversarial training within the training loop{.}}
\label{alg:adv-train}
\end{algorithm}
% \begin{figure}[h!]
%   \fbox{
%     \parbox{.9\textwidth}{
% % \includegraphics[width=\linewidth]{images/podnn-adv-burger-graph-means.pdf}
% \includegraphics[width=\linewidth]{images/podnn-adv-burger-graph-samples.pdf}
%     }
%   }
% \\ \parbox{0.75\textwidth}{
% \caption{This time, Burgers' equation solution has been handled with adversarial training. The same setup and quantities as in Figure \ref{fig:podnn-burger-graph} are shown} 
% \label{fig:podnn-adv-burger-graph}
% }
% \end{figure}

%% file: content/fig-vardnn-pod.tex
\tikzset{%
  every neuron/.style={
    circle,
    draw,
    minimum size=1cm
  },
  neuron missing/.style={
    draw=none, 
    scale=3,
    text height=0.333cm,
    execute at begin node=\color{black}$\vdots$
  },
}

\begin{figure}[t]
  \centering
  % \fbox{
    \resizebox{.8\textwidth}{!}{
\begin{tikzpicture}[x=1.5cm, y=1.5cm, >=stealth]

\foreach \m/\l [count=\y] in {1,2,3,missing,4}
  \node [every neuron/.try, neuron \m/.try] (input-\m) at (0,2.5-\y) {};

\foreach \m [count=\y] in {1,missing,2}
  \node [every neuron/.try, neuron \m/.try ] (hidden1-\m) at (2,2-\y*1.25) {};

\foreach \m [count=\y] in {1,missing,2}
\node [every neuron/.try, neuron \m/.try ] (hidden1-\m) at (2,2-\y*1.25) {};

\foreach \m [count=\y] in {1,missing,2}
  \node [every neuron/.try, neuron \m/.try ] (hidden2-\m) at (5,2-\y*1.25) {};

\foreach \m [count=\y] in {1,missing,2}
  \node [every neuron/.try, neuron \m/.try ] (output-\m) at (7,2.5-\y*0.8) {};

\foreach \m [count=\y] in {1,missing,2}
  \node [every neuron/.try, neuron \m/.try ] (outputrho-\m) at (7,-0.0-\y*0.8) {};

\foreach \l [count=\i] in {1,2,3,S}
  \draw [<-] (input-\i) -- ++(-1,0)
    node [above, midway] {${X}_\l$};

\foreach \l [count=\i] in {1, l^{(1)}}
\node [above] at (hidden1-\i.north) {$h_{\l, 1}$};

\foreach \l [count=\i] in {1,l^{(d)}}
\node [above] at (hidden2-\i.north) {$h_{\l, d}$};

\foreach \l [count=\i] in {1,L}
  \draw [->] (output-\i) -- ++(2,0)
    node [above, midway] {$\mu^v_\l$};
\foreach \l [count=\i] in {1,L}
  \node [{
    circle,
    draw,
    minimum size=0.1cm
  }] (output-rho-step-\l) at (8,0.8-2*\i*0.8) {};
\foreach \l [count=\i] in {1,L}
\node [below] at (output-rho-step-\l.south) {$\textrm{softplus}$};

\foreach \l [count=\i] in {1,L}
  \draw [->] (outputrho-\i) -- (output-rho-step-\l)
    node [above, midway] {$\rho^v_\l$};

\foreach \l [count=\i] in {1,L}
\draw [->] (output-rho-step-\l) -- ++(1,0)
  node [above, midway] {$(\sigma^v_\l)^2$};

\foreach \i in {1,...,4}
  \foreach \j in {1,...,2}
    \draw [->] (input-\i) -- (hidden1-\j);

\foreach \i in {1,...,2}
  \foreach \j in {1,...,2}
    \draw [->] (hidden1-\i) -- (hidden2-\j);

    \foreach \i in {1,...,2}
    \foreach \j in {1,...,2}
    \draw [->] (hidden2-\i) -- (output-\j);

    \foreach \i in {1,...,2}
    \foreach \j in {1,...,2}
    \draw [->] (hidden2-\i) -- (outputrho-\j);

\node [align=center, above] at (0,2) {Input\\layer};
\node [align=center, above] at (2,2) {Hidden \\layer $(1)$};
\node [align=center, above] at (5,2) {Hidden \\layer $(d)$};
\node [align=center, above] at (7,2) {Output \\layer};

\node[fill=white,scale=4,inner xsep=0pt,inner ysep=5mm] at ($(hidden1-1)!.5!(hidden2-2)$) {$\dots$};

\end{tikzpicture}
  }
  % }
  \caption{$\hat{u}_{DB}(\bm{X}; \bm{w}, \bm{b})\sim\mathcal{N}\left(\bm{\mu}^v(\bm{X}), \bm{\sigma}^v(\bm{X})^2\right)$, a Deep Neural Network regression with a dual mean and variance output{.}}
  \label{fig:vardnn-reg}
\end{figure}

%% file: content/podbnn.tex
\FloatBarrier
\section{Bayesian Neural Networks and Variational Inference as an Alternative}
\label{sec:podensnn-learning-bnn}
\input{content/fig-bnn-pod.tex}
Making a model aware of its associated uncertainties can ultimately be achieved by adopting the Bayesian view. Recently, it has become easier to include a fully Bayesian treatment within Deep Neural Networks \cite{Blundell2015}, designed to be compatible with backpropagation. In this section, we implement this version of Bayesian Neural Networks within the POD-NN framework, which we will refer to as POD-BNN, and compare it to the Deep Ensembles approach.  \par
\subsection{Overview}
To address the \emph{aleatoric uncertainty}, arising from noise in the data, Bayesian Neural Networks can make use of the same dual-output setting as the NNs we used earlier for Deep Ensembles, $(\bm{\mu}^v, \bm{\rho}^v)$ in our context, with the variance ${\bm{\sigma}^v}^2$ subsequently retrieved with the softplus function defined in (\ref{eq:softplus}). \par
However, in the \emph{epistemic uncertainty} treatment the process and issues are very different. Earlier, even though the NNs were providing us with a mean and variance, they were still deterministic, and variability was obtained by assembling randomly initialized models. 
The Bayesian treatment instead aims to assign distributions to the network's weights, and they therefore have a probabilistic output by design (see {Fig.}\ \ref{fig:bnn-reg}). In this context, it is necessary to make multiple predictions, instead of numerous {(parallel) trainings}, in order to obtain data on uncertainties.  \par
Considering a dataset $\mathcal{D}=\{\bm{X}_i, \bm{v}_i\}$, a \emph{likelihood} function $p(\mathcal{D}|\bm{w})$ can be built, with $\bm{w}$ denoting both the weights $\bm{w}$ and the biases $\bm{b}$ for simplicity. The goal is then to construct a \emph{posterior distribution} $p(\bm{w}|\mathcal{D})$ to achieve the following \emph{posterior predictive distribution} on the target $\bm{v}$ for a new input $\bm{X}$
\begin{equation}
    p(\bm{v}|\bm{X},\mathcal{D}) = \int p(\bm{v}|\bm{X},\bm{w})p(\bm{w}|\mathcal{D})\,d\bm{w},\label{eq:podensnn-posterior}
\end{equation}
which cannot be achieved directly in a NN context, due to the infinite possibilities for the weights $\bm{w}$, leaving the posterior $p(\bm{w}|\mathcal{D})$ intractable as explained in \cite{Blundell2015}. A few observations can be made on this formula. First, the initial term in the integral, $p(\bm{v}|\bm{X}, \bm{w})$, stands for the distribution of the target $\bm{v}$ for the input $\bm{X}$ according to a weight configuration $\bm{w}$. It directly describes the noise in the data and is handled by the NN's dual-output setting. Second, the posterior distribution $p(\bm{w}|D)$ accounts for the distribution on the weights given the dataset $\mathcal{D}$, which bundles the uncertainty on the weights since they are sampled in a finite setting \cite{williamhsieh2009}. This decomposition shows the power of the approach{. Y}et the bottleneck resides in the intractability of the posterior. \par
While various attempts have been made at approximating this integral in a NN context, such as Markov Chains methods \cite{Neal1993, Neal1995Thesis}, the most common way is through \emph{Variational Inference}, first presented by \cite{Hinton1993varinf}, which ultimately led to trainable BNNs in \cite{Blundell2015}. The idea is to construct a new $\bm{\theta}$-parametrized distribution $q(\bm{w}|\bm{\theta})$ as an approximation of $p(\bm{w}|\mathcal{D})$, by minimizing their Kullback-Leibler divergence, with the goal being computing (\ref{eq:podensnn-posterior}). The KL measures the difference between two distributions and can be defined for two continuous densities $a(x)$ and $b(x)$ as
\begin{equation}
\textrm{KL}(a(x)||b(x))=\int a(x) \log \dfrac{a(x)}{b(x)}\ dx,
\end{equation}
and has the property of being non-negative. In our case, it writes as $\textrm{KL}(q(\bm{w}|\bm{\theta})||p(\bm{w}|\mathcal{D}))$ with respect to the new parameters $\bm{\theta}$ called \emph{latent variables}, such as

\begin{align}\label{eq:podbnn-kl-def}
\textrm{KL}(q(\bm{w} | \bm{\theta}) || p(\bm{w} | \mathcal{D})) &=
\int q(\bm{w} | \bm{\theta}) \log 
    \dfrac{q(\bm{w} | \bm{\theta})}{p(\bm{w}|\mathcal{D})}\, d\bm{w}=\mathbb{E}_{q(\bm{w} | \bm{\theta})}
    {
    \left[
    \log 
    \dfrac{q(\bm{w} | \bm{\theta})}{p(\bm{w}|\mathcal{D})}\right]
    }.
\end{align}

Applying Bayes rule, the posterior $p(\bm{w}|{\mathcal{D}})$ can be rewritten as $p(\mathcal{D}|\bm{w})p(\bm{w})/p(\mathcal{D})$, and so
\begin{align}
\textrm{KL}(q(\bm{w} | \bm{\theta}) || p(\bm{w} | \mathcal{D}))
    &=\mathbb{E}_{q(\bm{w} | \bm{\theta})}
    {
    \left[
    \log \dfrac{q(\bm{w} | \bm{\theta})p(\mathcal{D})}{p(\mathcal{D}|\bm{w})p(\bm{w})}
    \right]
    }
    \\
    &=\mathbb{E}_{q(\bm{w} | \bm{\theta})}
    \left[\log
    \dfrac{q(\bm{w} | \bm{\theta})}{p(\bm{w})}
    - \log
    p(\mathcal{D}|\bm{w})
    + \log p(\mathcal{D})\right].
\end{align}
Recognizing a KL difference between the approximated distribution $q(\bm{w}|\bm{\theta})$ and the prior distribution on the weights $p(\bm{w})$, and the non-dependence on the weights of the \emph{marginal likelihood} $p(\mathcal{D})$:
\begin{align}
  \textrm{KL}(q(\bm{w} | \bm{\theta}) || p(\bm{w} | \mathcal{D}))
    &=\textrm{KL}(q(\bm{w}|\bm{\theta})||p(\bm{w})) -\mathbb{E}_{q(\bm{w} | \bm{\theta})} {\left[ \log p(\mathcal{D}|\bm{w})\right]} + \log p(\mathcal{D})\label{eq:podbnn-kl-2}\\
    &=:\mathcal{F}(\mathcal{D}, \bm{\theta}) + \log p(\mathcal{D}).\label{eq:kl-f}
\end{align}
The term $\mathcal{F}(\mathcal{D}, \bm{\theta})$ is commonly known as the \emph{variational free energy}, and minimizing it with respect to the weights does not involve the last term $\log p(\mathcal{D})$, and so it is equivalent to the goal of minimizing $\textrm{KL}(q(\bm{w}|\bm{\theta}),||p(\bm{w}|\mathcal{D}))$. If an appropriate choice of $q$ is made, (\ref{eq:kl-f}) can be computationally tractable, and the bottleneck is worked around. In any case, this term acts as a lower bound on the likelihood, tending to an exact inference case where $\mathcal{F}(\mathcal{D}, \bm{\theta})$ would become the log-likelihood $\log p(\mathcal{D}|\bm{w}$), \cite{Goodfellow-et-al-2016}. \par
By drawing $N_\textrm{mc}$ samples $\bm{w}^{(i)}$ from the distribution $q(\bm{w}|\bm{\theta})$ at the layer level, it is possible to construct a tractable Monte-Carlo approximation of the variational free energy, such as
\begin{equation}\label{eq:f-loss}
\mathcal{F}(\mathcal{D},\bm{\theta}) \approx
% {1 \over N}
\sum_{i=1}^{N_\textrm{mc}}
{
\left[
\log q(\bm{w}^{(i)} | \bm{\theta}) -
\log p(\bm{w}^{(i)}) -
\log p(\mathcal{D} | \bm{w}^{(i)})
\right]
},
% \sum_{i=1}^{N_\textrm{mc}} \left[
% \log q(\bm{w}^{(i)} | \bm{\theta}) -
% \log p(\bm{w}^{(i)}) -
% \log p(\mathcal{D} | \bm{w}^{(i)})\right] =: \mathcal{L}_{\textrm{ELBO}}(\mathcal{D},\bm{\theta}),
\end{equation}
with $p(\bm{w}^{(i)})$ denoting the \emph{prior} on the drawn weight $\bm{w}^{(i)}$, which is chosen by the user, with an example given in (\ref{eq:podensnn-prior}). The variational free energy is a sum of two terms, the first being linked to the prior, named \emph{complexity cost}, while the latter is related to the data and referred to in \cite{Blundell2015} as the \emph{likelihood cost}. The latter shows to be approximated by summing on the $N_{\textrm{mc}}$ samples at the output level (for each training input). \par 
Equation (\ref{eq:f-loss}) defines our new loss function $\mathcal{L}_\textrm{ELBO}$. This name comes from the \emph{Evidence Lower Bound} function, commonly known in the literature, and corresponding to the opposite maximizing objective. The third term in (\ref{eq:f-loss}) may be recognized as a Negative Log-Likelihood, which was used in the training of Deep Ensembles, and will be evaluated from the NN’s outputs. The first two are issued from an approximation of the KL divergence at the layer level.
% An important observation right away is the Monte-Carlo summation presented here to compute the integral and expectation terms in (\ref{eq:podbnn-kl-2}), leading to the questioning on how it will perform in presence of discontinuities. This will be further discussed in Section \ref{sec:podensnn-bench} and \ref{sec:podensnn-applications}.

% \begin{figure}[t]
%   \centering
%   \fbox{
%     \parbox{.9\textwidth}{
% \includegraphics[width=\linewidth]{images/podensnn-burger-graph-meansamples.pdf}
%     }
%   }
%   \\ \parbox{0.75\textwidth}{
% \caption{From left to right: comparing the predicted $\hat{u}_D$ and the observed data $u_D$ from the dataset across three time-steps, and vertically across two random snapshots for Burgers' equation solution (1D, time-dependent) respectively in and out of the training bounds, and this time we can see the uncertainty increasing when we exit the training bounds} 
% \label{fig:podbnn-burger-graph-samples}
% }
% \end{figure}

\subsection{Choice of prior distributions}\label{sec:podbnn-prior}
The Bayesian view differs that of the frequentists with its ability to reduce the overall uncertainty by observing new data points. The initial shape is described by a prior distribution, representing the previously known information to encode in the model. \par
In our case, the prior distribution of the NN weights, $p(\bm{w})$. For simplicity, in this work we start by reusing the \emph{fixed} Gaussian mixture proposed in \cite{Blundell2015}, defined for three positive hyperparameters $\pi_0$, $\pi_1$, and $\pi_2$, such that

\begin{equation}
    p(\bm{w}) = \pi_0 \mathcal{N}(\bm{w} | 0,\pi_1^2) + (1 - \pi_0) \mathcal{N}(\bm{w} | 0,\pi_2^2).
    \label{eq:podensnn-prior}
\end{equation}

% While it led to good results in the original paper as well as in the smooth physical solutions cases presented in Section \ref{sec:podensnn-bench-ackley} and \ref{sec:app-river-results}, in which it will be further discussed, convergence wasn’t reached in all cases with this hardcoded prior. As a different take, one can define a \emph{trainable} prior as
% \begin{equation}\label{eq:trainable-prior}
%   p(\bm{w})=\mathcal{N}(\bm{w} | \bm{\mu}_\pi,\bm{1}),
% \end{equation}
% with its parameters $\bm{\mu}_\pi$ added to the list of model parameters $\bm{\theta}$ to be trained by minimizing $\mathcal{L}_\textrm{ELBO}(\mathcal{D}, \bm{\theta})$. Here, as recommended by TensorFlow Probability, we train a number of means corresponding to the sum of the weights and biases for each layer, and we leave their corresponding standard deviations unitary. This allows from infering prior knowledge directly from the data, instead of specifying additional hyperparameters as in the fixed mixture case.
\subsection{Training}
The idea behind the work of \cite{Blundell2015} was to have a fully Bayesian treatment of the weights while providing it in a form compatible to the usual \emph{backpropagation} algorithm, mentioned in Section \ref{sec:podensnn-learning}. One of the blockers is the forward pass that requires gradients to be tracked, allowing their derivatives to be backpropagated. \par 
At the $j$-th \emph{variational layer}, we consider a Gaussian distribution for the approximated distribution $q(\bm{w}^{(j)}|\bm{\theta}^{(j)})$, effectively parameterizing the weights and the biases by a mean $\bm{\theta}_\mu^{(j)}$ and raw variance $\bm{\theta}_\rho^{(j)}$, acting as local latent variables. This setting leads the total number of trainable parameters of the network to be twice that in a standard NN, as each $\bm{w}^{(j)}$ is sampled from the approximated two-parameter Gaussian distribution $q(\bm{w}^{(j)}|\bm{\theta}^{(j)})\sim\mathcal{N}(\bm{\theta}_\mu^{(j)}, \bm{\theta}_\rho^{(j)})$. \par
In the forward pass, to keep track of the gradients, each operation must be differentiable. To sample the weights $\bm{w}^{(j)}$, we construct a function $f(\bm{\theta}_\mu^{(j)}, \bm{\theta}_\rho^{(j)})=\bm{\theta}_\mu^{(j)} + \bm{\theta}_\rho^{(j)} \odot \bm{\epsilon}^{(j)}=:\bm{w}^{(j)}$, with $\bm{\epsilon}^{(j)}$ sampled from a parameter-free normal distribution, $\bm{\epsilon} \sim \mathcal{N}(\bm{0}, \bm{I})$. This is known as the \emph{reparametrization trick} \cite{Kingma2014}.  \par
The true variance of the weights $\bm{\theta}_\sigma^{(j)}$ is not the direct parameter, but as stated earlier, to ensure positivity and numerical stability, it is defined through a softplus function, with $\bm{\theta}_\sigma^{(j)}=\log(1+\exp(\bm{\theta}_\rho^{(j)}))$.  \par
Going back to (\ref{eq:f-loss}), it can be observed that the Monte-Carlo summation is actually going to be two-fold while training. Firstly, at each layer $j$, the same number of weights in $\bm{w}^{(j)}$ as the number of neurons $l^{(j)}$ are going to be produced, creating the summation, and enabling the approximated posterior $q(\bm{w}^{(j)}|\bm{\theta}^{(j)})$ and the prior $p(\bm{w}^{(j)})$ distributions to be contributed in the logarithm form to the loss $\mathcal{L}_\textrm{ELBO}$. Secondly, a full forward pass is required to compute the Negative Log-Likelihood of the outputs $-\log p(\mathcal{D}|\bm{w})$, and contributes to the loss as well. The practical implementation steps for one training epoch are summarized in Algorithm \ref{alg:bnn-train}. \par
The activation function has been chosen to be ReLU by default, as for the ensembles approach in Section \ref{sec:podensnn-learning}. However, reaching convergence for some discontinuous time-dependent problems was achieved with the $\phi: x \mapsto \tanh(x)$ activation, known to perform better in probabilistic models contexts \cite{Goodfellow-et-al-2016}.

\begin{algorithm}[h]
  \DontPrintSemicolon
  Feed the model with the dataset $\mathcal{D}$\;
  \For{each variational layer $1\leq j \leq d$}{
    $\bm{\epsilon}^{(j)} \sim \mathcal{N}(\bm{0}, \bm{I})$\;
    $\bm{w}^{(j)} = f(\bm{\theta}_\mu^{(j)}, \bm{\theta}_\rho^{(j)}, \bm{\epsilon}^{(j)})$\;
    $\bm{\theta}^{(j)}_\sigma = \textrm{softplus}(\bm{\theta}^{(j)}_\rho)$\;
    Sample the variational posterior $q(\bm{w}^{(j)}| \bm{\theta}^{(j)})\ {\sim}\ \mathcal{N}(\bm{\theta}_\mu^{(j)}, \bm{\theta}_\sigma^{(j)})$\;
    Sample the prior $p(\bm{w}^{(j)})$\;
    Contribute the posterior and prior values to the loss, $\mathcal{L}_{\textrm{ELBO}}\mathrel{{+}{=}}\log q(\bm{w}^{(j)}| \bm{\theta}^{(j)}) + \log p(\bm{w}^{(j)})$\;
    Perform the forward pass $\bm{h}^{(j)} = \phi(\bm{w}^{(j)}\bm{h}^{(j-1)} + \bm{b}^{(j)})$\;
  }
  Retrieve {each outputs pair} $\bm{\mu}^v, {\bm{\sigma}^v}^2$ from the NN\;
  Compute the likelihood from the outputs, $p(\mathcal{D}|\bm{w})\sim\mathcal{N}(\bm{\mu}^v, {\bm{\sigma}^v}^2)$\;
  Contribute the NLL to the loss, $\mathcal{L}_{\textrm{ELBO}}\mathrel{{+}{=}}-\log p(\mathcal{D}|\bm{w})$\;
  Backpropagate the gradients $\dfrac{\partial \mathcal{L}_\textrm{ELBO}}{\partial \bm{\theta}}$ to update the latent variables $\bm{\theta}$\;
\caption{Epoch training of a BNN via \emph{Bayes by Backprop} \cite{Blundell2015}.}
\label{alg:bnn-train}
\end{algorithm}

\subsection{Predictions}
Applying Algorithm \ref{alg:bnn-train} for each training epoch produces an optimal value of the variational parameters, referred to as $\bm{\theta}_\textrm{ELBO}$, which minimizes the loss function $\mathcal{L}_\textrm{ELBO}(\mathcal{D}, \bm{\theta})$ and defines the approximated posterior, $q(\bm{w}|\bm{\theta}_\textrm{ELBO})$. From this distribution, regular NN weights $\bm{w}$ can be drawn, and sample predictions can be produced by evaluating the network with a forward pass as in (\ref{eq:podensnn-hidden-states}), for any new input data $\bm{X}$. If new targets $\bm{v}$ are now considered to be predicted, it is possible to approximate the predictive posterior distribution (\ref{eq:podensnn-posterior}) as

\begin{align}
  p(\bm{v}|\bm{X}, \mathcal{D})&=\int p({\bm{v}}|\bm{X}, \bm{w})q(\bm{w}|\bm{\theta}_\textrm{ELBO})\ d\bm{w}\label{eq:posterior-theta}
  % &=\int p(t|x, f(\bm{\theta}))q(f(\bm{\theta})|\bm{\theta})\ d\bm{\theta}\label{eq:posterior-theta}.
\end{align}
It can be observed that considering one weights' configuration $\bm{w}_b$ sampled from the inferred distribution $q(\bm{w}|\bm{\theta}_\textrm{ELBO})$ from the optimal latent variables $\bm{\theta}_\textrm{ELBO}$, $p(\bm{v}|\bm{X}, \bm{w}_b)=p(\bm{v}|\bm{X}, f(\bm{\theta}_\textrm{ELBO}))$ represents the network output distribution with moments $(\bm{\mu}^v_{\bm{w}_b}(\bm{X}), \bm{\sigma}^v_{\bm{w}_b}(\bm{X})^2)$. Therefore, (\ref{eq:posterior-theta}) shows that the posterior predictive distribution is equivalent to averaging predictions from an ensemble of NNs, weighted by the posterior probabilities of their weights, $\bm{w}_b$. While each output distribution accounts for the variability in the data, or aleatoric uncertainty, (\ref{eq:posterior-theta}) tracks the variability in the model configuration, the epistemic uncertainty, via the $\bm{\theta}$-parametrized distribution and the integral. The mean of the predictions is hence given by
\begin{equation}
\mu_{\bm{X}} = \int \bm{v}\, p(\bm{v}|\bm{X}, \mathcal{D})\, d\bm{X} = \iint \bm{v} \,p(\bm{v}|\bm{X}, {\bm{\theta}}) q({\bm{\theta}}|\mathcal{D})\,d\bm{X} d{\bm{\theta}} = \int q(
{\bm{\theta}}|\mathcal{D}) \mu({\bm{\theta}})\, d|{\bm{\theta}}.
\end{equation}
By drawing $B$ samples $\bm{w}_b$ from $q(\bm{w}|\bm{\theta}_\textrm{ELBO})$, the mean of the predictions in the reduced space is approximated by
% we can effectively achieve (\ref{eq:posterior-theta}) and equivalently (\ref{eq:podensnn-posterior}) via a Monte-Carlo summation. The retrieved mixture is approximated here as a Normal distribution in the same way as Section \ref{sec:podensnn-learning}, with the moments
\begin{equation}
    \bm{\mu}_*^v(\bm{X}) = \dfrac{1}{B} \sum_{b=1}^{B}\bm{\mu}^v_{\bm{w}_b}(\bm{X}).
\end{equation}
As for the ensembles approach in Section \ref{sec:podensnn-learning}, we approximate each NN variance in one distribution, with the following, which allows for a fast estimation of the mixture in a single Gaussian,
\begin{equation}
    \bm{\sigma}_*^v(\bm{X})^2 = \dfrac{1}{B} \sum_{b=1}^{B} \left[\bm{\sigma}_{\bm{w}_b}^v(\bm{X})^2 + \bm{\mu}^v_{\bm{w}_b}(\bm{X})^2\right] - \bm{\mu}_*(\bm{X})^2.
\end{equation}
Expanded space predictions $(\bm{\mu}_*(\bm{X}), \bm{\sigma}_*(\bm{X})^2)$ are performed then to retrieve the full solution $\hat{u}_D(\bm{s})$, just as for the ensembles approach, with (\ref{eq:decompress-mu}) and (\ref{eq:decompress-var}).

%% file: content/fig-bnn-pod.tex
\tikzset{%
  every neuron/.style={
    circle,
    draw,
    minimum size=1cm
  },
  neuron missing/.style={
    draw=none, 
    scale=3,
    text height=0.333cm,
    execute at begin node=\color{black}$\vdots$
  },
}

\begin{figure}[t]
  \centering
  % \fbox{
    \resizebox{.8\textwidth}{!}{
\begin{tikzpicture}[x=1.5cm, y=1.5cm, >=stealth,
  pics/graph/.style={code={\draw[double=orange,white,thick,double distance=1pt,shorten
  >=0pt]      plot[variable=\t,domain=-0.5:0.5,samples=51] 
  ({\t},{#1});}}]

\foreach \m/\l [count=\y] in {1,2,missing,3}
  \node [every neuron/.try, neuron \m/.try] (input-\m) at (0,2.1-\y) {};

\foreach \m [count=\y] in {1,missing,2}
  \node [every neuron/.try, neuron \m/.try ] (hidden1-\m) at (2,2-\y*1.25) {};

\foreach \m [count=\y] in {1,missing,2}
\node [every neuron/.try, neuron \m/.try ] (hidden1-\m) at (2,2-\y*1.25) {};

\foreach \m [count=\y] in {1,missing,2}
  \node [every neuron/.try, neuron \m/.try ] (hidden2-\m) at (5,2-\y*1.25) {};

\foreach \m [count=\y] in {1,missing,2}
  \node [every neuron/.try, neuron \m/.try ] (output-\m) at (7,2.5-\y*0.8) {};

\foreach \m [count=\y] in {1,missing,2}
  \node [every neuron/.try, neuron \m/.try ] (outputrho-\m) at (7,-0.0-\y*0.8) {};

\foreach \l [count=\i] in {1,2,S}
  \draw [<-] (input-\i) -- ++(-1,0)
    node [above, midway] {$s_\l$};

\foreach \l [count=\i] in {1, l^{(1)}}
\node [above] at (hidden1-\i.north) {$h_{\l, 1}$};

\foreach \l [count=\i] in {1,l^{(d)}}
\node [above] at (hidden2-\i.north) {$h_{\l, d}$};

\foreach \l [count=\i] in {1,L}
  \draw [->] (output-\i) -- ++(2,0)
    node [above, midway] {$\mu^v_\l$};
\foreach \l [count=\i] in {1,L}
  \node [{
    circle,
    draw,
    minimum size=0.1cm
  }] (output-rho-step-\l) at (8,0.8-2*\i*0.8) {};
\foreach \l [count=\i] in {1,L}
\node [below] at (output-rho-step-\l.south) {$\textrm{softplus}$};

\foreach \l [count=\i] in {1,L}
  \draw [->] (outputrho-\i) -- (output-rho-step-\l)
    node [above, midway] {$\rho^v_\l$};

\foreach \l [count=\i] in {1,L}
\draw [->] (output-rho-step-\l) -- ++(1,0)
  node [above, midway] {$(\sigma^v_\l)^2$};

\foreach \i in {1,...,3}
  \foreach \j in {1,...,2}
    \draw [->] (input-\i) -- (hidden1-\j);

\foreach \i in {1,...,2}
  \foreach \j in {1,...,2}
    \draw [->] (hidden1-\i) -- (hidden2-\j);

    \foreach \i in {1,...,2}
    \foreach \j in {1,...,2}
    \draw [->] (hidden2-\i) -- (output-\j);

    \foreach \i in {1,...,2}
    \foreach \j in {1,...,2}
    \draw [->] (hidden2-\i) -- (outputrho-\j);

\node [align=center, above] at (0,2) {Input\\layer};
\node [align=center, above] at (2,2) {Hidden \\layer $(1)$};
\node [align=center, above] at (5,2) {Hidden \\layer $(d)$};
\node [align=center, above] at (7,2) {Output \\layer};

\path (input-2) -- (hidden1-1) pic[midway]{graph={-0.3+0.6*exp(-6*(\t+0.3)*\t)}};
\path (input-2) -- (hidden1-2) pic[midway]{graph={-0.3+0.6*exp(-25*(\t+0.15)*(\t+0.15))}};

\path (hidden2-1) -- (output-2) pic[midway]{graph={-0.3+0.6*exp(-25*(\t-0.12)*(\t+0.15))}};
\path (hidden2-2) -- (outputrho-1) pic[midway]{graph={-0.3+0.6*exp(-6*\t*\t)}};
% \path (input-1) -- (hidden1-1) node[graph={-0.3+0.6*exp(-6*\t*\t)}];
% \path (input-2) -- (hidden1-1) node[graph={-0.3+0.6*exp(-25*(\t+0.15)*(\t+0.15))}];

\node[fill=white,scale=4,inner xsep=0pt,inner ysep=5mm] at ($(hidden1-1)!.5!(hidden2-2)$) {$\dots$};
% \path (hidden1-1) -- (hidden2-1) node[above, midway] {$\bm{w} \sim p(\bm{w})$};
\path (hidden1-1) -- (hidden2-1) node[above, midway] {$\bm{\epsilon}\sim\mathcal{N}(\bm{0}, \bm{I})$};
\path (hidden1-1) -- (hidden2-1) node[below, midway] {$\bm{w} = f(\bm{\theta}_\mu, \bm{\theta}_\rho, \bm{\epsilon})$};

\end{tikzpicture}
  }
  % }
  \caption{$\hat{u}_{DB}(\bm{X}; \bm{\theta})\sim\mathcal{N}\left(\bm{\mu}^v(\bm{X}), \bm{\sigma}^v(\bm{X})^2\right)$, a probabilistic Bayesian Neural Network regression with a dual mean and variance output, and distributions on the weights{.}}
  \label{fig:bnn-reg}
\end{figure}
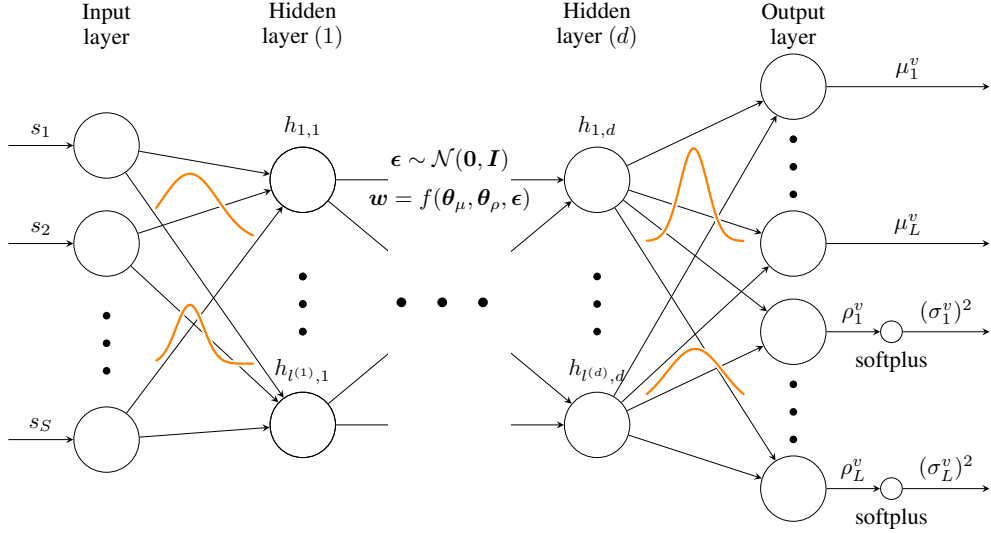

%% file: content/benchmarks-uq.tex
\FloatBarrier

\section{Benchmark with Uncertainty Quantification}\label{sec:podensnn-bench}
% \begin{figure}[t]
%   \centering
%   \fbox{
%     \parbox{\textwidth}{
% \includegraphics[width=.9\linewidth]{images/podensnn-ackley-graph-means.pdf}
%     }
%   }
%   \\ \parbox{0.75\textwidth}{
% \caption{The two graphs are contour plots of the prediction over the mean of the training samples $u_D(\bar{s_{\textrm{tst}}})$} 
%     \label{fig:podensnn-ackley-graph}
%   }
% \end{figure}
\par In this section, we assess the uncertainty propagation component of our framework against {a} steady and two-dimensional {benchmark}, known as the Ackley Function.
\subsection{Setup}
\par The library TensorFlow version 2.2.0 \cite{Abadi2016TensorFlow} is used for all results, while the SVD algorithm and various matrix operations are performed {using} NumPy, all in Python 3.8. To implement variational layers, we used the new TensorFlow Probability module in version 0.10.0, which allows for greater interoperability with regular networks \cite{Dillon2017}. Its source code and the corresponding results were validated in-house against a custom adaptation of the code presented in \cite{Krasser2020blog}. Documented source code {is} available at  \href{https://github.com/pierremtb/POD-UQNN}{https://github.com/pierremtb/POD-UQNN}, on both \texttt{POD-EnsNN} and \texttt{POD-BNN} branch{es}. \par
In almost all benchmarks, the \textit{activation function} on all hidden layers is the default ReLU nonlinearity $\phi:\,x \mapsto \max(0, x)$. At the same time, a linear mapping is applied to the output layer, since in a regression case, real-valued variables are needed as outputs. 
We perform normalization on all non-spatial parameters $\bm{s}$, to build the inputs $\bm{X}$ as
\begin{equation}
    \bm{X} = \dfrac{\bm{s} - \bar{\bm{s}}}{\bm{s}_{\textrm{std}}},
\end{equation}
with $\bar{\bm{s}}$ and $\bm{s}_{\textrm{std}}$ respectively the empirical mean and standard deviation over the dataset{. The two quantities are computed on each column to maintain the physical meaning, e.g., the time would be normalized against the mean time step and the standard deviation of the time steps, and not against space quantities.}

% \par Our dataset generation for these simulations have been optimized in Python through multi-threading and machine code Just-In-Time compilation, using the library Numba (\cite{Lam:2015:Numba}).

% \par As it was used earlier in physics-informed machine learning research of \cite{Raissi2019PINNs}, the quasi-newton LBGFS optimizer has been tried on the benchmark problems and proved to be overall less effective than Adam, \cite{kingma2014adam}, as far as convergence speed and result accuracies where concerned. It is, however, to note that a combination of the two—starting with Adam for a few iterations and then finishing the optimization via LBGFS, performed very well, yet required fine-tuning.

% \par Simple benchmarks like the following are functions that we can call, and that output a result. One or more of their parameters is non-spatial and stochastic, i.e., to get a reference solution, we need to run it a rather large number of times. Our baseline for these simulations is a sampling size of $N_{\textrm{val}}=500$ for the training/validation sets, and $N_{\textrm{tst}}$ for the test sets (which can be bigger for time-dependent cases), leading to long computation times if it is done without optimization in Python. One way to drastically reduce it is through multi-threading and machine code Just-In-Time compilation, using the library Numba, \cite{Lam:2015:Numba}. \par

To achieve GPU parallel training, we used the Horovod library \cite{Sergeev2018horovod}, which allowed us to efficiently train $M=5$ models on $M=5$ GPUs at the same time. This number is recommended as a good starting point in \cite{lakshminarayanan2017simple}. \par

% For instance, for the Ackley function, Section \ref{sec:podnn-bench-ackley}, the high-fidelity generation time shrunk from about 1 hour without Numba acceleration, to 10 minutes with it.

Numba optimizations have also been used for both POD and data generation, which allows for multiple threading and native code compilation within Python, and is especially useful for loop-based computations {\cite{Lam:2015:Numba}}. \par
It is also important to note that in practice, a hyperparameter can be added to ensure the stability of the output variance when going through the softplus function for positivity requirements in both approaches (POD-BNN and POD-EnsNN). Denoted as $\kappa$ with a default value of $1$, this hyperparameter is involved in the softplus function calls as
\begin{equation}
  \textrm{softplus}(x)=\log(1+\exp(\kappa x)).
\end{equation}

\begin{theorem}
For the following benchmarks and the subsequent applications in Section \ref{sec:podensnn-applications}, we chose a constant $20\%$ \emph{validation} split $\mathcal{D}_\textrm{val}$ of the generated dataset $\mathcal{D}$ from Equations (\ref{eq:podensnn-svd}–\ref{eq:podensnn-pod-red-basis}). The relative error $RE$ defined in (\ref{eq:re}) is computed at each training epoch for both the training set and the validation set. By keeping track of both, we try to avoid \emph{overfitting}. A manual \emph{early stopping} is therefore performed, in case the validation error might increase at some epoch $N_e$ in the training. No mini-batch split is performed, as our dataset is small enough to be fully handled in memory, and no improvement for using a mini-batch split was shown in our experiments. The final results are reported as well on a testing set generated for $N_\textrm{tst}$ different points in the domain $\Omega$.
\end{theorem}

\subsection{Stochastic Ackley function}\label{sec:podensnn-bench-ackley}
\begin{figure}[t]
  \centering
    \parbox{\textwidth}{
\includegraphics[width=1.0\linewidth]{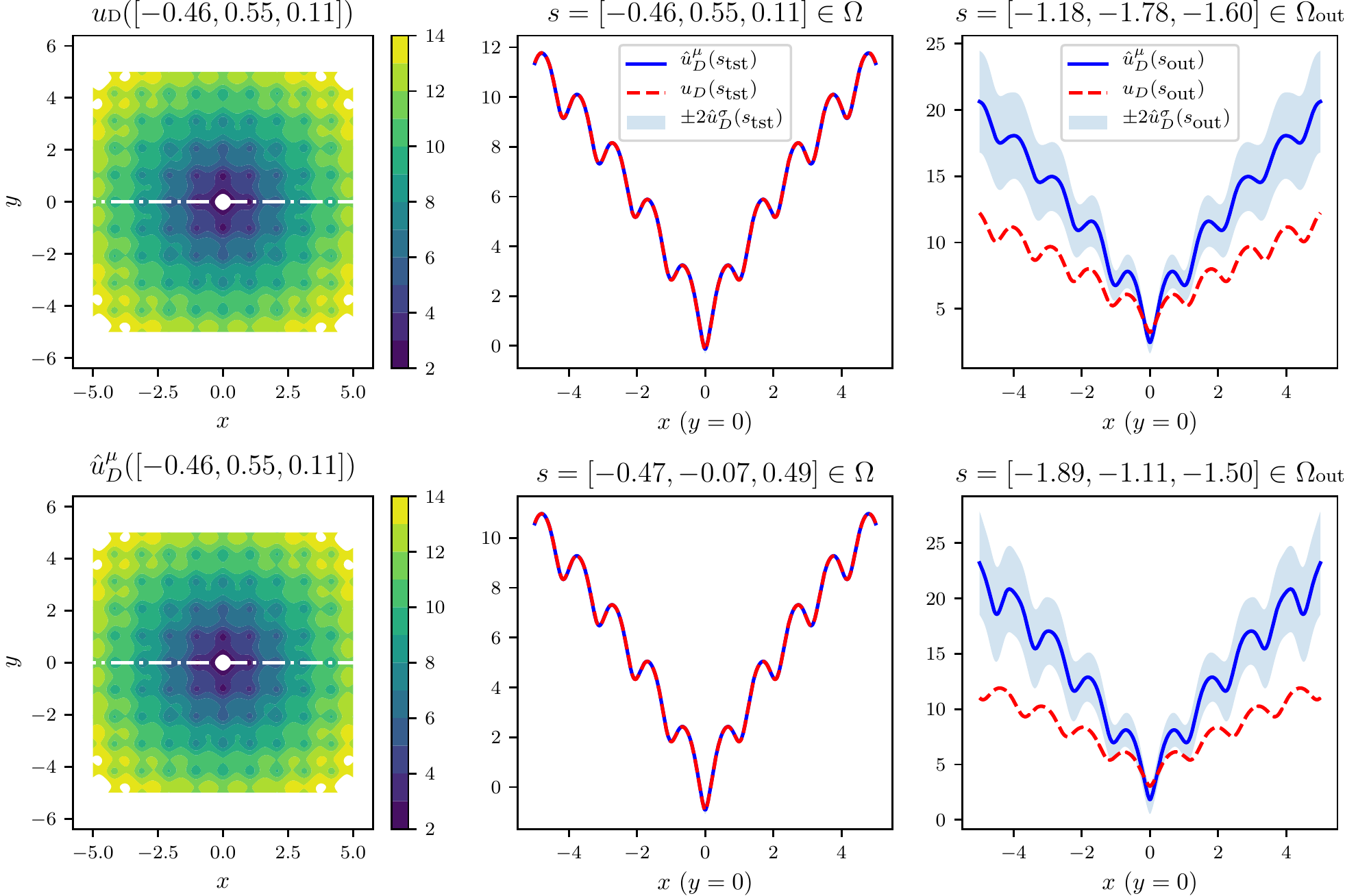}
    }
  \caption{Ackley Function (2D). The first column show{s} the contour plots of a random test sample with the predicted mean $\hat{u}_D^\mu$ on the bottom, and the true solution $u_D$ on top. The second column shows the predicted mean $\hat{u}_D^\mu$ and standard deviation $\hat{u}_D^\sigma$, and the true data $u_D$ {for} two random {samples} inside the training bounds and within the test set (top/bottom). The third column shows the results for {two} samples $s_{\textrm{out}}$, that are taken outside the dataset bounds and thus have more substantial uncertainties.
  }
    \label{fig:podensnn-ackley-graph-samples}
\end{figure}
\begin{figure}[t]
\includegraphics[width=1.0\linewidth]{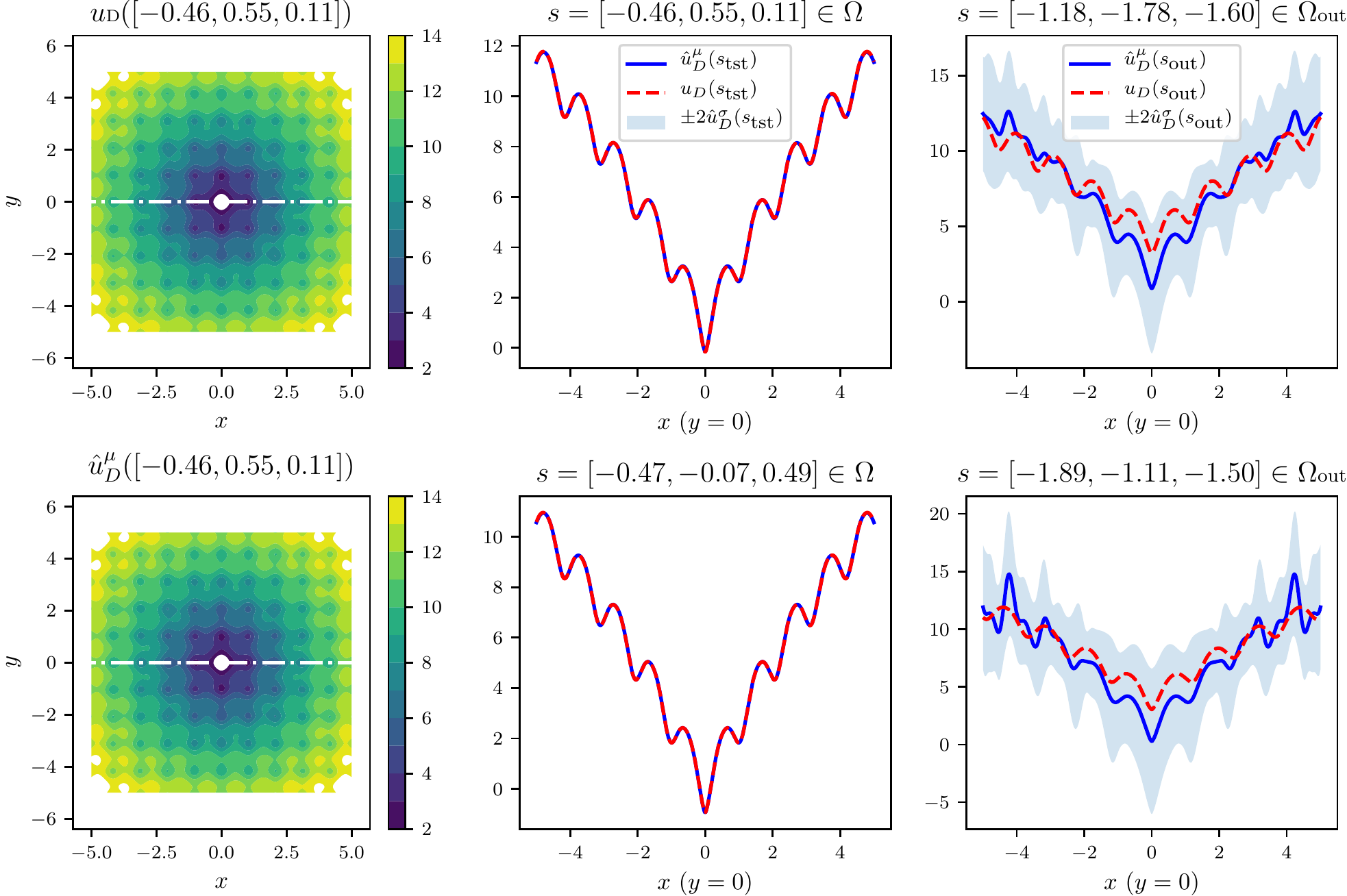}
\caption{Identical setup as in {Fig.}\ \ref{fig:podensnn-ackley-graph-samples} {but for the} Bayesian Neural Network regression{.}} 
    \label{fig:podbnn-ackley-graph-samples}
\end{figure}
As a first test case, we introduce a stochastic version of the Ackley function, a highly irregular baseline with multiple extrema presented in \cite{Sun2019}, which takes $P=3$ parameters. Being real-valued ($D=1$) and two-dimensional in space ($n=2$), it is defined as
\begin{align}
    u:\ \realspace^{2 + P} \to &\realspace\\
(x, y; \bm{s}) \mapsto &-20\left(1+0.1s_3\right) \exp\left(-0.2(1+0.1s_2) \sqrt{0.5(x^2+y^2)}\right) \nonumber\\
                     &- \exp\left(0.5(\cos(2\pi (1+0.1s_1) x) + \cos(2\pi(1+0.1s_1)y))\right) \nonumber\\ 
                     & + 2{1} \nonumber,
\end{align}
with the non-spatial parameters vector $\bm{s}$ of size $P=3$, and each element $s_i$ randomly sampled over $\Omega=[-1, 1]$, as in \cite{Sun2019}.

\par The 2D space domain $\Omega_{xy}=[-5, 5]\times[-5, 5]$ is {uniformly} discretized {with} $400$ {grid points per dimension}, leading {to} $H = 160,000$ {DOFs}. With $S = N_S = 500$ as our default number of samples of the parameters $\bm{s}$, we use a Latin Hypercube Sampling (LHS) strategy to sample each non-spatial parameter on their domain $\Omega=[-1,1]$ and generate the matrix of snapshots $\bm{U} \in \realspace^{H\times S}$, as well as $N_\textrm{tst}=100$ testing points to make a separate $\bm{U}_\textrm{tst}$. Selecting $\epsilon=10^{-10}$, $L=14$ coefficients are produced and matched by half of the final layer. The rest of the NN topology is chosen to include $d=3$ hidden layers, of widths $l^{(1)}=l^{(2)}=l^{(3)}=128$. 
A fixed learning rate of $ \tau=0.001$ is set for the Adam optimizer, as well as an L2 regularization with the coefficient $\lambda=0.01$. The training epochs count is $N_e={120,000}$, and a softplus coefficient of $\kappa=0.01$ is used.  \par
The training of each model in the ensemble took {on average 4 minutes and 5} seconds on each of the 5 GPUs, and the total, real-time of the parallel process was {6 minutes and 23} seconds. {This experiment and the following were performed on Compute Canada's Graham cluster (V100 GPUs), as well as Calcul Québec's Hélios cluster (K80 GPUs), depending on their respective availability.} To picture the random initialization of each model in the ensemble, the training losses were: {$\mathcal{L}=3.8457\times10^{-3}$, $3.1084\times10^{-3}$, $ 2.6536\times10^{-3}$, $4.1969\times10^{-3}$, and $2.2483\times10^{-3}$}, down from the initial losses:  {$\mathcal{L}_0=7.0313\times10^{9}$, $9.9997\times10^{0}$, $1.0000\times10^{0}$, $9.9999\times10^{-1}$, and $9.9997\times10^{-1}$}. 
The overall relative errors reached were {$RE_{\textrm{val}} = 0.17\%$ and $RE_{\textrm{tst}} = 0.16\%$}, for validation and testing, respectively. {For the following experiments, these low-level details will be gathered in \ref{apx:table}.}\par
The first column of {Fig.}\  \ref{fig:podensnn-ackley-graph-samples} shows two contour plots of the predicted mean across the testing set as well as the analytical solution, making it easy to quickly visualize the Ackley function, its irregularity and its various local extrema. The second column shows two different random samples within the same testing set with predicted and analytical values, while the third column contains \emph{out-of-distribution} cases, sampled in $\Omega_\textrm{out}$, defined as
\begin{equation}
  \Omega_\textrm{out}=[-2, -1]\cup[1, 2].
\end{equation}
The most important information revealed in {the} last column of {Fig.}\  \ref{fig:podensnn-ackley-graph-samples} is that the two parameters are sampled \emph{out-of-distribution}, meaning they are outside of the dataset bounds. We can see that the predicted mean, represented by the continuous blue line, is performing poorly compared to the red dashed line, which represents the true value. This predicted mean should be approximately the same as the point estimate prediction of a regular Deep Neural Network. And, even though our out-of-distribution mean prediction is indeed "off", thanks to the wide confidence zone defined by {two times the} standard deviations of the prediction, we get a warning that the model \emph{{does not} know}, and therefore it does not try to make a precise claim. To picture the difference in confidence between in- and out-of-scope predictions quantitatively, we computed {$MPIW_\textrm{tst}=0.15$ and $MPIW_\textrm{out}=10.0$}. \par
A similar experiment was then performed with the POD-BNN approach on the same dataset; the results are shown in {Fig.}\ \ref{fig:podbnn-ackley-graph-samples}. Two hidden variational layers of sizes $l^{(1)}=l^{(2)}=40$ were set up, with a number of epochs $N_e=120,000$ and a fixed learning rate of $ \tau=0.01$, as well as the softplus coefficient $\kappa=0.01$. The prior distribution was chosen to have the standard parameters $\pi_0=0.5$ and $\pi_2=0.1$, and we selected $\pi_1=4.0$. The trainable parameters $\bm{\theta}^{(j)}$ (weight or bias) of the $j$-th layer were randomly initialized, with
\begin{equation}\label{eq:podensnn-ackley-init}
\bm{\theta}^{(j)} = (\bm{\theta}^{(j)}_\mu, \bm{\theta}^{(j)}_\sigma) \sim \mathcal{N}\left(\bm{0}, \sqrt{\pi_0 \pi_1^2+(1-\pi)\pi_2^2}\bm{I}\right).
\end{equation}
The training time for the BNN approach on a single GPU was 5 minutes and 5 seconds, to reach overall relative errors of $RE_{\textrm{val}} = 0.68\%$ and $RE_{\textrm{tst}} = 1.11\%$, for validation and testing, respectively.  \par
 The same behavior can be observed from the Bayesian approach as for the ensembles, with tiny uncertainties predicted for the sample inside the training scope, which is expected because the data is not corrupted by noise. However, when predictions are made out-of-distribution, they are correctly pictured by a significant uncertainty revealed by the {shadow} around the predicted mean. Quantitatively, we report mean prediction interval widths of $MPIW_\textrm{tst}=0.11$ and $MPIW_\textrm{out}=3.22$, which in both cases is in the same order as in the Ensembles case. {In this first experiment, the same number of training epochs has willingly been used in both cases for comparison purpose. We have noticed that POD-EnsNN reached a proper convergence much earlier, and more generally, we may further stress that POD-BNN was computationally heavier.}

% \subsection{Comments}
% These two benchmarks,the stochastic Ackley function and the Burgers' equation, have shown the high performance of the models in three of the total of four cases. Evaluating with these benchmarks proved the flexibility of the ensembles approach with various types of problems, including multi-dimensional, time-dependent, smooth, or  discontinuous physical solutions.
% However, this evaluation also helped to reveal the difficulties involved in the Bayesian approach when discontinuities are issued for the underlying physical phenomenon. This approach is general in its essence, yet difficult to implement due to its inherent intractability involving approximations via Variational Inference. In its simplest version with the approximated posterior distribution $q(\bm{w}|\bm{\theta})$ considered as a uniform distribution, it corresponds to the ensembles approach, which  achieves excellent results. At the same time, the Bayesian approach, as first presented in \cite{Blundell2015} had more difficulty converging when discontinuities appeared in the physical solutions, which is not a trivial problem for Neural Networks in general, as discussed extensively in \cite{Llanas2008}. The  action taken to overcome this issue was to use the common but less wide-spread hyperbolic tangent activation function. It should be noted that increasing the expressivity of the network in the Bayesian case was achieved by going deeper with relatively {thin} layers (40), given that expanding their width would not allow us to train them correctly.

%% file: content/applications.tex
\FloatBarrier
\section{Flood Modeling Application: the Mille Îles River}
\label{sec:podensnn-applications}
After assessing how both the Deep Ensembles and the BNN version of the POD-NN model performed on {a 2D benchmark problem}, here we aim at a real-world engineering problem: flood modeling. The goal is to propose a methodology to predict probabilistic flood maps. Quantification of the uncertainties in the flood zones is assessed through the propagation of the input parameters' aleatoric uncertainties via the numerical solver of the Shallow Water equations.
\subsection{Background}\label{sec:podensnn-app-background}
\begin{figure}
  \centering
  % \fbox{
    \begin{tikzpicture}
      \draw (0,0) -- (8,0) node[midway, above]{$b=z=0$};
      \draw[blue] (8,1) -- (4,1) -- (4,2) -- (0,2);
      \draw[blue,->] (-1,1) node[anchor=east]{Inflow $Q_0$} -- (0,1); 
      \draw[blue,->] (8,0.5) -- (9,0.5)
        node[anchor=west] {Outflow $Q_\textrm{out}$}; 
      \draw[gray, thick, <->] (4.5, 1) -- (4.5, 2)
        node[above] {$\Delta h$};
      \draw[gray, thick, <->] (2, 0) -- (2, 2)
        node[above] {$h$};
    \end{tikzpicture}
  % }
 \vspace*{4pt}
    \caption{Simple representation of the water flow and main quantities before a dam break ($\Delta h>0$){.}}\label{fig:1d-case-schema}
\end{figure}
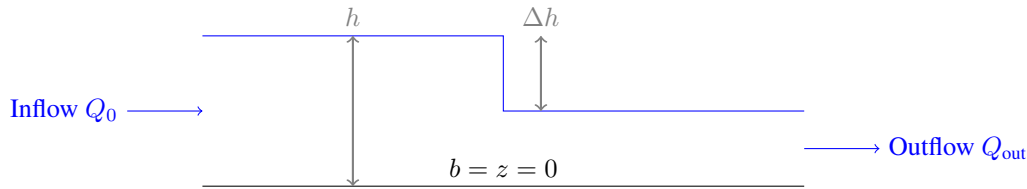
Just like wildfires or hurricanes, floods are natural phenomena that can be devastating, especially in densely populated areas. Around the globe, floods have become more and more frequent, and ways to predict them should be found in order to deploy safety services and evacuate areas when needed.  \par
The primary physical phenomenon in flooding predictions involves \emph{free surface flows} and is usually described by the Shallow Water equations  for rivers and lakes, extensively studied in \cite{Toro2001}, which, in their inviscid form, are defined as follows
\begin{align}
  \dfrac{\partial}{\partial t} \int_{\Omega_{xy}} \bm{U}\ d\Omega_{xy} + \int_{\partial \Omega_{xy}} ([\bm{G(\bm{U})}\ \bm{H(\bm{U})}]\cdot\bm{n})\ d\Gamma
  = \int_{\Omega_{xy}} \bm{S}(\bm{U})\,d\Omega_{xy}
  \quad \textrm{on}\ [0, T_s],
  % \dfrac{\partial}{\partial t} \int_{\Omega_{xy}} \bm{U}\ d\Omega_{xy} + \int_{\partial \Omega_{xy}} ([\bm{G(\bm{U})}\ \bm{H(\bm{U})}]\cdot\bm{n})\ d\Gamma \quad \textrm{on}\ [0, T_s],
\end{align}
with $T_s$ denoting the time duration, and 
\begin{align*}
  \bm{U} = \begin{bmatrix}
    h \\
    hv_x \\
    hv_y \\
  \end{bmatrix},\quad 
  \bm{G}(\bm{U}) = \begin{bmatrix}
    hv_x \\
    hv_x^2 + \frac{1}{2} g h^2 \\
    hv_xv_y \\
  \end{bmatrix},\quad 
  \bm{H}(\bm{U}) = \begin{bmatrix}
    hv_{y} \\
    {hv_xv_y} \\
    {hv_y^2 + \frac{1}{2} g h^2} \\
  \end{bmatrix},
\end{align*}
\begin{align*}
  \bm{S}(\bm{U}) = \begin{bmatrix}
    0 \\
    gh(S_{0_x}-S_{f_x}) \\
    gh(S_{0_x}-S_{f_y}) \\
  \end{bmatrix},\quad 
  \begin{bmatrix}
    S_{0_x} \\
    S_{0_y} \\
  \end{bmatrix}=-\nabla b,\
\end{align*}
\begin{align*}
  \textrm{and }  
  \bm{S}_f =
  \begin{bmatrix}
    S_{f_x} \\
    S_{f_y} \\
  \end{bmatrix}=
  \begin{bmatrix}
    \dfrac{m^2 v_x \sqrt{v_x^2+v_y^2}}{h^{4/3}} \\ 
    \dfrac{m^2 v_y \sqrt{v_x^2+v_y^2}}{h^{4/3}} \\ 
  \end{bmatrix}{.} 
\end{align*}
{Here} $h=\eta - b$ {is} the water depth, $\eta$ the free surface elevation of the water, $(v_x, v_y)$ the velocity components, $m$ the Manning roughness, $g$ the gravity {acceleration}, $\bm{S}_f$ the friction vector, and $b$ the bottom depth, or bathymetry, for a reference level. \par  
These equations can be discretized using finite volumes, as detailed in \cite{Toro2001} and \cite{zok2012cmame}. And, while we do already have decent numerical simulation programs to make these predictions, with well-validated software like \emph{TELEMAC} \cite{Galland1991telemac} or \emph{CuteFlow} \cite{zok2012cmame}, these are both computational- and time-expensive for multi-query simulations such as those used in uncertainties propagation. Therefore, it is difficult to run them in real-time, as they depend on various stochastic parameters. The POD-NN model, enriched with uncertainty quantification via Deep Ensembles and BNN, is designed to address this type of problem.

\subsection{In-context validation with a one-dimensional discontinuous test case}
\label{sec:app-riemann}
\begin{figure}[t]
  \centering
  \includegraphics[width=1.0\linewidth]{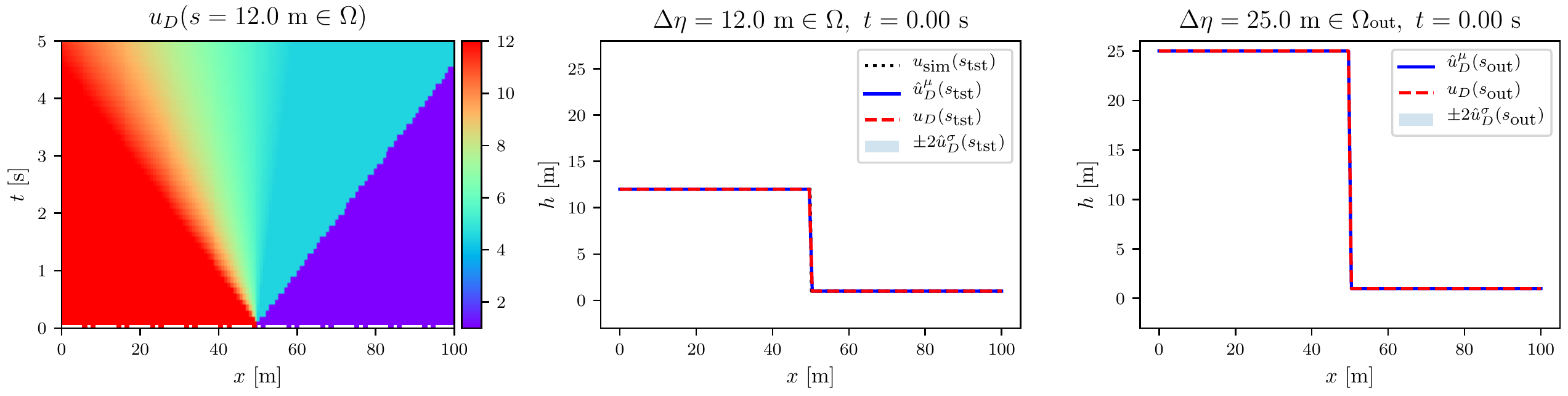}
  % \noindent\rule{0pt}{8pt}
  \includegraphics[width=1.0\linewidth]{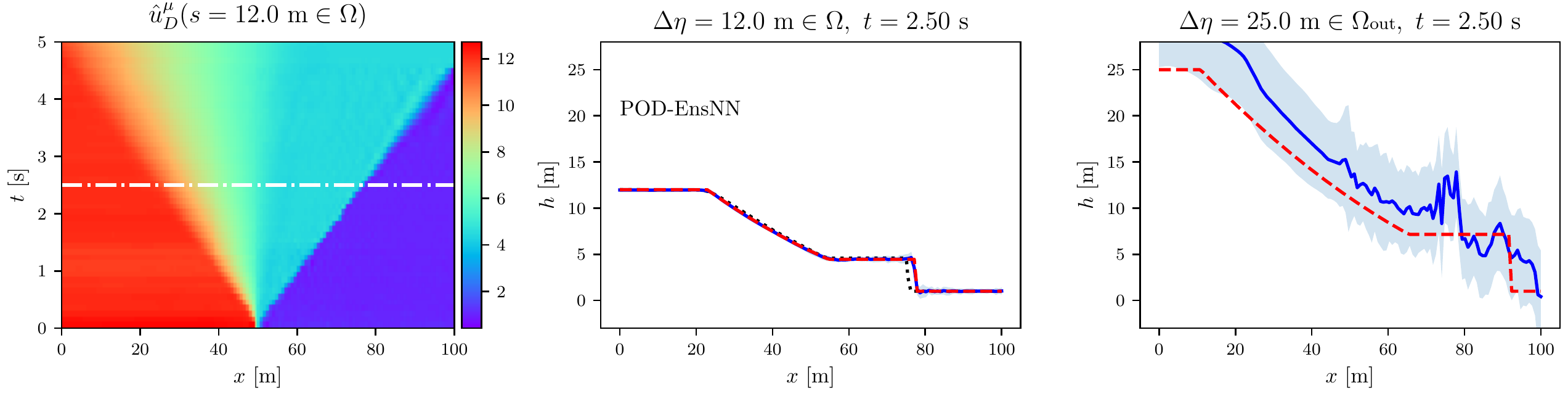}
  % \noindent\rule{0pt}{8pt}
  \includegraphics[width=1.0\linewidth]{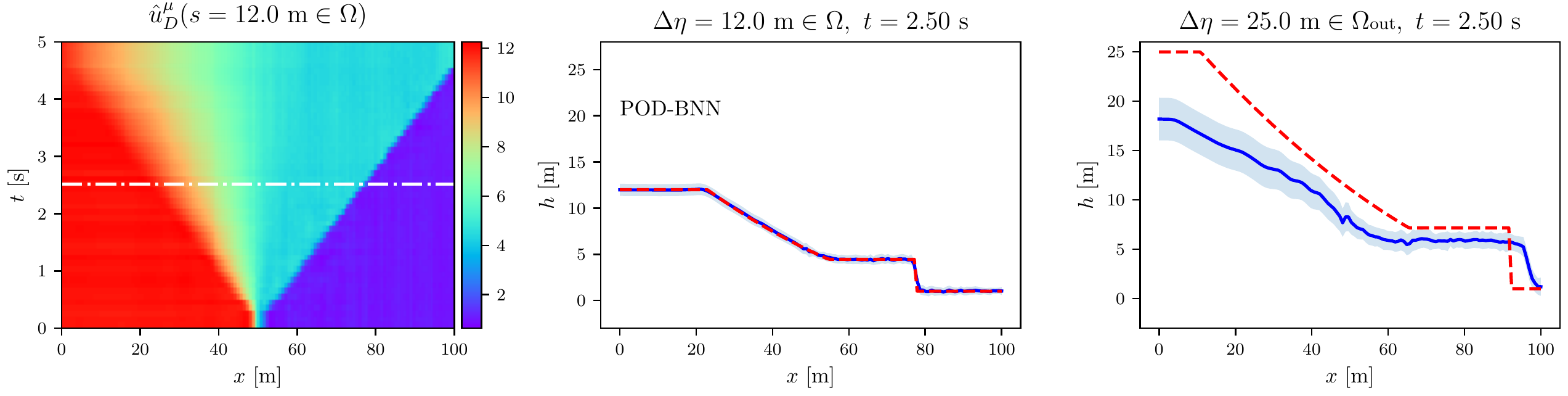}
  % \noindent\framebox[\linewidth]{\rule{0pt}{4cm}}
  \caption{1D test case for SWE, water elevation results. The first two columns show results for a random sample in the test set, while the last column shows a random sample taken out-of-distribution. The white lines on the color maps denote the time steps of the last two columns. The lines $u_\textrm{sim}$ are computed numerically by CuteFlow, and compared to the predicted mean $\hat{u}_D$ as well as the analytical value $u_D$. Ensembles are used on the second row, and BNNs on the third.} \label{fig:podensnn-shock-h}
\end{figure}
\begin{figure}[t]
  \centering
  \includegraphics[width=1.0\linewidth]{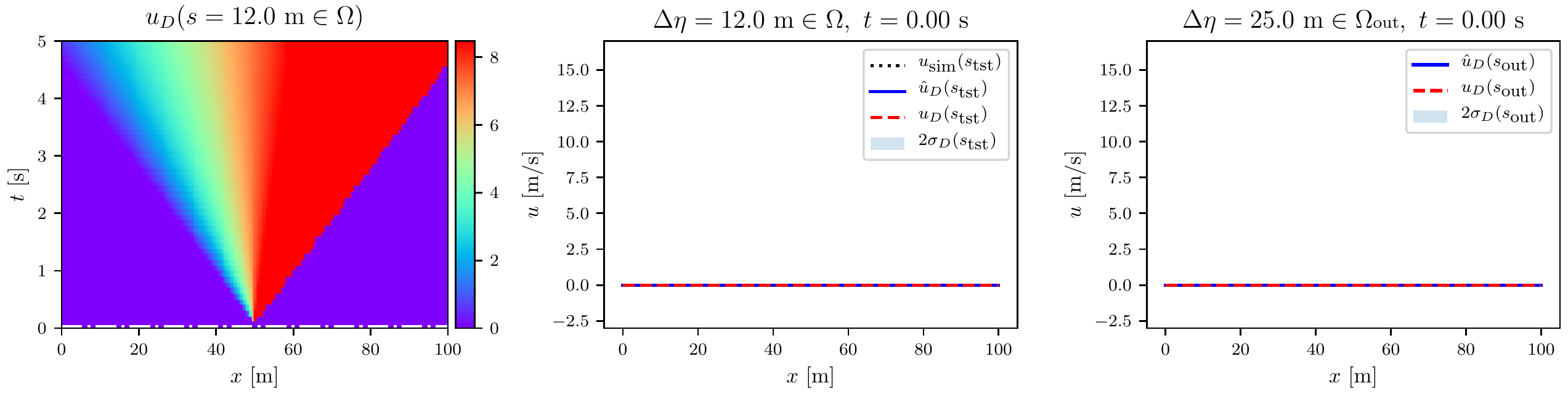}
  \includegraphics[width=1.0\linewidth]{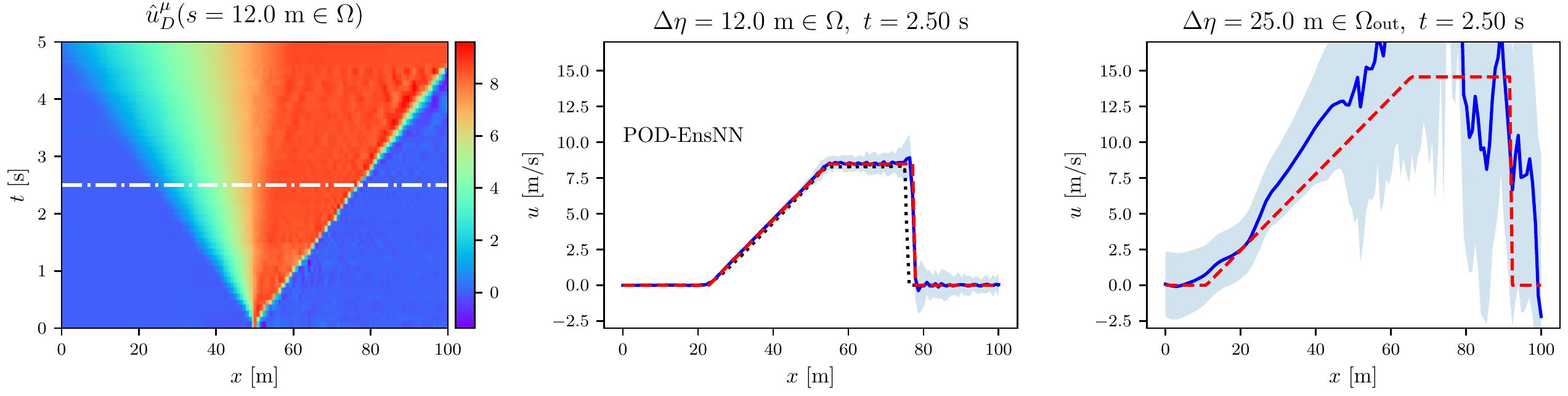}
  \includegraphics[width=1.0\linewidth]{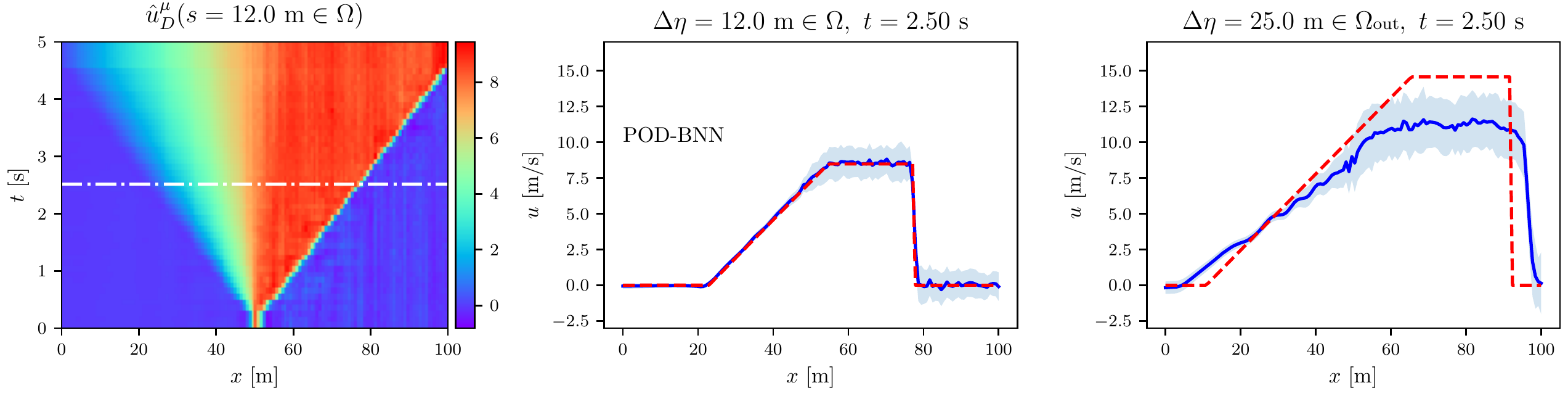}
  % \noindent\framebox[\linewidth]{\rule{0pt}{4cm}}
  \caption{1D test case for SWE, velocity results. The first two columns show results for a random sample in the test set, while the last column shows a random sample taken out-of-distribution. The white lines on the color maps denote the time steps of the last two columns. The lines $u_\textrm{sim}$ are computed numerically by CuteFlow, and compared to the predicted mean $\hat{u}_D$ as well as the analytical value $u_D$. Ensembles are used on the second row, and BNNs on the third.} \label{fig:podensnn-shock-u}
\end{figure}
We first put forward a one-dimensional test case in the Shallow Water equations  application, with two goals in mind. The first is to have a reproducible benchmark on the same equations that will be used for flood modeling, with an analytically available solution, and, therefore, generable data. The second is to make sure that the solver CuteFlow performs correctly with respect to the analytical solution, since in future experiments, it will be our only data source. \par
The 1D domain $\Omega_{x}=[0, 100]\ \textrm{m}$ is considered, with $N_x=132$ points, uniformly distributed. An initial condition is set up, with two levels of water depth, $s=\Delta h$ denoting the difference, that will act as our stochastic parameter in this study, with the water depth in the outflow fixed at $h=1\ \textrm{m}$. Following the initial discontinuity at $t=0$, we consider $N_t=50$ time-steps for snapshots sampling, separated by $\Delta t=0.1\ \textrm{s}$, in the domain $\Omega_t=[0,5]\ \textrm{s}$. There are ${D}=2$ DOFs per node, the water depth $h$ and the velocity $u$, leading to the total number of DOFs $H=264$.\par 
The dataset for the training/validation $\mathcal{D}=\{\bm{X}, \bm{v}\}$ of size $N=40$ was generated from an analytical solution {presented in \cite{Wu1999}, with a }uniform sampling $s$ in $\Omega=[2, 20]${. The same process generates }a testing dataset $\mathcal{D}_\textrm{tst}=\{\bm{X}_\textrm{tst},\bm{v}_\textrm{tst}\}$ of size $N_\textrm{tst}$, with $\bm{s}_\textrm{tst}=[2, 3, \ldots, 20]^\intercal\ \textrm{m}$. Additionally, the numerical finite volume solver CuteFlow was used to generate corresponding test solutions, from which we also exported $N_t=50$ solutions corresponding to the uniform analytical sampling after the initial condition. This solver was run with a 2D dedicated mesh of $25551$ nodes and $50000$ triangular elements specifically designed to represent this 1D problem in a compatible way for the solver. \par
{This problem is time-dependent, {making the matrix of snapshots $\bm{U}$ growing substantially}. We therefore make use of the two-step POD algorithm, presented in Section \ref{sec:podnn-reducing-dualpod}. The results are comforting: on a dataset of size $S=10,000$, with $N_t=100$, the time to compute the SVD decomposition shrunk from $0.63$ seconds to $0.51$ by switching from the regular POD to the two-step POD algorithm, which could result in a significant gain on more massive datasets. }
% The Python and TensorFlow implementation involves a topology of three layers of $l^{(1)}=l^{(2)}=l^{(3)}=256$ neurons for each network of the ensemble to account for nonlinearities. The POD manages the water depth $h$ as well as the velocity $u$, and its truncation is performed with $\epsilon=10^{-5}$, producing $L=79$ coefficients to be matched by half of the final layer. $N_e=100,000$ epochs are set for training, with a learning rate of $\tau=0.005$. L2 regularization is used with a coefficient of $\lambda=10^{-4}$, while {the} adversarial training {coefficient} is set to $\zeta=0.001$, and the softplus coefficient to its default value $\kappa=1$. \par
% The training of each model in the ensemble took 49, 50, 50, 51, and 51 seconds on each GPU, and the total duration of the parallel process was 1 minute and 15 seconds. This model training realized the following losses $\mathcal{L}=-1.9059\times10^{0}$, $-1.8365\times10^{0}$, $-1.0905\times10^{0}$, $-2.3819\times10^{0}$, and $-1.8352\times10^{0}$, down from the initial $\mathcal{L}_0=1.3699\times10^{2}$, $1.3085\times10^{2}$, $1.3238\times10^{2}$, $1.3352\times10^{2}$, and $1.3545\times10^{2}$ reported, indicating the variance within the ensemble due to the random initialization.
The overall relative errors were $RE_{\textrm{val}} = 3.56\%$ and $RE_{\textrm{tst}} = 3.93\%$, for validation and testing, respectively.  %\par
The results are displayed in {Fig.}\ \ref{fig:podensnn-shock-h} for the water depth, and in {Fig.}\ \ref{fig:podensnn-shock-u} for the velocity. On both figures, two samples are visible, with one within the testing set, pictured on the first column as a color map for graphic visibility, and plotted for two time-steps on the second column. The first time-step is the initial condition, which is well handled by the POD compression-expansion. A black line in the second column, representing the corresponding solution computed by the numerical solver CuteFlow, is very close to the analytical solution, and thus validates it for later use in more complex cases. \par 
A second out-of-distribution sample from $\Omega_\textrm{out}=[20, 30]\ \textrm{m}$ is plotted for the same two time-steps on the third column. The model performance within the training set was very good considering the nonlinearities involved, with relatively small uncertainties, and decreases when going out-of-distribution, as expected. We report mean predictive interval width values of $MPIW_\textrm{tst} = 1.64$ and $MPIW_\textrm{out} = 3.97$, {aligning with the qualitative comparison}.\par
% MPIW_tst: 9.6567e-01
% MPIW_tst_out: 5.5610e+01
% MPIW_tst: 3.6507e+00
% MPIW_tst_out: 1.2035e+01

% MPIW_tst: 1.6456e+00
% MPIW_tst_out: 3.9661e+00
% bnn
% MPIW_tst: 1.2252e+00
% MPIW_tst_out: 2.0490e+0

The Bayesian approach was also applied to this discontinuous problem; the results are indicated in the last row of {Fig.}\ \ref{fig:podensnn-shock-h} and \ref{fig:podensnn-shock-u}.
%The training parameters were as follows: $N_e=70,000$, $\tau=0.01$, $\kappa=0.01$, $\zeta=0.001$, and the prior settings $\pi_0=0.5, \pi_1=0.2, \pi_2=0.1$. 
{W}e had to resort to a tanh activation function to reach a decent convergence, but the out-of-distribution warning is not present, as shown in the third column of both figures. 
%Three hidden variational layers were used, each the same size $l^{(1)}=l^{(2)}=l^{(3)}=256$, just as with the Ensembles. The initialization of the weights is achieved using the method proposed in (\ref{eq:podensnn-ackley-init}). Using one GPU, the training took 17 minutes and 22 seconds to complete. 
The overall relative errors  were $RE_{\textrm{val}} = 6.30\%$ and $RE_{\textrm{tst}} = 5.32\%$, for validation and testing, respectively. {Values for the mean predictive interval width are  $MPIW_\textrm{tst} = 1.23$ and $MPIW_\textrm{out} = 2.05$.}\par
{Here,} the POD-BNN {performance is a bit less striking compared to POD-EnsNN, in connection with the computational costs and despite our best tuning efforts.} 
{ This evaluation helped to reveal the difficulties involved in the Bayesian approach when discontinuities are issued for the underlying physical phenomenon. This approach is general in its essence, yet difficult to implement due to its inherent intractability involving approximations via Variational Inference. In its simplest version with the approximated posterior distribution $q(\bm{w}|\bm{\theta})$ considered as a uniform distribution, it corresponds to the ensembles approach, which  achieves excellent results. At the same time, the Bayesian approach, as first presented in \cite{Blundell2015} had more difficulty converging when discontinuities appeared in the physical solutions, which is not a trivial problem for Neural Networks in general, as discussed extensively in \cite{Llanas2008}. The  action taken to overcome this issue was to use the common but less wide-spread hyperbolic tangent activation function.}
% It should be noted that increasing the expressivity of the network in the Bayesian case was achieved by going deeper with relatively {thin} layers (40), given that expanding their width would not allow us to train them correctly.} 
\par
Nonetheless, this test case allowed for a great benchmark of the numerical simulator and is another example showcasing the flexibility of the ensembles approach. We move on to real-world examples with probabilistic flooding predictions, involving first a steady context.

\subsection{Probabilistic flooding maps}
\subsubsection{River model setup}
\label{sec:app-river-setup}
\begin{figure}[t]
  \centering
  \parbox{\linewidth}{
    \centering
    \includegraphics[width=.5\linewidth]{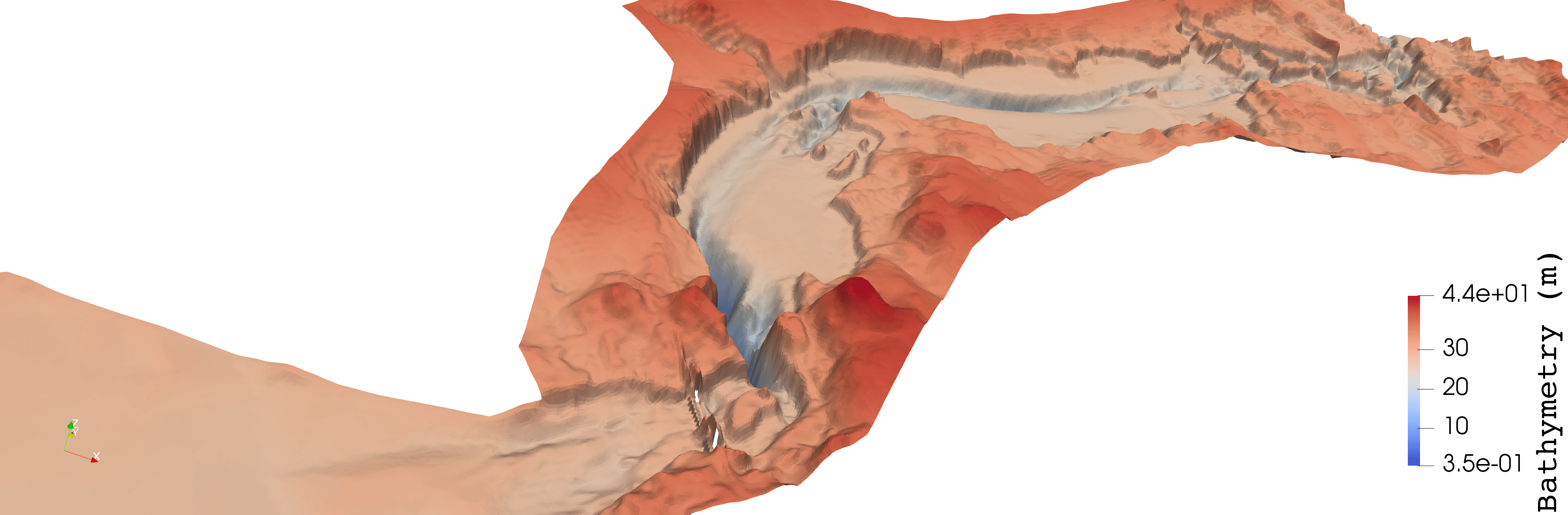}
    \vspace*{2pt}
  }
  \parbox{\linewidth}{
    \centering
    \includegraphics[width=.5\linewidth]{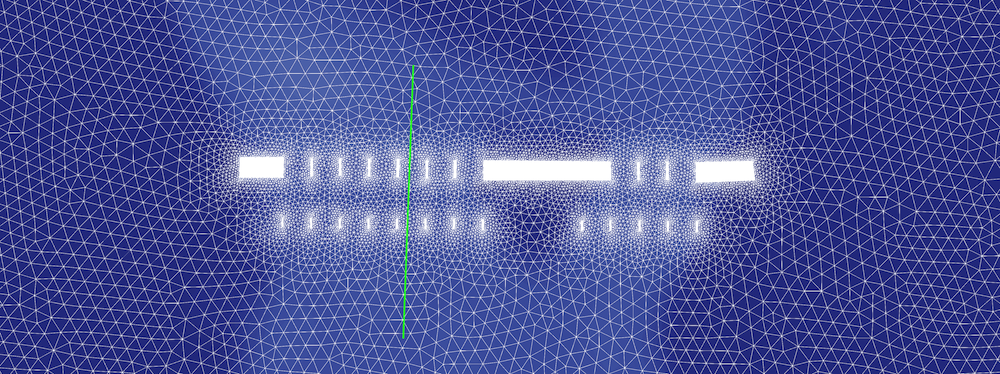}
    \vspace*{1pt}
  }
  \caption{Setup for the Mille Îles river in Laval, QC, Canada. On top is a representation of the river's bathymetry, given by the \emph{Communauté Métropolitaine de Montréal}, and below, the portion of the triangle-based mesh around the piers of a bridge, which features refinements. The green line indicates  a cross-section $\bm{x}’$, studied later.} \label{fig:podensnn-river}
\end{figure}

Our domain $\Omega_{xy}$ is composed of an unstructured mesh of $N_{xy}=24361$ nodes, connected in 481930 triangular elements. It is represented in {Fig.}\ \ref{fig:podensnn-river}. Each node has in reality 3 degrees of freedom, but only $N_{\textrm{val}}=1$ degree of freedom, the water depth $h$, will be considered in this study, leading to the global number of DOFs to be $H=24361$ for the POD snapshots.
\par For this first study, we will consider the time-independent case, and have at our disposal a dataset of $S=180$ samples for different inflow discharge ($Q_0$) values used for training{, the varying parameter here}, and another  of $S=20$ used for testing, with the solution computed numerically with the software CuteFlow. Both datasets were  uniformly sampled before being split into the domain $\Omega=[800, 1200]\ \textrm{m}^3\textrm{s}^{-1}$. This domain was chosen to be just  above the regular flow in the river of $Q_r=780\ \textrm{m}^3\textrm{s}^{-1}$, \cite{zokagoa}.
\subsubsection{Results}
\label{sec:app-river-results}
\par We selected a POD truncating criterion of $\epsilon=10^{-10}$, producing $L=81$ coefficients to be matched by half of the final layer. The ReLU activation function is chosen. No mini-batching is performed, i.e., the whole dataset is run through at once for each epoch. 
% For the ensembles approach, we chose a number of epochs $N_e=120,000$, a learning rate of $\tau=0.03$, a low regularization coefficient $\lambda=10^{-8}$, a default softplus coefficient of $\kappa=1.0$, and disabled adversarial training. Each network featured three hidden layers of equal size $l^{(1)}=l^{(2)}=l^{(3)}=128$. \par
% The training of each model in the ensemble took 4 minutes 19 seconds, 4 minutes 20 seconds, 4 minutes 20 seconds, 4 minutes 21 seconds, and 4 minutes 21 seconds on each GPU, and the total duration of the parallel process was 4 minutes and 45 seconds. Again, to show the diversity in the five models, here are the final training losses: $\mathcal{L}=-2.2240\times10^{0}$, $-3.5421\times10^{0}$, $-3.5199\times10^{0}$, $-2.3894\times10^{0}$, and $-3.5003\times10^{0}$, down from the initial $\mathcal{L}_0=2.6522\times10^{4}$, $2.6075\times10^{4}$, $2.6522\times10^{4}$, $2.5357\times10^{4}$, and $2.6958\times10^{4}$. 
The overall relative errors reached were $RE_{\textrm{val}} = 1.90\%$ and $RE_{\textrm{tst}} = 1.46\%$, for validation and testing, respectively.  \par
\begin{figure}[t]
  \centering
  \parbox{\textwidth}{
    \centering
    \parbox{0.81\linewidth}{
      \hspace*{6pt}
    \subfloat[View from afar of a random test sample $Q_0=884.4\ \textrm{m}^3/\textrm{s}$. This shows an iso-contour at $h=0.05\ m$, with its boundary being the flooding lines. For illustration purposes, the overall predicted relative water height $h_\textrm{pred}$ has been pictured throughout. The green box shows the location of the close-up shots presented below in (b) and (c).]{
      \fbox{
         \includegraphics[width=\linewidth]{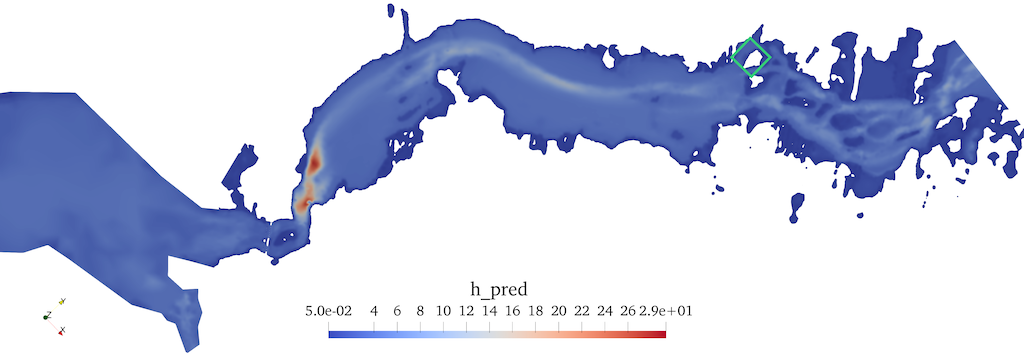}
      }
    }
  }
    \parbox{0.9\linewidth}{
  \subfloat[Random test sample with two levels of zoom, incoming flow of $Q_0=884.4\ \textrm{m}^3/\textrm{s}$]{
    \includegraphics[width=\linewidth]{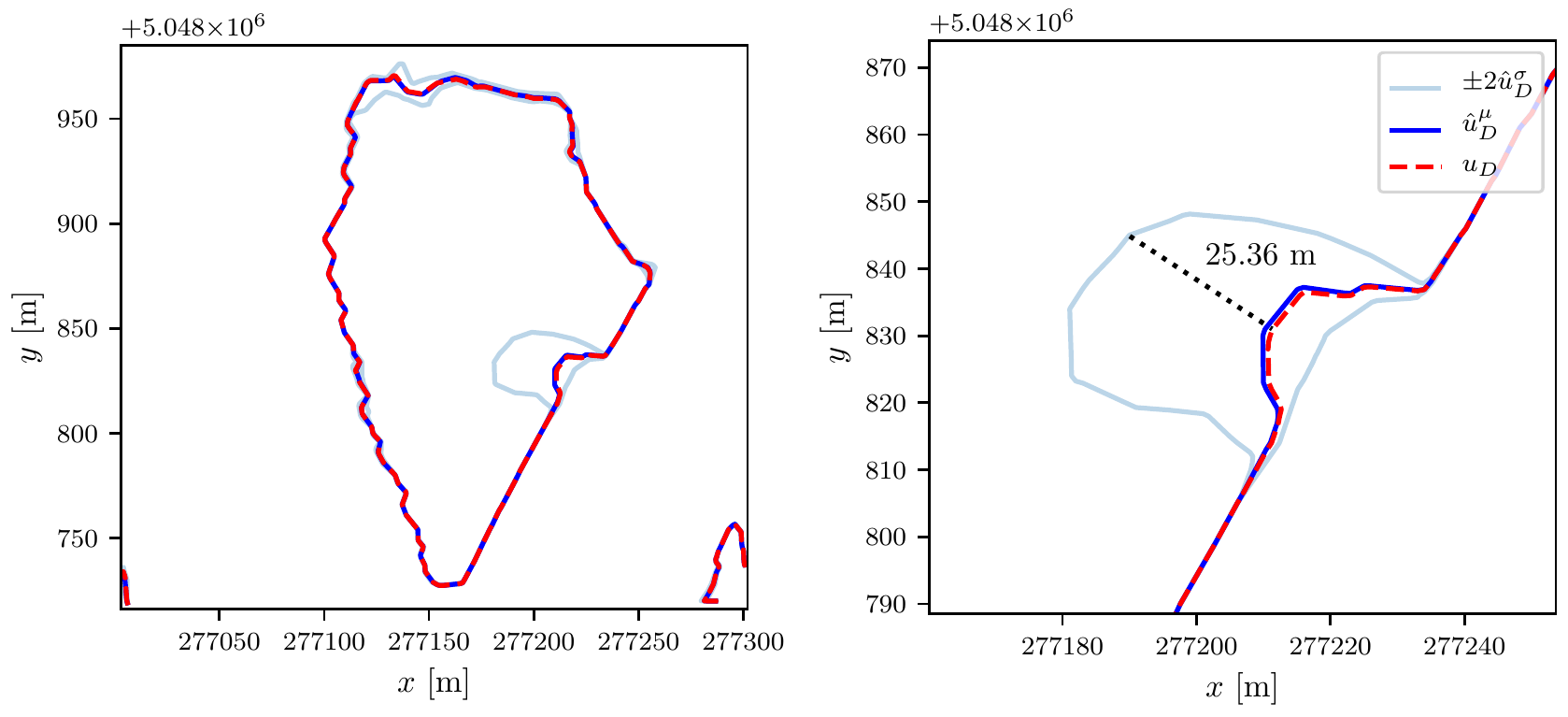}
    }
  }
    \parbox{0.9\linewidth}{
  \subfloat[Random test sample with two levels of zoom, incoming flow of $Q_0=1159.8\ \textrm{m}^3/\textrm{s}$]{
    \includegraphics[width=\linewidth]{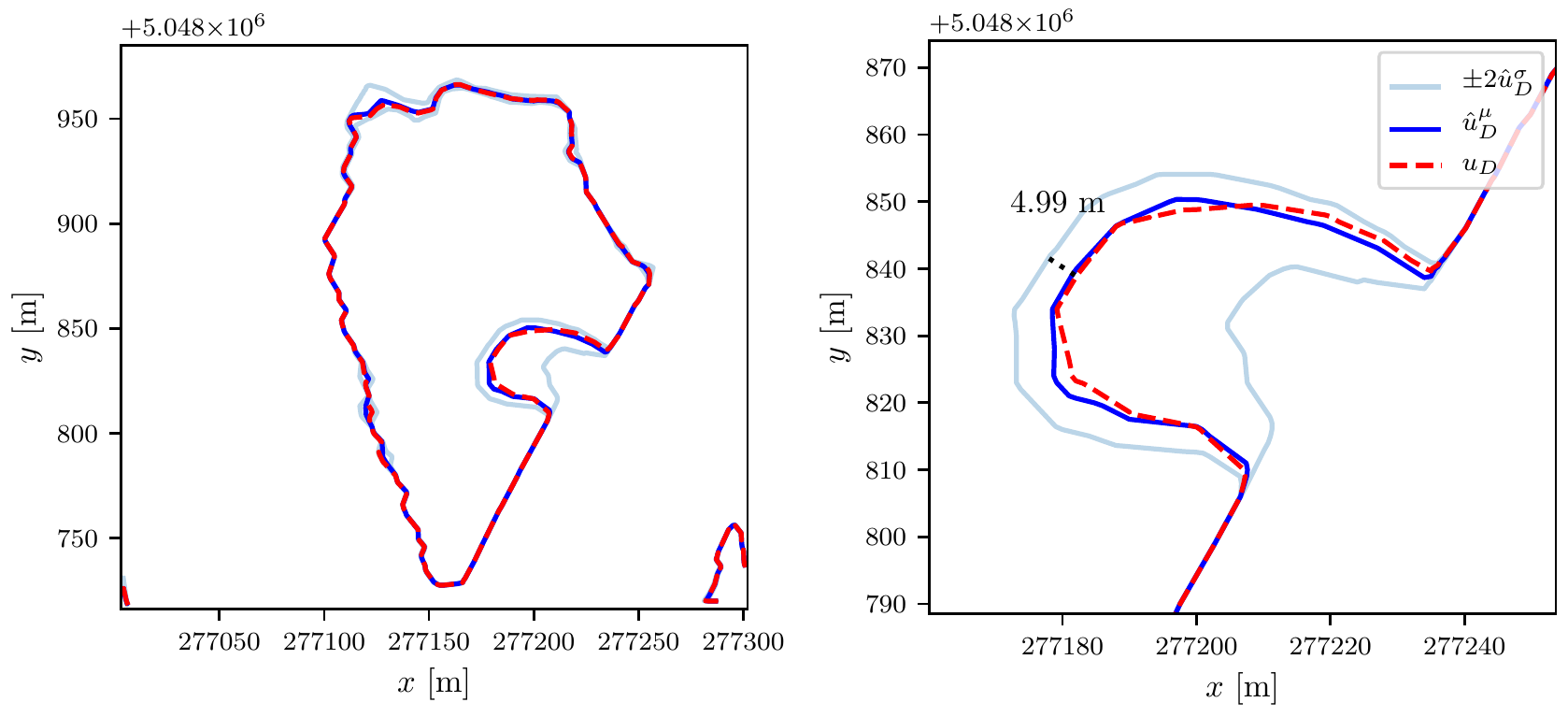}
    }
  }
  }
  \caption{POD-EnsNN application: flood modeling on the Mille Îles river, represented in (a). Flooding lines at $h=0.05\ m$ are shown on the close-up shots (b-c), with the red lines for the CuteFlow solution, and the light blue lines representing the end of the predicted confidence interval $\pm2\sigma_D$. The distance between the simulated value and the upper bound is measured.} \label{fig:podensnn-sw-map}
\end{figure}

{Fig.}\ \ref{fig:podensnn-sw-map} shows random test predictions using the open-source visualization software Paraview \cite{ahrens2005paraview} on two random samples for the water depth $h$. We can see that the flooding limits, achieved by slicing at $h=0.05\ m$ of water depth—in place of $0$ for stability, are very well predicted when compared to the simulation results from CuteFlow (red line). The light blue lines can be retrieved {by} adding $\pm2$ {times the} standard deviations on top of the mean predictions, depicted by the blue body of water {These additional light blue lines}  define the confidence interval of the predicted flood lines. We consider that having this probability-distribution outcome instead of the usual point-estimate prediction of a regular network in the POD-NN framework is a step forward for practical engineering.  \par
\begin{figure}[t]
  \vspace{-11pt}
  \centering
  \parbox{\textwidth}{
    \centering
    \parbox{0.81\linewidth}{
      \hspace*{6pt}
    \subfloat[View from afar of a random test sample $Q_0=884.4\ \textrm{m}^3/\textrm{s}$. This shows an iso-contour at $h=0.05\ m$, with its boundary being the flooding lines. For illustration purposes, the overall predicted relative water height $h_\textrm{pred}$ has been pictured throughout. The green box shows the location of the  close-up shots presented below in (b) and (c).]{
      \fbox{
    \includegraphics[width=\linewidth]{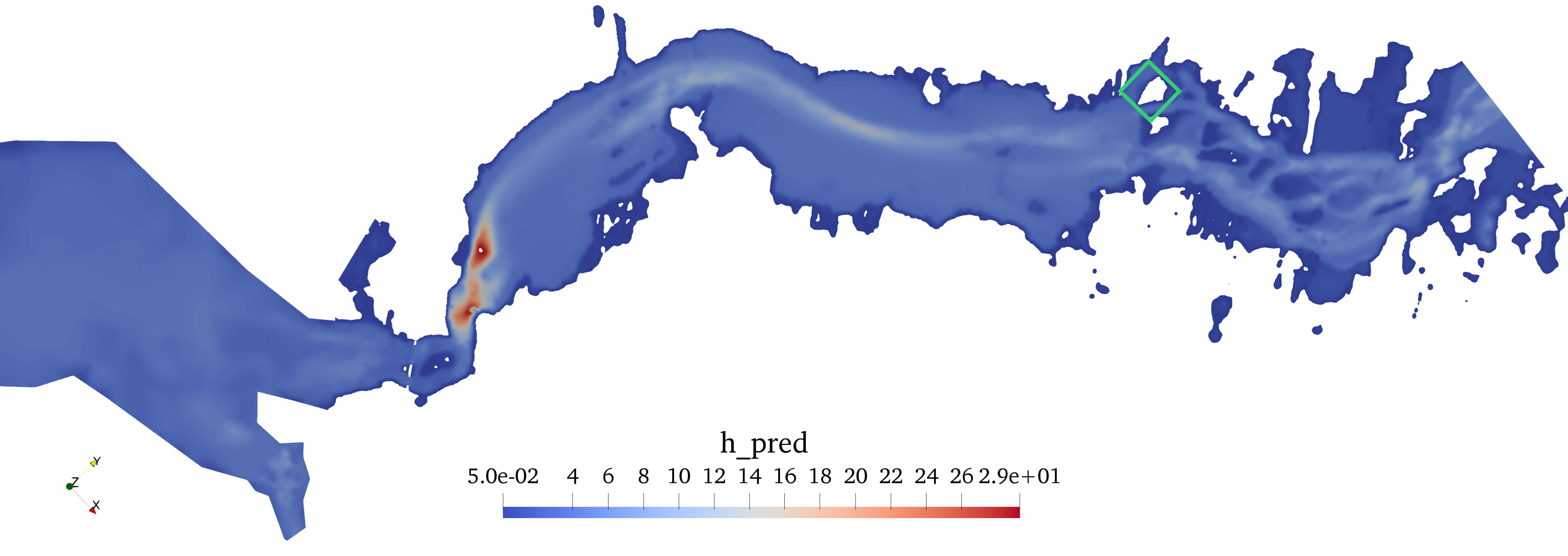}
      }
    }
  }
  % \vspace{5mm}
    \parbox{0.9\linewidth}{
  \subfloat[Random test sample with two levels of zoom, incoming flow of $Q_0=884.4\ \textrm{m}^3/\textrm{s}$]{
    \includegraphics[width=\linewidth]{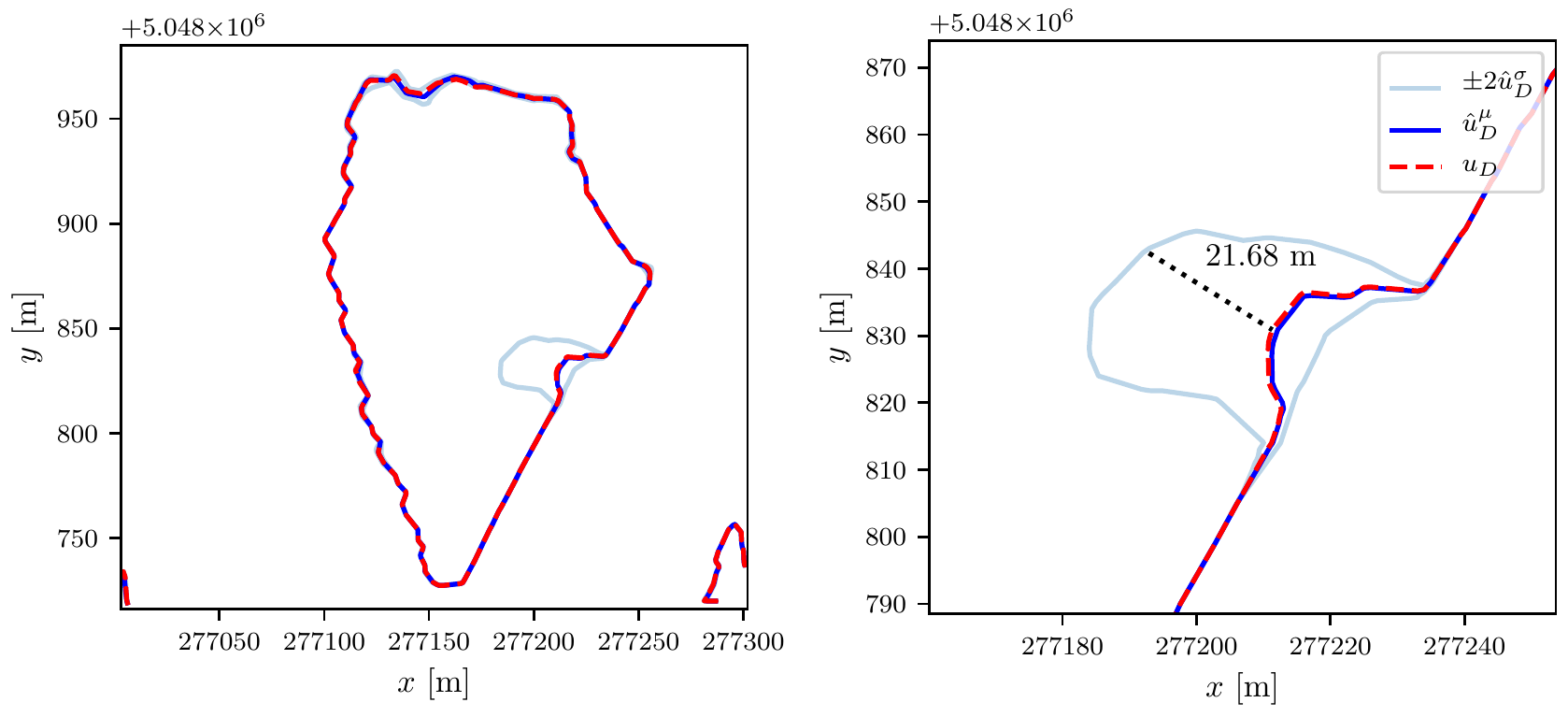}
    }
  }
  % \vspace{-5mm}
    \parbox{0.9\linewidth}{
  \subfloat[Random test sample with two levels of zoom, incoming flow of $Q_0=1159.8\ \textrm{m}^3/\textrm{s}$]{
    \includegraphics[width=\linewidth]{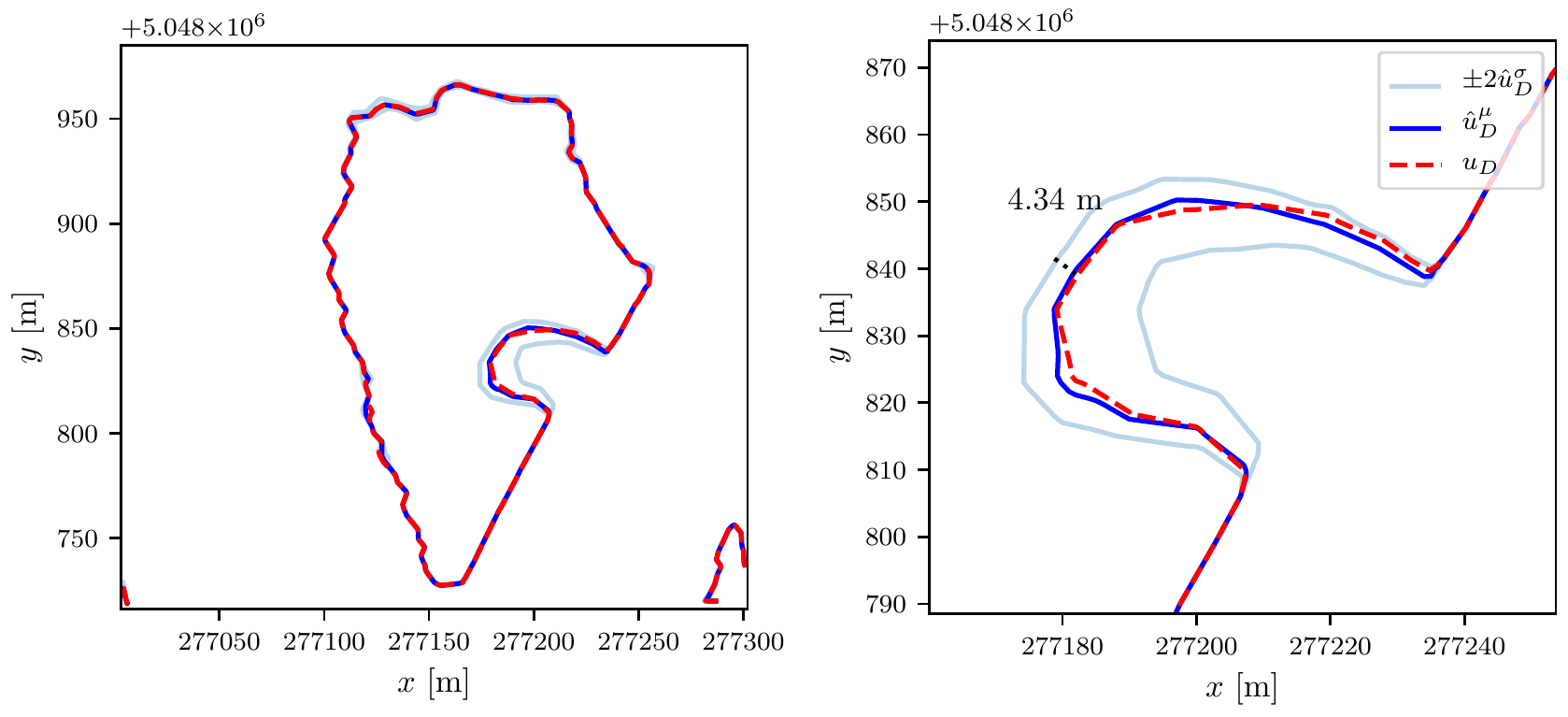}
    }
  }
  }
  \caption{POD-BNN application: flood modeling on the Mille Îles river, represented in (a). Flooding lines at $h=0.05\ m$ are shown on the close-up shots (b-c), with the red lines for the CuteFlow solution, and the light blue lines representing the end of the predicted confidence interval $\pm2\sigma_D$. The distance between the simulated value and the upper bound is measured.} \label{fig:podbnn-sw-map}
\end{figure}

%For the Bayesian approach, we picked a number of epochs $N_e=300,000$, a learning rate of $\tau=0.01$, a softplus coefficient of $\kappa=0.01$, and the prior parameters $\pi_0=0.5$, $\pi_1=4$, $\pi_2=0.1$. Each network featured three hidden layers of equal size $l^{(1)}=l^{(2)}=l^{(3)}=40$.
{As in the previous experiments, we settled on a thinner architecture in the POD-BNN, which facilitates training considering the computational burden of POD-BNN.} %\par
{Fig.}\ \ref{fig:podbnn-sw-map} depicts the same random test predictions as {Fig.}\ \ref{fig:podensnn-sw-map}. The flooding limits are also very well predicted when compared to the simulation results from CuteFlow (red line). The confidence interval around these predictions is very similar to the one predicted by the POD-EnsNN, and as for the distances measured to verify it: we found a distance between the predicted mean value and the upper confidence bound of $d_{2\sigma}=25.36\ \textrm{m}$ for the POD-EnsNN results compared to $d_{2\sigma}=24.58\ \textrm{m}$ for the POD-BNN results on the first close-up shot (b), and $d_{2\sigma}=4.99\ \textrm{m}$ versus $d_{2\sigma}=4.34\ \textrm{m}$, respectively, for the second close-up shot (c). While not being exactly equal, we assume that having the same order of magnitude is a solid accomplishment. In this application, no convergence issues for the Bayesian approach have been observed with the default configuration of a mixture prior and ReLU activation function, compared to previous attempts. These earlier efforts were notably performed on highly nonlinear and time-dependent test cases, where the Variational Inference steps were certainly facing harder circumstances. \par
\begin{figure}[t]
 % \vspace{-11pt}
  \centering
  \subfloat[POD-EnsNN]{
  \includegraphics[width=0.3\textwidth]{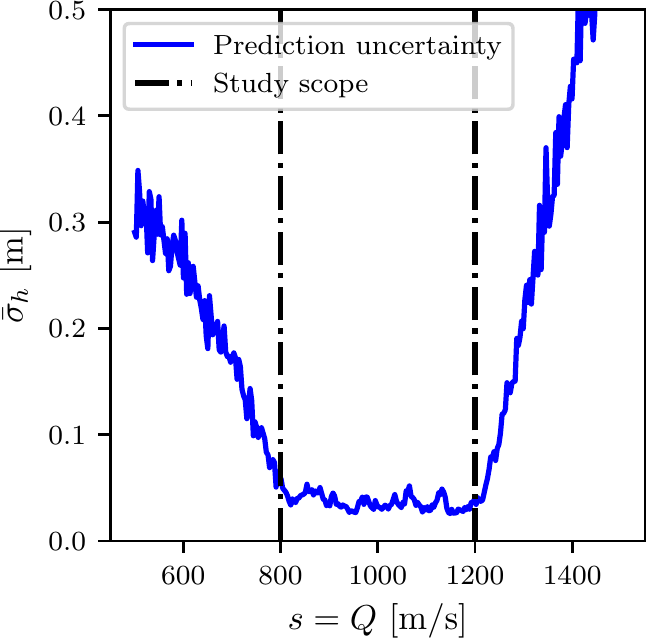}
  }
  \hspace{10pt}
  \subfloat[POD-BNN]{
  \includegraphics[width=0.3\textwidth]{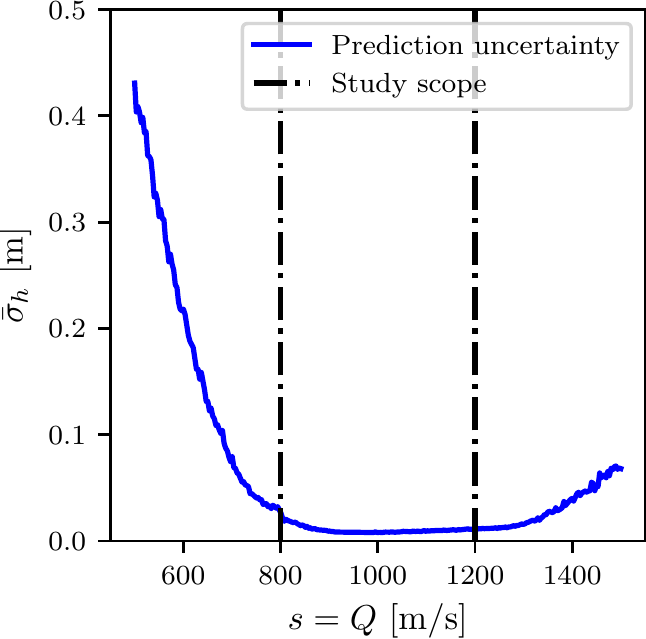}
  }
  \caption{Uncertainties on the flooding case. Visualization of the average uncertainties for a range of inputs with the two approaches. The two vertical black lines denote the boundaries of the training and testing scopes.} \label{fig:podensnn-sw-in-n-out}
\end{figure}

Finally, to make sure that our \emph{out-of-distribution} predictions were not just coincidences in the previous benchmark (see Section \ref{sec:podensnn-bench-ackley}), we also sampled new parameters from the whole $\Omega_{\textrm{out}} \cup \Omega$ domain, retrieved the mean across all DOFs of the predicted standard deviation, and rendered it in {Fig.}\ \ref{fig:podensnn-sw-in-n-out}. We observe that uncertainties snowball as soon as we exited the space where the model \emph{knows}, just as expected. Nonetheless, it is easy to see  the difference in the magnitude of increase when leaving the training bounds, which is much higher in the case of the POD-EnsNN when compared to the POD-BNN. The choice of the prior in the latter has shown to have an impact on this matter, {and a more thorough choice could further improve the performance of POD-BNN}. \par
% \FloatBarrier
\clearpage
\clearpage
\subsubsection{Contribution to standard uncertainty propagation}
\label{sec:app-river-up}
Instead of considering the domain of the sampled inflow $\Omega$ as simply a dataset, in the field it is often used as the source of random inputs around a central, critical point for \emph{uncertainty propagation} tasks, as performed in a similar context in \cite{zokagoa}. For this purpose, the use of a surrogate model is mandatory, since we wish to approximate the statistical moments of the output distributions to the model, i.e., the mean $\mu_\textrm{up}$ and the standard deviation $\sigma_\textrm{up}$. \par
In the flood modeling problem for the Mille Îles river, the regular inflow is estimated to be on the order of $Q_r=780\ \textrm{m}^3\textrm{s}^{-1}$. Our snapshots were sampled uniformly in $\Omega=[800, 1200]\ \textrm{m}^3\textrm{s}^{-1}$, targeting a critical mean value of $Q_\textrm{crit}=1000\ \textrm{m}^3\textrm{s}^{-1}$, which corresponds to an extreme flood discharge. \par
\begin{figure}[t]
  \centering
  \parbox{0.81\linewidth}{
    % \hspace*{6pt}
  \subfloat[View from afar of the mean over the whole predicted domain $\Omega$. This perspective shows an iso-contour at $h=0.05\ m$, with the flooding lines as its boundary. For illustration purposes, the mean predicted relative water height $h_\textrm{mean}$ has been pictured throughout. The green boxes show the two locations of the close-up shots (shown below in (b) and (c).]{
      \fbox{
  \includegraphics[width=\linewidth]{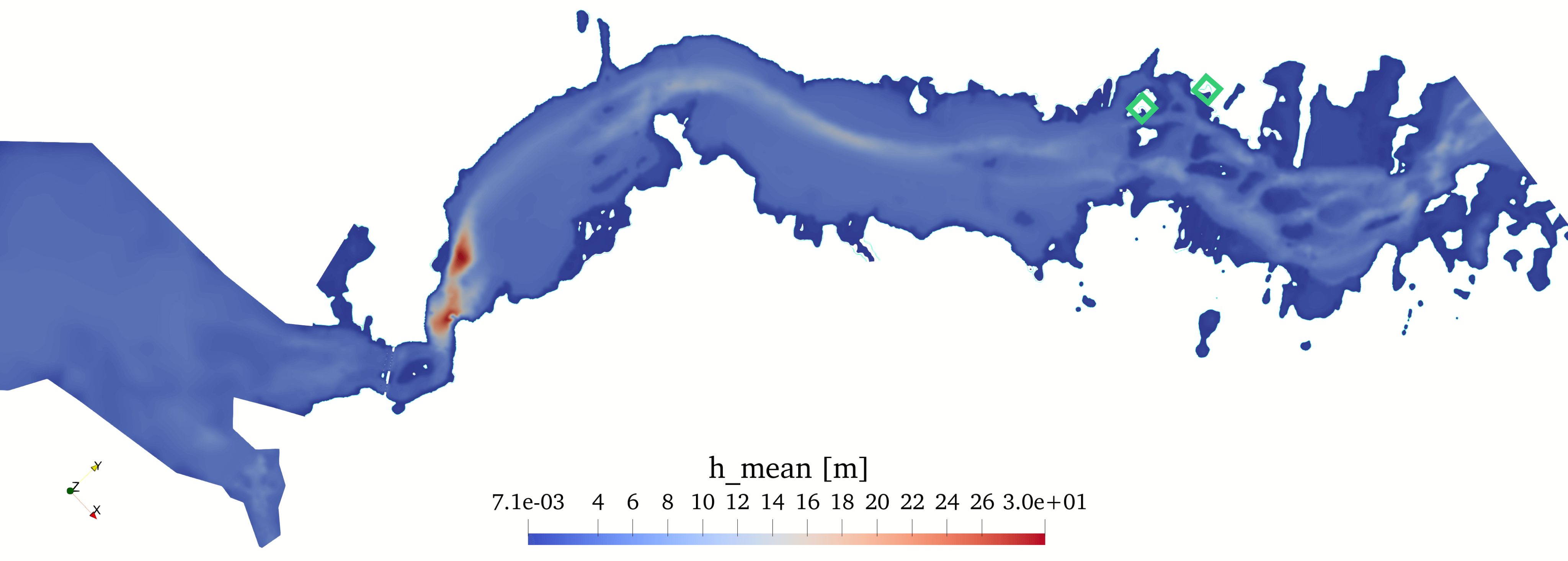}
  }
  }
}
  \parbox{\linewidth}{
    \centering
\subfloat[POD-EnsNN: Two close-up shots, showing the differences in the uncertainty around the mean water level (in blue).]{
  \includegraphics[width=0.78\linewidth]{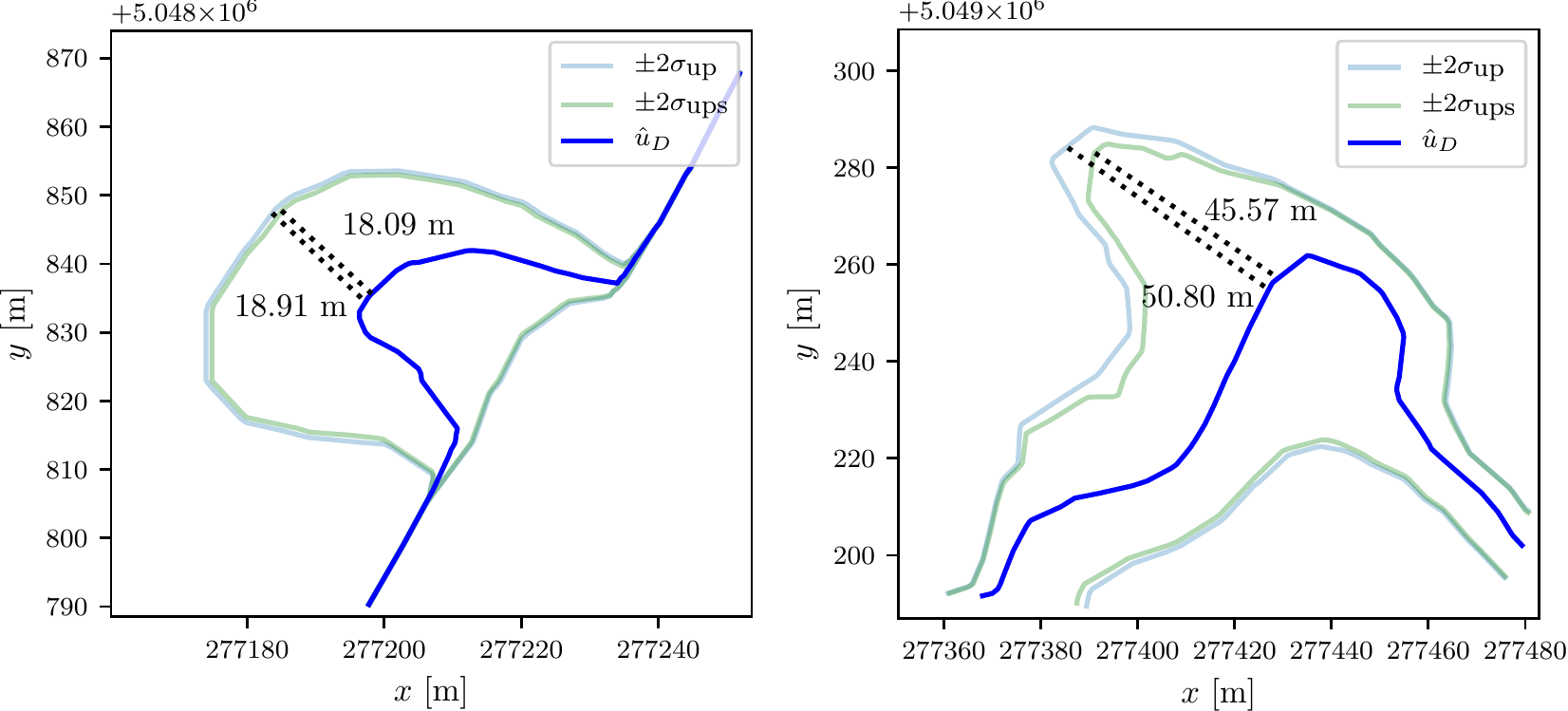}
  }
}
  \parbox{\linewidth}{
    \centering
\subfloat[POD-BNN: Two close-up shots, showing the differences in the uncertainty around the mean water level (in blue).]{
  \includegraphics[width=0.78\linewidth]{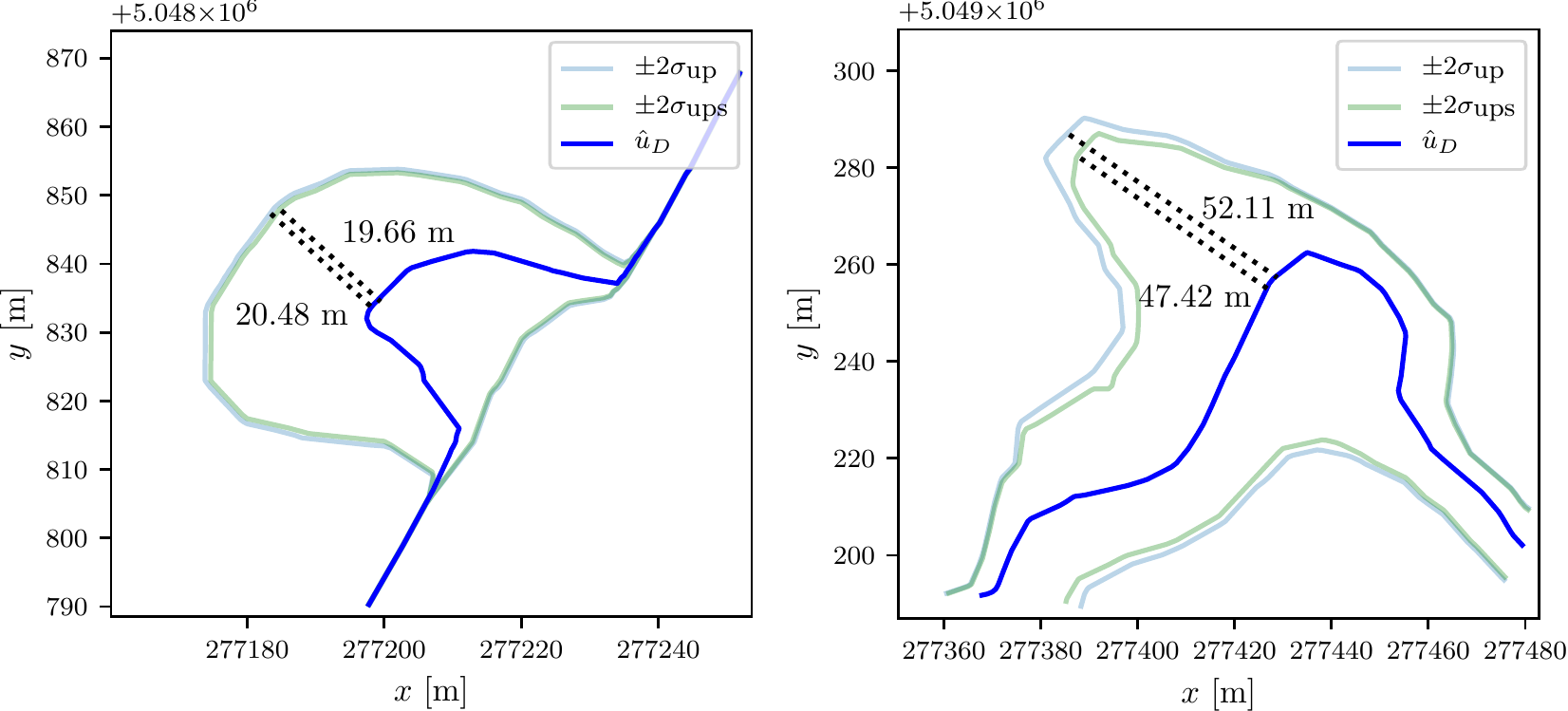}
  }
}
  \caption{Uncertainty propagation on the Mille Îles river. Flooding lines at $h=0.05\ m$ are shown in the close-up shots, with the green lines indicating  $\pm2\sigma_\textrm{ups}$, the standard deviation over each predicted mean, and the {light blue} lines representing $\pm2\sigma_\textrm{up}$, the approximation over each predicted mean and variance. Distances are measured between the mean, represented by the blue lines, and each of these quantities.}
  \label{fig:podensnn-sw-map-up}
\end{figure}

After having successfully trained and validated the model in Section \ref{sec:podensnn-app-background}, we now uniformly generate a new set of inputs $\bm{X}_\textrm{up}$ of size $N_\textrm{up}=10^3$  on $\Omega$. Running the full POD-EnsNN model, we obtain the outputs $\bm{U}_\textrm{up}$, with the quantity of interest being the water depth $h$. Since our model provides a local uncertainty for each sample point, we can approximate the statistical moments using the same mixture formulas as for sample prediction $(\mu_{*_i}, \sigma_{*_i})$, 
\begin{align}
    \mu_\textrm{up} &= \dfrac{1}{N_\textrm{up}} \sum_{i=1}^{N_\textrm{up}}\mu_{*_i},\label{eq:up-mu}\\
    \sigma^2_\textrm{up} &= \dfrac{1}{N_\textrm{up}} \sum_{i=1}^{N_\textrm{up}} (\sigma_{*_i}^2 + \mu_{*_i}^2) - \mu_\textrm{up}^2.
\end{align}
Additionally, we monitor the regular statistical standard deviation $\sigma_\textrm{ups}$ on the means, as a point of comparison, defined as
\begin{equation}
    \sigma^2_\textrm{ups} = \dfrac{1}{N_\textrm{up}} \sum_{i=1}^{N_\textrm{up}}(\mu_{*_i}-\mu_\textrm{up})^2.
\end{equation}
As a test case, the trained model of Section \ref{sec:app-river-results} produced two probabilistic flooding maps, depicted in {Fig.}\ \ref{fig:podensnn-sw-map-up}. On the very top, a broad view of the flooding at $h=0.05\ \textrm{m}$ is visible, with the predicted $h_\textrm{mean}=\mu_\textrm{up}$ from (\ref{eq:up-mu}), depicted as a color map throughout, and the green box locating the two close-up shots. These are displayed in the second row for the ensembles approach, and in the third row for the Bayesian approach, for comparison purposes. On both approaches' close-ups, there are four lines on top of the mean blue water level: two green lines, showcasing two bands of the standard deviation over the predicted means only, $\pm2\sigma_\textrm{ups}$, and two light blue lines, representing two bands of a standard deviation $\pm2\sigma_\textrm{up}$ obtained by averaging across each mean and variance predicted locally by either the POD-EnsNN or the POD-BNN framework. \par
While these lines are very close in both cases, as is well-represented by the {first} close-up shot{s (left column)} in {Fig.}\ \ref{fig:podensnn-sw-map-up}, where the measured distance is tiny, the gap does increase sometimes, for instance, in the case of the {second} close-up shot{s (right column)}, where the measured difference is somewhat significant. This attests to the potential usefulness of our approach in the realm of uncertainty propagation, as it effectively combines aleatoric (due to the distribution of $Q_0$) and epistemic (due to the modeling step) sources of uncertainty. Nonetheless, the epistemic uncertainty remains relatively minor in this case, as averaging over the quite broad domain $\Omega$ mostly wipes away the predicted local variances. \par 

\clearpage
\subsection{An unsteady case: the failure of a fictitious dam}
\label{sec:app-river-damb}
While flooding prediction in the sense of generating flooded/non-flooded limits is a handy tool for public safety, it seemed promising to apply the same framework to a time-dependent case: the results of a fictitious dam break on the same river, whose model was presented in Section \ref{sec:app-river-setup}, and which is also of interest to dam owners in general.  \par
The setup involves the same Shallow Water equations as described in Section \ref{sec:podensnn-app-background}. The domain of study is a sub-domain of the previous domain $\Omega_{xy}$, with only $N_{xy}=9734$ nodes and $18412$ elements, registering one degree of freedom per node{—the}  water elevation $\eta$. However, for this case, we consider $N_t=100$ time-steps, after the initial {time} $t=0\ \textrm{s}$ with a sampling step of $\Delta t=0.3\ \textrm{s}$ — which is different from the adaptive time-steps in the numerical solver. $N_s=100$ samples are considered for the non-spatial parameter: the water surface elevation of the inflow cross-section, considering a dried out outflow ($\eta=b$) at the moment of the dam break $s=\eta_0$, as pictured in {Fig.}\ \ref{fig:1d-case-schema}, sampled uniformly on $\Omega=[30, 31]\ \textrm{m}$. These samples comprise the training/validation dataset $\mathcal{D}$, while we consider one random test snapshot $s_\textrm{tst}$. %\par
%As training hyperparameters for the POD-EnsNN framework, we settled on a number of epochs $N_e=70,000$, a learning rate of $\tau=0.001$, L2 regularization {with} $\lambda=0.001$, and adversarial training with $\zeta=0.001$. A softplus factor of $\kappa=0.01$ had to be set for proper convergence. 
Dual POD was performed with $\epsilon_0=10^{-6}$ and $\epsilon=10^{-6}$, producing $L=60$ coefficients to be matched by half of the final layer{.}\par
\begin{figure}[t]
  \centering
  \includegraphics[width=.77\linewidth]{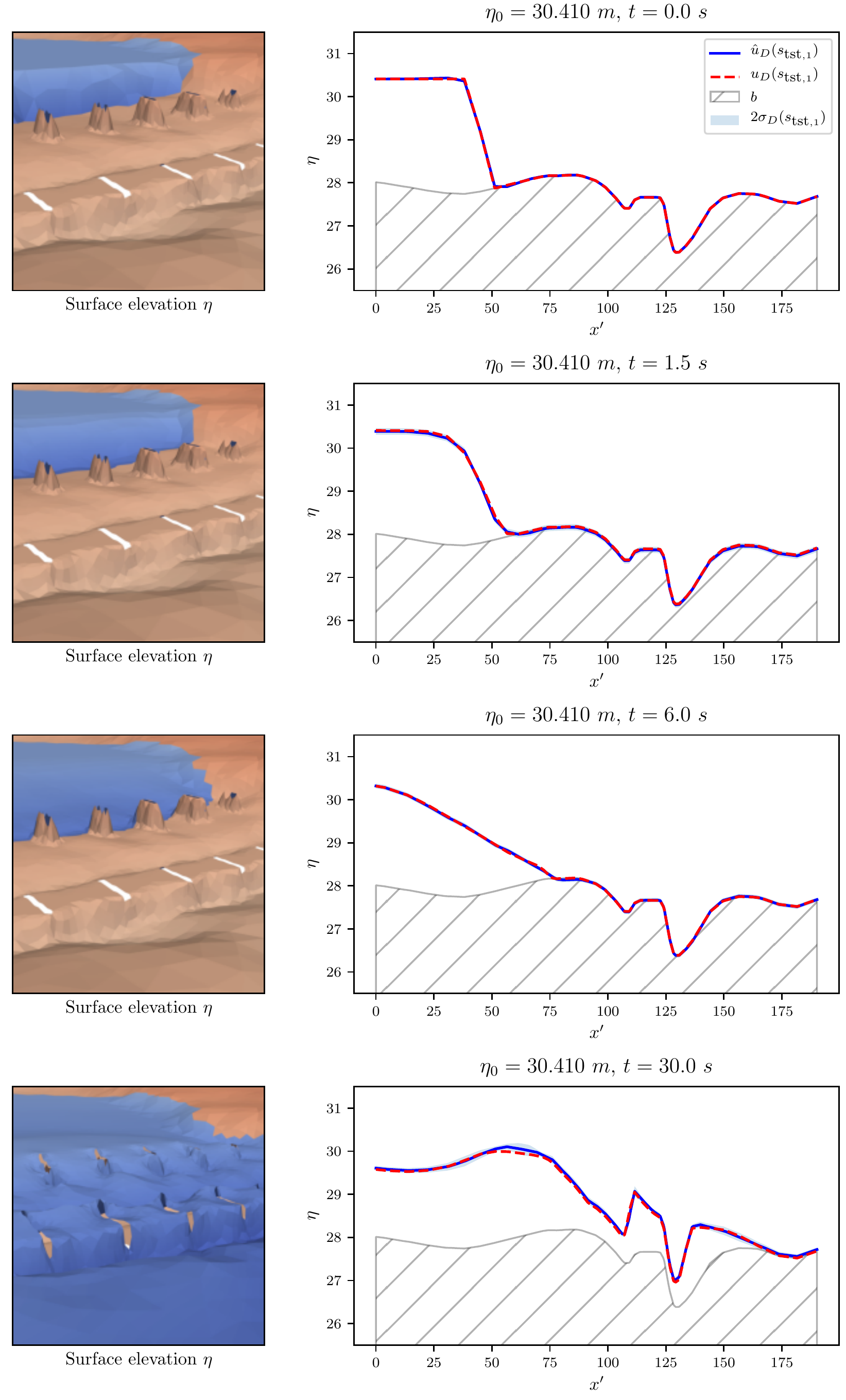}
  \caption{Dam break with POD-EnsNN. Left: color maps {of} $\eta$. Right: plots of the water elevations in the cross-section from {Fig.}\ \ref{fig:podensnn-river} of a random test snapshot on {four} time-steps, with the predictions $\hat{u}_D$, true values $u_D$, and confidence intervals, as well as the bathymetry levels in gray. The water in the river is flowing from left to right.} 
  \label{fig:podensnn-swt-samples}
\end{figure}

%, and the NN topology {consisted of} four hidden layers of $l^{(1)}=l^{(2)}=l^{(3)}=l^{(4)}=128$ neurons.  \par
The training of each model in the ensemble took close to 31 minutes on each GPU. The total, real time of the parallel process was 32 minutes.
The results are displayed in {Fig.}\ \ref{fig:podensnn-swt-samples}, in which, from top to bottom, there are  representations of four time-steps, $t=0\ \textrm{s}$, $t=1.5\ \textrm{s}$, $t=6.0\ \textrm{s}$, and $t=30.0\ \textrm{s}$. On the left, a 3D rendering of the blue river on the orange bed is displayed to help understand  the problem visually. The subsequent time-steps picture the intense dynamics that follow the initial discontinuity. The investigated cross-section, which was depicted as a green line in {Fig.}\ \ref{fig:podensnn-river}, is on the right. There is clearly a decent approximation performed by the model, considering the high nonlinearity of the problem. The uncertainty associated, obtained from (\ref{eq:article-deepens-sig}), is represented by the light blue area around the predicted blue line. 
The relative errors reached in the POD-EnsNN case were $RE_{\textrm{val}} = 9.8\%$ and $RE_{\textrm{tst}} = 2.8\%$, for validation and testing, respectively. \par
\begin{figure}[t]
  \centering
  \includegraphics[width=.77\linewidth]{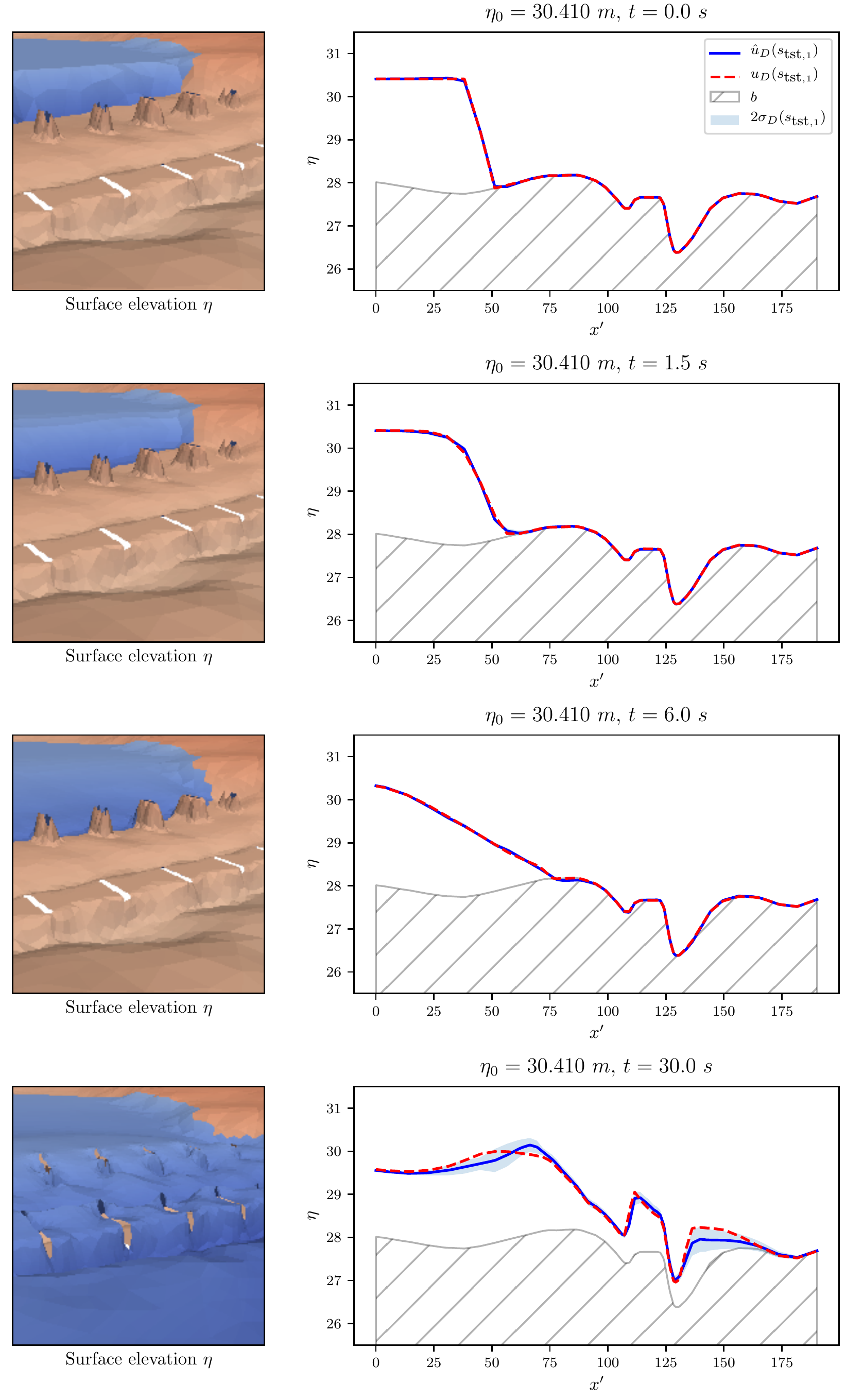}
  \caption{Dam break with POD-BNN.  Left: color maps {of} $\eta$. Right: plots of the water elevations in the cross-section from {Fig.}\ \ref{fig:podensnn-river} of a random test snapshot on {four} time-steps, with the predictions $\hat{u}_D$, true values $u_D$, and confidence intervals, as well as the bathymetry levels in gray. The water in the river is flowing from left to right.} 
  \label{fig:podbnn-swt-samples}
\end{figure}

%Subsequently, the POD-BNN framework was applied as well, with three hidden variational layers of sizes $l^{(1)}=l^{(2)}=l^{(3)}=128$, and the following hyperparameters: $N_e=150,000$ epochs, a learning rate of $\tau=0.003$, a softplus coefficient of $\kappa=0.01$, a low adversarial training {coefficient} of $\zeta=10^{-5}$, and the default ReLU activation function. The prior parameters were selected  as $\pi_0=0.5$, $\pi_1=0.2$, and $\pi_2=0.1$. 
The relative errors in the POD-BNN case were {$RE_{\textrm{val}} = 11.65\%$ and $RE_{\textrm{tst}} = 9.35\%$}, for validation and testing, respectively. %\par
% #: 150000  L: 0.0000e+00 RE_val: 0.0010 MPIW_val: 0.6364 std: 83.2990 std_val: 88.4994 T: 25.4
% RE_tst: 0.000935
% Training finished (epoch 150000): duration = 01:03:58
% As previously, focusing on the $t\in [0, 6]\ \textrm{s}$ domain of interest, these are significantly reduced to $RE_{\textrm{val}} = 0.25\%$ and $RE_{\textrm{tst}} = 0.11\%$.
The BNN training was completed by a single GPU in 1 hour 3 minutes, and the results are displayed in {Fig.}\ \ref{fig:podbnn-swt-samples}.
We can observe comparable results with those of the POD-EnsNN framework, except for  a decrease in the curve-fitting performance, as well as more considerable uncertainties, notably near the end of the simulation time. \par
% One can also observe that the first time-step shows a more substantial uncertainty overall, due to the model having a harder time fitting the sharp edges induced by the water difference, just before the fictitious dam break. Even if it is, in fact, an initial condition that is well-known, the framework does not allow for this kind of information to be provided to the model.
{To illustrate the time efficiency of the method, we measured an average computation time of 91.02 seconds per snapshot running the numerical solver CuteFlow on 2 parallel V100 GPUs. Unsurprisingly, evaluating the models in the Uncertainty Quantification context on a standard CPU took 2.72 seconds per snapshot in the POD-EnsNN case, and 6.94 seconds per snapshot in the POD-BNN case, reinforcing a crucial advantage of the offline/online approach.}

%% file: content/wrapup.tex
\section{Conclusion}
\label{sec:wrapup}
The excellent regression power of Deep Neural Networks has proved to be an asset to deploy along with Proper Orthogonal Decomposition to build reduced-order models. Their advantage is most notable when extended with recent progress in Deep Learning for a Computational Fluid Dynamics application. \par
Utilizing  1D and 2D benchmarks, we have shown that {this approach achieved excellent results in terms of accuracy, and the training times were very reasonable, even on regular computers. It combines several state-of-the-art techniques from the reduced-order modeling and machine learning fields}. Deep Ensembles and Bayesian Neural Networks were presented and compared as a way to bundle uncertainty quantification within the model. While Deep Ensembles require multiple training times, {which} can easily be done in parallel, Bayesian Neural Networks are trained only once, which can be a decisive advantage,  especially in terms of the available computational resources. However, {one has to consider} the time spent finding the right hyperparameters for the Bayesian approach, { its computational cost that led to longer training time while often including fewer neurons per layer (or even less layers),} and in some cases resulted in less accurate results, notably in time-dependent settings, compared to the relatively \emph{plug-and-play} behavior of  ensembles, which we strongly recommend. \par

It has also been shown that while standard NNs were rapidly predicting inaccurate quantities when brought out of the training scope, adopting an uncertainty-enabled approach {could produce} the intended warning. This is where the uncertainty-enabled approach especially shows its worth, since the models are capable of producing flooding lines within a predicted confidence interval, either in a local prediction manner, such as a real-time context where these lines need to be computed for a new parameter, or in a more global, uncertainty propagation case, where there is an unknown extreme and critical inflow, and thus the consequences of profound changes in this quantity need to be assessed. Instead of computing the statistical moments of the output distribution from the point estimates of a surrogate model, such as a standard Neural Network, the model  considers the contribution of each local uncertainty and, therefore, produces a more extensive and safer confidence area around the predicted flooding line. \par

Future work will focus on stabilizing the Bayesian Neural Networks approach, which still requires a much finer tuning compared to the flexibility of Deep Ensembles. Additionally, applying both methods to refined meshes will require the POD step to be performed on a sub-domain basis to avoid memory issues{. This will allow to} better assess the performance of the uncertainties-aware POD-NN framework in a more complicated engineering problem. While the reduced-basis compression helped in the handling of the relatively large space domain of the river, the number of POD modes still has to grow with the problem's size {for a fixed tolerance}; hence additional research is needed to better understand the impact of the curse of dimensionality on this framework. The Bayesian approach also faced convergence issues for problems showcasing discontinuities in  time-dependent settings, and decent results could only be reached by using a different activation function in the test case of Section \ref{sec:app-riemann}. 
For long time-dependent simulations, error accumulation is known to corrupt results over time in the standard POD. However, using a multiple POD basis can enhance the accuracy and reduce the computing resources needed to apply the SVD algorithm on high-dimensional snapshot matrices \cite{zokagoa}. The multi-POD {method} can easily be implemented in the framework presented in the current paper. 
Flood modeling offers many future exploration directions, as  various other parameters have a direct influence on the results, such as the Manning roughness of the bed, as well as its elevation, and are also complicated by measurement uncertainties.

%% file: main.bbl
\begin{thebibliography}{10}

\bibitem{szegedy2017inception}
Christian Szegedy, Sergey Ioffe, Vincent Vanhoucke, and Alexander~A Alemi.
\newblock {Inception-v4, inception-resnet and the impact of residual
  connections on learning}.
\newblock In {\em Thirty-First AAAI Conference on Artificial Intelligence},
  2017.

\bibitem{Mikolov2013word2vec}
Tomas Mikolov, Ilya Sutskever, Kai Chen, Greg Corrado, and Jeffrey Dean.
\newblock {Distributed Representations of Words and Phrases and Their
  Compositionality}.
\newblock In {\em Proceedings of the 26th International Conference on Neural
  Information Processing Systems - Volume 2}, NIPS'13, pages 3111--3119, Red
  Hook, NY, USA, 2013. Curran Associates Inc.

\bibitem{karras2019analyzing}
Tero Karras, Samuli Laine, Miika Aittala, Janne Hellsten, Jaakko Lehtinen, and
  Timo Aila.
\newblock {Analyzing and improving the image quality of stylegan}.
\newblock {\em arXiv preprint arXiv:1912.04958}, 2019.

\bibitem{Stokes2020}
Jonathan~M. Stokes, Kevin Yang, Kyle Swanson, Wengong Jin, Andres
  Cubillos-Ruiz, Nina~M. Donghia, Craig~R. MacNair, Shawn French, Lindsey~A.
  Carfrae, Zohar Bloom-Ackerman, Victoria~M. Tran, Anush Chiappino-Pepe,
  Ahmed~H. Badran, Ian~W. Andrews, Emma~J. Chory, George~M. Church, Eric~D.
  Brown, Tommi~S. Jaakkola, Regina Barzilay, and James~J. Collins.
\newblock {A Deep Learning Approach to Antibiotic Discovery}.
\newblock {\em Cell}, 180(4):688--702.e13, feb 2020.

\bibitem{Benner2015}
Peter Benner, Serkan Gugercin, and Karen Willcox.
\newblock {A survey of projection-based model reduction methods for parametric
  dynamical systems}.
\newblock {\em SIAM Review}, 57(4):483--531, 2015.

\bibitem{Holmes1997}
Philip~J. Holmes, John~L. Lumley, Gal Berkooz, Jonathan~C. Mattingly, and
  Ralf~W. Wittenberg.
\newblock {Low-dimensional models of coherent structures in turbulence}.
\newblock {\em Physics Report}, 287(4):337--384, 1997.

\bibitem{sirovich1987turbulence}
Lawrence Sirovich.
\newblock {Turbulence and the dynamics of coherent structures. I. Coherent
  structures}.
\newblock {\em Quarterly of applied mathematics}, 45(3):561--571, 1987.

\bibitem{Burkardt2006}
John Burkardt, Max Gunzburger, and Hyung~Chun Lee.
\newblock {Centroidal voronoi tessellation-based reduced-order modeling of
  complex systems}.
\newblock {\em SIAM Journal on Scientific Computing}, 28(2):459--484, 2006.

\bibitem{Couplet2005}
M.~Couplet, C.~Basdevant, and P.~Sagaut.
\newblock {Calibrated reduced-order POD-Galerkin system for fluid flow
  modelling}.
\newblock {\em Journal of Computational Physics}, 207(1):192--220, jul 2005.

\bibitem{zok2012cmame}
Jean~Marie Zokagoa and Azzeddine Soulaimani.
\newblock {A POD-based reduced-order model for free surface shallow water flows
  over real bathymetries for Monte-Carlo-type applications}.
\newblock {\em Computer Methods in Applied Mechanics and Engineering}, s
  221–222:1--23, 2012.

\bibitem{zokagoa}
Jean~Marie Zokagoa and Azzeddine Soulaimani.
\newblock {A POD-based reduced-order model for uncertainty analyses in shallow
  water flows}.
\newblock {\em International Journal of Computational Fluid Dynamics}, pages
  1--15, 2018.

\bibitem{Hestaven2018}
J.S. Hesthaven and S.~Ubbiali.
\newblock {Non-intrusive reduced order modeling of nonlinear problems using
  neural networks}.
\newblock {\em Journal of Computational Physics}, 363:55--78, jun 2018.

\bibitem{Wang2019}
Qian Wang, Jan~S. Hesthaven, and Deep Ray.
\newblock {Non-intrusive reduced order modeling of unsteady flows using
  artificial neural networks with application to a combustion problem}.
\newblock {\em Journal of Computational Physics}, 384:289--307, may 2019.

\bibitem{Ijzerman2000}
Wl~Ijzerman.
\newblock {Signal Representation and Modeling of Spatial Structures in Fluids}.
\newblock 2000.

\bibitem{Brunton2016DiscoveringGov}
Steven~L. Brunton, Joshua~L. Proctor, and J.~Nathan Kutz.
\newblock {Discovering governing equations from data by sparse identification
  of nonlinear dynamical systems}.
\newblock {\em Proceedings of the National Academy of Sciences},
  113(15):3932--3937, apr 2016.

\bibitem{Kutz2017}
J~Nathan Kutz.
\newblock {Deep learning in fluid dynamics}.
\newblock {\em Journal of Fluid Mechanics}, 814:1--4, 2017.

\bibitem{Carlberg2019}
Kevin~T. Carlberg, Antony Jameson, Mykel~J. Kochenderfer, Jeremy Morton, Liqian
  Peng, and Freddie~D. Witherden.
\newblock {Recovering missing CFD data for high-order discretizations using
  deep neural networks and dynamics learning}.
\newblock {\em Journal of Computational Physics}, 395:105--124, oct 2019.

\bibitem{Tao2019}
Jun Tao and Gang Sun.
\newblock {Application of deep learning based multi-fidelity surrogate model to
  robust aerodynamic design optimization}.
\newblock {\em Aerospace Science and Technology}, 92:722--737, sep 2019.

\bibitem{Hanna2020}
Botros~N. Hanna, Nam~T. Dinh, Robert~W. Youngblood, and Igor~A. Bolotnov.
\newblock {Machine-learning based error prediction approach for coarse-grid
  Computational Fluid Dynamics (CG-CFD)}.
\newblock {\em Progress in Nuclear Energy}, 118:103140, jan 2020.

\bibitem{Despres2020}
Bruno Despr{\'{e}}s and Herv{\'{e}} Jourdren.
\newblock {Machine Learning design of Volume of Fluid schemes for compressible
  flows}.
\newblock {\em Journal of Computational Physics}, 408:109275, may 2020.

\bibitem{Brunton2019}
Steven~L Brunton and J~Nathan Kutz.
\newblock {\em {Data-Driven Science and Engineering}}.
\newblock Cambridge University Press, jan 2019.

\bibitem{owhadi2015bayesian}
Houman Owhadi.
\newblock Bayesian numerical homogenization, 2015.

\bibitem{Raissi2017MLODE}
Maziar Raissi, Paris Perdikaris, and George~Em Karniadakis.
\newblock {Machine learning of linear differential equations using Gaussian
  processes}.
\newblock {\em Journal of Computational Physics}, 348:683--693, nov 2017.

\bibitem{Raissi2019PINNs}
M.~Raissi, P.~Perdikaris, and G.E. Karniadakis.
\newblock {Physics-informed neural networks: A deep learning framework for
  solving forward and inverse problems involving nonlinear partial differential
  equations}.
\newblock {\em Journal of Computational Physics}, 378:686--707, feb 2019.

\bibitem{Raissi2019DeepVIV}
Maziar Raissi, Zhicheng Wang, Michael~S. Triantafyllou, and George~Em
  Karniadakis.
\newblock {Deep learning of vortex-induced vibrations}.
\newblock {\em Journal of Fluid Mechanics}, 861:119--137, feb 2019.

\bibitem{rumelhart1985learning}
David~E Rumelhart, Geoffrey~E Hinton, and Ronald~J Williams.
\newblock {Learning internal representations by error propagation}.
\newblock Technical report, California Univ San Diego La Jolla Inst for
  Cognitive Science, 1985.

\bibitem{lstm1997}
Sepp Hochreiter and J{\"{u}}rgen Schmidhuber.
\newblock {Long Short-Term Memory}.
\newblock {\em Neural Computation}, 9(8):1735--1780, 1997.

\bibitem{Hu2019}
R.~Hu, F.~Fang, C.C. Pain, and I.M. Navon.
\newblock {Rapid spatio-temporal flood prediction and uncertainty
  quantification using a deep learning method}.
\newblock {\em Journal of Hydrology}, 575:911--920, aug 2019.

\bibitem{McDermott2019}
Patrick~L. McDermott and Christopher~K. Wikle.
\newblock {Bayesian recurrent neural network models for forecasting and
  quantifying uncertainty in spatial-temporal data}.
\newblock {\em Entropy}, 21(2), 2019.

\bibitem{williamhsieh2009}
William~W Hsieh.
\newblock {\em {Machine Learning Methods in the Environmental Sciences: Neural
  Networks and Kernels}}.
\newblock Cambridge University Press, aug 2009.

\bibitem{lakshminarayanan2017simple}
Balaji Lakshminarayanan, Alexander Pritzel, and Charles Blundell.
\newblock {Simple and scalable predictive uncertainty estimation using deep
  ensembles}.
\newblock In {\em Advances in Neural Information Processing Systems}, pages
  6402--6413, 2017.

\bibitem{Valdenegro-Toro2019}
Matias Valdenegro-Toro.
\newblock {Deep Sub-Ensembles for Fast Uncertainty Estimation in Image
  Classification}.
\newblock (NeurIPS), 2019.

\bibitem{snoek2019trust}
Jasper Snoek, Yaniv Ovadia, Emily Fertig, Balaji Lakshminarayanan, Sebastian
  Nowozin, D~Sculley, Joshua Dillon, Jie Ren, and Zachary Nado.
\newblock {Can you trust your model's uncertainty? Evaluating predictive
  uncertainty under dataset shift}.
\newblock In {\em Advances in Neural Information Processing Systems}, pages
  13969--13980, 2019.

\bibitem{Mackay1995}
David J~C Mackay.
\newblock {Probable networks and plausible predictions — a review of
  practical Bayesian methods for supervised neural networks}.
\newblock {\em Network: Computation in Neural Systems}, 6(3):469--505, 1995.

\bibitem{Barber1998}
David Barber and Christopher Bishop.
\newblock {Ensemble learning in Bayesian neural networks}.
\newblock {\em Nato ASI Series F Computer and Systems Sciences}, (Bishop
  1995):215--237, 1998.

\bibitem{NIPS2011_4329}
Alex Graves.
\newblock {Practical Variational Inference for Neural Networks}.
\newblock In J~Shawe-Taylor, R~S Zemel, P~L Bartlett, F~Pereira, and K~Q
  Weinberger, editors, {\em Advances in Neural Information Processing Systems
  24}, pages 2348--2356. Curran Associates, Inc., 2011.

\bibitem{pmlr-v37-hernandez-lobatoc15}
Jose~Miguel Hernandez-Lobato and Ryan Adams.
\newblock {Probabilistic Backpropagation for Scalable Learning of Bayesian
  Neural Networks}.
\newblock In Francis Bach and David Blei, editors, {\em Proceedings of the 32nd
  International Conference on Machine Learning}, volume~37 of {\em Proceedings
  of Machine Learning Research}, pages 1861--1869, Lille, France, 2015. PMLR.

\bibitem{Blundell2015}
Charles Blundell, Julien Cornebise, Koray Kavukcuoglu, and Daan Wierstra.
\newblock {Weight Uncertainty in Neural Networks}.
\newblock In {\em Proceedings of the 32nd International Conference on
  International Conference on Machine Learning - Volume 37}, ICML'15, pages
  1613--1622. JMLR.org, 2015.

\bibitem{Hinton1993varinf}
Geoffrey~E Hinton and Drew van Camp.
\newblock {Keeping the Neural Networks Simple by Minimizing the Description
  Length of the Weights}.
\newblock In {\em Proceedings of the Sixth Annual Conference on Computational
  Learning Theory}, COLT '93, pages 5--13, New York, NY, USA, 1993. Association
  for Computing Machinery.

\bibitem{reviewer1-a}
Liang Yan and Tao Zhou.
\newblock An adaptive surrogate modeling based on deep neural networks for
  large-scale bayesian inverse problems, 2019.

\bibitem{reviewer1-b}
Liu Yang, Xuhui Meng, and George~Em Karniadakis.
\newblock B-pinns: Bayesian physics-informed neural networks for forward and
  inverse pde problems with noisy data, 2020.

\bibitem{nix1994estimating}
David~A Nix and Andreas~S Weigend.
\newblock {Estimating the mean and variance of the target probability
  distribution}.
\newblock In {\em Proceedings of 1994 ieee international conference on neural
  networks (ICNN'94)}, volume~1, pages 55--60. IEEE, 1994.

\bibitem{UncertaintiesBayesianDV}
Alex Kendall and Yarin Gal.
\newblock {What Uncertainties Do We Need in Bayesian Deep Learning for Computer
  Vision?}
\newblock In I~Guyon, U~V Luxburg, S~Bengio, H~Wallach, R~Fergus,
  S~Vishwanathan, and R~Garnett, editors, {\em Advances in Neural Information
  Processing Systems 30}, pages 5574--5584. Curran Associates, Inc., 2017.

\bibitem{Krogh1992WeightDecay}
Anders Krogh and John~A Hertz.
\newblock {A Simple Weight Decay Can Improve Generalization}.
\newblock In J~E Moody, S~J Hanson, and R~P Lippmann, editors, {\em Advances in
  Neural Information Processing Systems 4}, pages 950--957. Morgan-Kaufmann,
  1992.

\bibitem{kingma2014adam}
Diederik~P Kingma and Jimmy Ba.
\newblock {Adam: A method for stochastic optimization}.
\newblock {\em arXiv preprint arXiv:1412.6980}, 2014.

\bibitem{Rumelhart1986}
David~E Rumelhart, Geoffrey~E Hinton, and Ronald~J Williams.
\newblock {Learning representations by back-propagating errors}.
\newblock {\em Nature}, 323(6088):533--536, 1986.

\bibitem{Linnainmaa1976}
Seppo Linnainmaa.
\newblock {Taylor expansion of the accumulated rounding error}.
\newblock {\em BIT Numerical Mathematics}, 16(2):146--160, jun 1976.

\bibitem{Sergeev2018horovod}
Alexander Sergeev and Mike {Del Balso}.
\newblock {Horovod: fast and easy distributed deep learning in TensorFlow}.
\newblock feb 2018.

\bibitem{Yao2019}
Jiayu Yao, Weiwei Pan, Soumya Ghosh, and Finale Doshi-Velez.
\newblock {Quality of Uncertainty Quantification for Bayesian Neural Network
  Inference}.
\newblock 2019.

\bibitem{Szegedy2014advtra}
Christian Szegedy, Wojciech Zaremba, Ilya Sutskever, Joan Bruna, Dumitru Erhan,
  Ian Goodfellow, and Rob Fergus.
\newblock {Intriguing properties of neural networks}.
\newblock In {\em International Conference on Learning Representations}, 2014.

\bibitem{Goodfellow2014advtra}
Ian~J. Goodfellow, Jonathon Shlens, and Christian Szegedy.
\newblock {Explaining and Harnessing Adversarial Examples}.
\newblock dec 2014.

\bibitem{Goodfellow2014GANs}
Ian~J. Goodfellow, Jean Pouget-Abadie, Mehdi Mirza, Bing Xu, David
  Warde-Farley, Sherjil Ozair, Aaron Courville, and Yoshua Bengio.
\newblock {Generative Adversarial Networks}.
\newblock jun 2014.

\bibitem{Neal1993}
Radford~M Neal.
\newblock {Probabilistic Inference Using Markov Chain Monte Carlo Methods}.
\newblock Technical report, 1993.

\bibitem{Neal1995Thesis}
Radford~M Neal.
\newblock {Bayesian Learning for Neural Networks}.
\newblock Technical report, 1995.

\bibitem{Goodfellow-et-al-2016}
Ian Goodfellow, Yoshua Bengio, and Aaron Courville.
\newblock {\em {Deep Learning}}.
\newblock MIT Press, 2016.

\bibitem{Kingma2014}
Diederik~P. Kingma and Max Welling.
\newblock {Auto-encoding variational bayes}.
\newblock In {\em 2nd International Conference on Learning Representations,
  ICLR 2014 - Conference Track Proceedings}. International Conference on
  Learning Representations, ICLR, dec 2014.

\bibitem{Abadi2016TensorFlow}
Martin Abadi.
\newblock {TensorFlow: A System for Large-Scale Machine Learning}, 2016.

\bibitem{Dillon2017}
Joshua~V. Dillon, Ian Langmore, Dustin Tran, Eugene Brevdo, Srinivas Vasudevan,
  Dave Moore, Brian Patton, Alex Alemi, Matt Hoffman, and Rif~A. Saurous.
\newblock {TensorFlow Distributions}.
\newblock nov 2017.

\bibitem{Krasser2020blog}
Martin Krasser.
\newblock {Variational inference in Bayesian neural networks - Martin Krasser's
  Blog}, 2019.

\bibitem{Lam:2015:Numba}
Siu~Kwan Lam, Antoine Pitrou, and Stanley Seibert.
\newblock {Numba: A LLVM-based Python JIT Compiler}.
\newblock In {\em Proceedings of the Second Workshop on the LLVM Compiler
  Infrastructure in HPC}, LLVM '15, pages 7:1----7:6, New York, NY, USA, 2015.
  ACM.

\bibitem{Sun2019}
Xiang Sun, Xiaomin Pan, and Jung-Il Choi.
\newblock {A non-intrusive reduced-order modeling method using polynomial chaos
  expansion}.
\newblock 2019.

\bibitem{Toro2001}
E.~F. Toro.
\newblock {\em {Shock-capturing methods for free-surface shallow flows}}.
\newblock John Wiley, 2001.

\bibitem{Galland1991telemac}
Jean-Charles Galland, Nicole Goutal, and Jean-Michel Hervouet.
\newblock {TELEMAC: A new numerical model for solving shallow water equations}.
\newblock {\em Advances in Water Resources}, 14(3):138--148, jun 1991.

\bibitem{Wu1999}
Chao Wu, Guofu Huang, and Yonghong Zheng.
\newblock {Theoretical solution of dam-break shock wave}.
\newblock {\em Journal of Hydraulic Engineering}, 125(11):1210--1214, 1999.

\bibitem{Llanas2008}
B.~Llanas, S.~Lantar{\'{o}}n, and F.~J. S{\'{a}}inz.
\newblock {Constructive approximation of discontinuous functions by neural
  networks}.
\newblock {\em Neural Processing Letters}, 27(3):209--226, 2008.

\bibitem{ahrens2005paraview}
James Ahrens, Berk Geveci, and Charles Law.
\newblock {Paraview: An end-user tool for large data visualization}.
\newblock 2005.

\end{thebibliography}
